\documentclass[aps,pra,twocolumn,superscriptaddress,10pt,article,nofootinbib,showpacs,longbibliography]{revtex4-1}
\usepackage{amsfonts}
\usepackage{amsmath}
\usepackage{amsthm}
\usepackage{braket}
\usepackage{graphicx}
\usepackage{hyperref}
\hypersetup{bookmarks,
            colorlinks,
           citecolor=red,
            filecolor=blue,
           urlcolor=blue,
           pdfpagemode=None}
\usepackage[dvipsnames]{xcolor}
\usepackage{amssymb}
\usepackage{dsfont}
\usepackage{verbatim}
\usepackage{amsmath,amscd}
\usepackage{multirow}
\usepackage[normalem]{ulem}
\usepackage{bm}
\usepackage{siunitx}
\usepackage{physics}
\def \ua{\uparrow}

\def \d{{\mathbf d}}
\def \k{{\mathbf k}}
\def \n{{\mathbf n}}
\def \p{{\mathbf p}}
\def \q{{\mathbf q}}
\def \Q{{\mathbf Q}}
\def \A{{\mathbf A}}
\def \r{{\mathbf r}}
\def \v{{\mathbf v}}
\def \rp{{\mathbf r}^\prime}
\def \s{{\mathbf s}}
\def \G{{\mathbf G}}
\def \R{{\mathbf R}}
\def \T{{\mathcal T}}
\def \K{{\mathbf K}}
\def \Kp{{\mathbf K^\prime}}
\def \R{{\mathbf R}}
\def \beq{\begin{eqnarray}}
\def \eeq{\end{eqnarray}}
\def \nn{\nonumber \\}

\begin{document}

\title{Inter-valley coherent order and isospin fluctuation mediated superconductivity in rhombohedral trilayer graphene}
\author{Shubhayu Chatterjee}
\affiliation{Department of Physics, University of California, Berkeley, CA 94720, USA}
\author{Taige Wang}
\affiliation{Department of Physics, University of California, Berkeley, CA 94720, USA}
\author{Erez Berg}
\affiliation{Department of Condensed Matter Physics, Weizmann Institute of Science, Rehovot 76100, Israel}
\author{Michael P. Zaletel}
\affiliation{Department of Physics, University of California, Berkeley, CA 94720, USA}
\affiliation{Materials Sciences Division, Lawrence Berkeley National Laboratory, Berkeley, California 94720}

\begin{abstract}
Superconductivity was recently discovered in rhombohedral trilayer graphene (RTG) in the absence of a moir\'e potential.
Superconductivity is observed proximate to a metallic state with reduced isospin symmetry, but it remains unknown whether this is a coincidence or a key ingredient for superconductivity. 
Using a Hartree-Fock analysis and constraints from experiments, we argue that the symmetry breaking is inter-valley coherent (IVC) in nature.
We evaluate IVC fluctuations as a possible pairing glue, and find that they  lead to chiral unconventional superconductivity when the fluctuations are strong. 
We further elucidate how the inter-valley Hund's coupling determines the spin-structure of the IVC ground state and breaks the degeneracy between spin-singlet and triplet superconductivity. 
Remarkably, if the normal state is spin-unpolarized, we find that a ferromagnetic Hund's coupling favors spin-singlet superconductivity, in agreement with experiments.
Instead, if the normal state is spin-polarized, then IVC fluctuations lead to spin-triplet pairing.
\end{abstract}

\maketitle

\section{Introduction}
The experimental discovery of robust superconductivity in graphene-based moir\'e heterostructures has placed graphene in the  spotlight for studying the physics of strong electronic correlations \cite{cao2018,yankowitz2019,lu2019,Arora2020,hao2021,park2021,AndreiMacdonaldreview,balentsReview}. 
Very recently, superconductivity was observed in an even simpler system --- ABC-stacked rhombohedral trilayer graphene (RTG) \textit{without any moir\'e pattern} \cite{Zhou_ABCSC}.
Near charge neutrality, just like monolayer graphene, the low-energy electrons of RTG are characterized by an isospin index that includes valley and spin \cite{Zhang2010,Jung2013}.
Superconductivity emerges on the cusp of isospin symmetry breaking transitions in hole-doped RTG in the presence of a perpendicular displacement field.
In particular, there are two superconducting phases (referred to as SC1 and SC2 in Ref.~\onlinecite{Zhou_ABCSC}) that flank two distinct isospin symmetry-broken phases [called a `partially isospin polarized' (PIP) phase in Ref.~\onlinecite{Zhou_ABCSC}].
While SC1 is suppressed by in-plane Zeeman fields and respects the Pauli paramagnetic limit \cite{Chandrasekhar,Clogston}, SC2 appears to strongly violate this limit.
Further, the low level of disorder in the sample, as evidenced by $\mu$m-scale mean-free path of electrons, leaves open the possibility for unconventional superconductors.

These remarkable observations naturally lead to important questions.
What is the nature of isospin symmetry-breaking in the metallic phases of RTG?
What are the pairing symmetries of SC1 and SC2 that emerge on the verge of isospin symmetry-breaking?
What role, if any, do electronic correlations play in aiding or suppressing superconductivity?

In this paper we propose isospin fluctuations as an all-electronic mechanism of superconductivity in RTG.
We first argue that the experimental data strongly constrains the nature of spontaneous symmetry-breaking in the correlated metallic states.
In particular, we demonstrate using self-consistent Hartree-Fock calculations that a promising candidate state near SC1 is an inter-valley coherent (IVC) metal that spontaneously breaks the U(1)$_v$ valley conservation symmetry, but lacks net spin or valley-polarization. 
Depending on the sign of the inter-valley Hund's coupling, such an IVC metal is either a time-reversal symmetric spin-singlet charge-density wave (CDW), or a collinear spin-density wave (SDW) that breaks time-reversal and global spin-rotation symmetry: both triple the unit cell \cite{HalperinRice,Aleiner2007}.
Near SC2, we propose that a spin-polarized IVC state, which microscopically corresponds to a ferromagnetic CDW, may be realized. 

Next, we investigate superconducting instabilities that arise from fluctuations of the IVC order parameter.
Interestingly, we find that the leading superconducting instability, as determined by solving a mean-field Bardeen-Cooper-Schrieffer (BCS) gap equation, shows a transition as a function of the IVC correlation length $\xi_{\textrm{IVC}}$. 
At large $\xi_{\textrm{IVC}}$, i.e, closer to criticality, the dominant instability is towards a chiral fully-gapped superconductor, while at smaller $\xi_{\textrm{IVC}}$ the dominant instability is towards a non-chiral nodal superconductor. 
Because of the presence of an additional valley degree of freedom, both these states could either be spin-singlet or triplet.
Within a model accounting only for intra-valley Coulomb scattering, spin-singlet and triplet superconductors are degenerate due to an enhanced SU(2)$_+ \times$ SU(2)$_-$ spin-rotation symmetry (valleys labeled by $\pm$). However, we argue that the inter-valley Hund's coupling arising from lattice-scale effects determines the spin-structure. 
The existence of  valley-unpolarized, spin-polarized phases in RTG implies that the Hund's coupling is ferromagnetic. 
Remarkably, we find that such a Hund's coupling prefers a \textit{spin-singlet} superconductor, consistent with SC1.
In contrast, SC2 is likely a non-unitary spin-triplet which inherits the spin-polarization of the ferromagnetic normal state. 

The rest of this paper is organized as follows.
In section \ref{sec:Model}, we introduce the interacting Hamiltonian for RTG and its symmetries.
In section \ref{sec:IVC}, we argue in favor of an IVC phase near SC1 using both Hartree-Fock and analytical calculations, and discuss its real-space and momentum space structures.
In section \ref{sec:Hunds}, we discuss how the inter-valley Hund's coupling has an unusual form which favors spin-triplet IVC over spin-polarization when ferromagnetic.
In section \ref{sec:SC}, we analyze superconducting instabilities arising from IVC fluctuations, and study the role of the Hund's term in splitting the degeneracy between spin-singlet and triplet superconductors.
We conclude in section \ref{sec:Outlook} with a summary of our main results, comparison to experimental data and recent theoretical work, and an outlook.

\section{Results}
\subsection{Hamiltonian and symmetries}
\label{sec:Model}

\begin{figure}[t]
\includegraphics[width=0.47\textwidth]{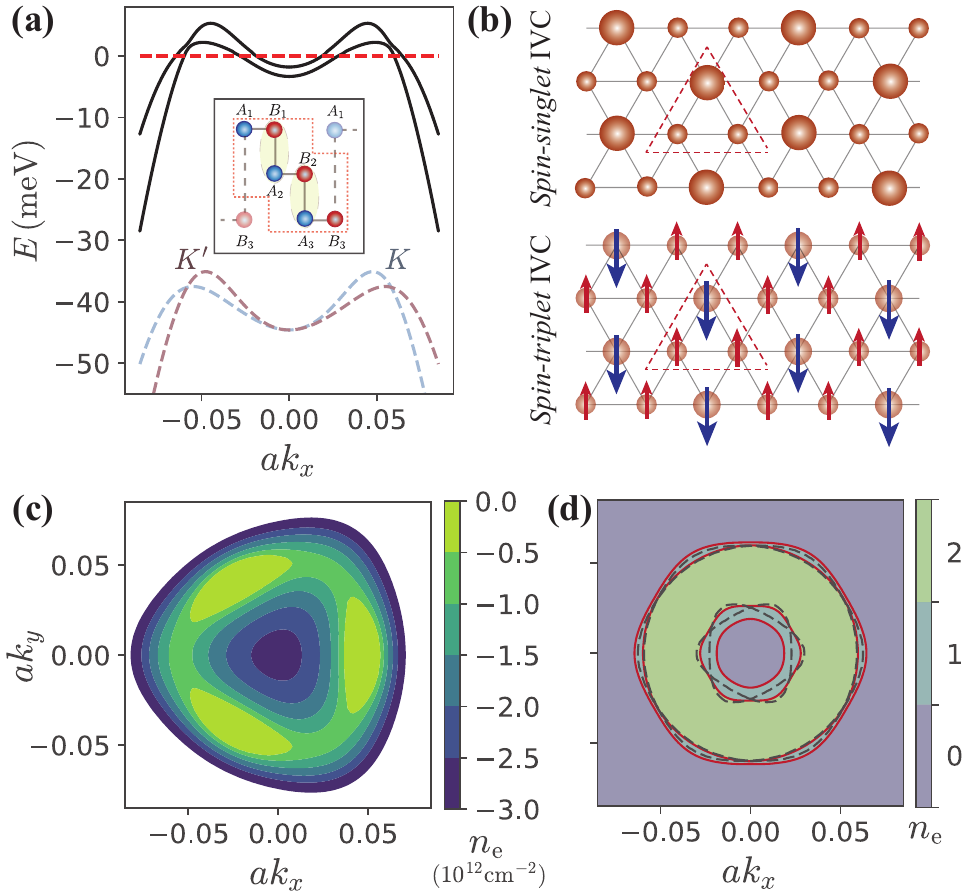}
\centering
\caption{Band structure and real space description of inter-valley coherent (IVC) phases (a) 1D cut of the Hartree-Fock band structure of the for a self-consistent HF IVC state close to the onset of the spin-singlet/triplet IVC phase, with the bare band structure in the two valleys shown by dotted lines below (displaced below for clarity). Inset shows the unit-cell of RTG, with $\{B_i, A_{i+1}\}$ being strongly hybridized ($i=1,2$) such that the active sublattices $A_1/B_3$ form a triangular lattice.
(b) Real space structure of the spin-singlet/CDW IVC (left) and the spin-triplet/SDW IVC (right) on the effective triangular lattice, with dotted lines showing the tripled unit cell in each case.
(c) Fermi surface of the single-particle band structure in the $K$ valley at different electron densities $n_e$, assuming fourfold isospin degeneracy.
(d) 2D depiction of the reconstructed Fermi surface of the same IVC state as panel (a), showing two annular pockets. Different colors indicate the number of filled HF bands. The dashed black curves are the Fermi surfaces of HF self-consistent \emph{symmetric} metal at identical filling.}
\label{fig:FS}
\end{figure}

\begin{figure*}[t]
    \centering
    \includegraphics[width=0.8\textwidth]{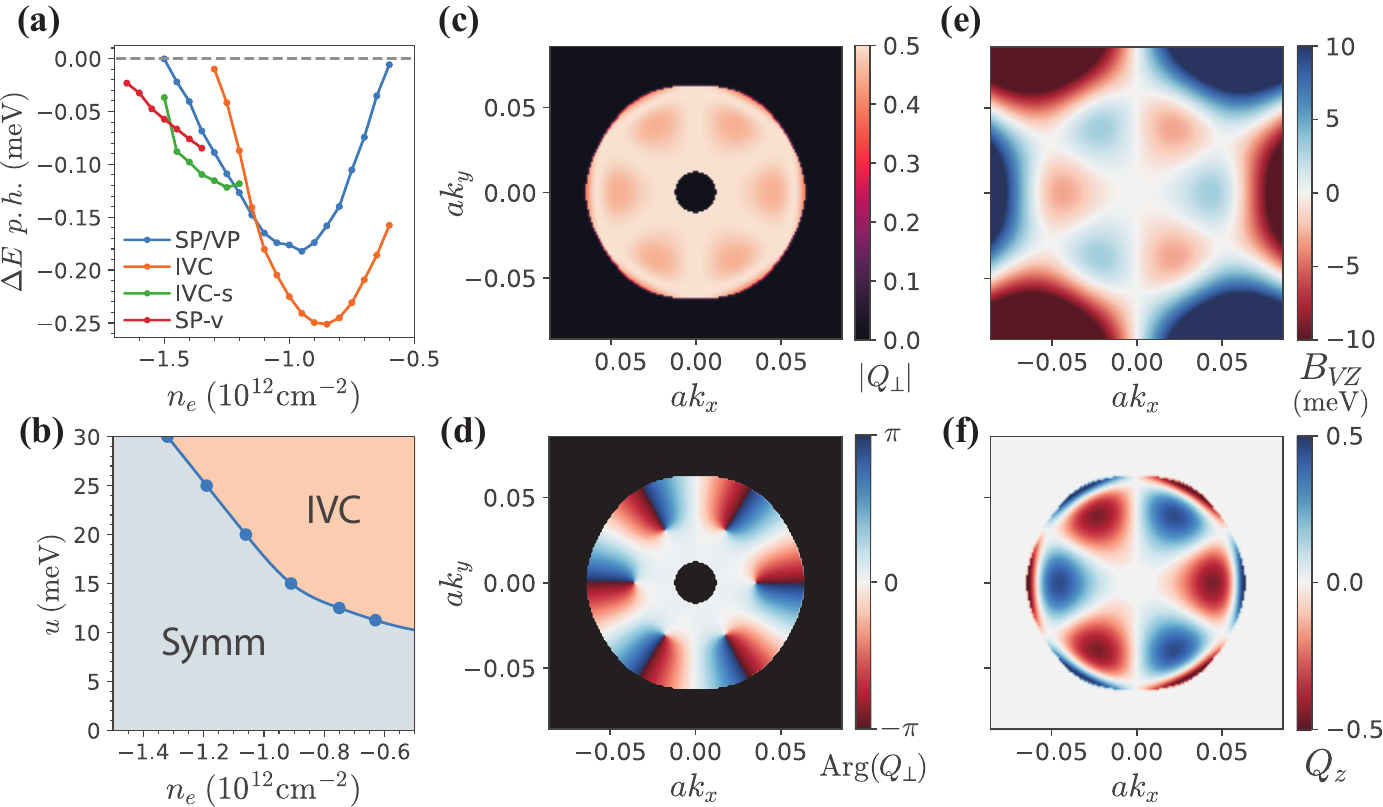}
    \caption{Phase diagram and the IVC order parameter (a) Self-consistent Hartree-Fock energetics of isospin symmetry-broken states for $u = \SI{30}{meV}$, including (i) spin or valley polarized (SP/VP), (ii) IVC, (iii) partially spin-polarized IVC (IVC-s), and (iv) partially spin and valley-polarized (SP-v) states. (See SM \cite{SM} for further details of SP-v and IVC-s). All energies are shown in meV per hole, relative to the fully symmetric metal. 
    (b) Hartree-Fock phase diagram as a function of hole-doping $n_e$ and displacement field $u$. Only the fully symmetric metal (Symm) and the spin-unpolarized IVC metal phases have been considered for clarity.
    (c) Magnitude and (d) phase of the self-consistent HF IVC order $Q_\perp(\k) = Q_x(\k) + i Q_y(\k)$ deep in the IVC phase where only the lower IVC band is filled ($n_e = \SI{-1.05E12}{\cm^{-2}}$).
    We have defined $Q_{\mu}(\k) =  \langle \psi_{\tau, s, \mathbf{k}}^{\dagger} \tau^{\mu}_{\tau \tau^\prime} \psi_{\tau^{\prime}, s, \mathbf{k}}\rangle$, normalized to unit magnitude. 
    The phase of $Q_\perp(\k)$ winds by $12 \pi$ around the outer Fermi surface. The region outside the Fermi surface is filled with black for clarity.
    (f) The valley polarization $Q_{z}(\k)$ in the self-consistent IVC solution follows (e) the local valley-Zeeman field $B_{VZ}(\k)$.}
    \label{fig:IVC}
\end{figure*}

ABC-stacked RTG is most accurately described using a six-band model per valley $(K/K^\prime)$ and spin \cite{Zhang2010,Jung2013}. 
All numerical calculations presented in this work use the six-band model with tight-binding parameters taken from Ref.~\onlinecite{Zhou2021_ABCmetals} (see \onlinecite{SM} for further details). 
However, it is useful to develop some intuition for the band structure within an approximate 2-band model which describes the low-energy physics in each valley.
The wave-functions of the two bands closest to the Fermi level reside mostly on the non-dimerized sites on the top/bottom layer (denoted by $\sigma = A_1/B_3$ respectively, see Fig.~\ref{fig:FS}(a)) \cite{SM}.
In this pseudospin basis, the effective Hamiltonian can be written as:
\beq
H = \sum_{\tau,s,\k} c^\dagger_{\tau,s,\k, \sigma} \left(  [h_{\tau}(\k)]_{\sigma \sigma^\prime} - \mu \, \delta_{\sigma \sigma^\prime} \right) c_{\tau,s, \k, \sigma^\prime} + H_{\rm C}, \nn 
\left[h_{\tau}(\k)\right]_{\sigma \sigma^\prime} = \begin{pmatrix} - u & \frac{v_0^3}{\gamma_1^2}\Pi^3 + \frac{\gamma_2}{2} \\ \frac{v_0^3}{\gamma_1^2}(\Pi^*)^3 + \frac{\gamma_2}{2} & u 
\end{pmatrix}_{\sigma \sigma^\prime}
\label{eq:Hfree}
\eeq
where $\Pi = \tau k_x + i k_y$, $\tau = \pm$ denotes valley, $s = \uparrow / \downarrow$ labels spin, and $\mu$ is the chemical potential. 
The band structure parameter 
$v_0$ is the Dirac velocity of monolayer graphene, $\gamma_1 \sim 300$ meV quantifies the strength of interlayer dimerization, $\gamma_2 \sim -15$ meV is the direct hopping between $A_1/B_3$ that contributes to trigonal warping, and $u \sim 10$s of meV is the potential difference between the two layers due to the perpendicular electric field.
When $\gamma_2 = 0$, the electric field $u$ gaps out the cubic-band touchings, leading to a  large density of states (DOS) at the band extrema centered on $K/K^\prime$. 
Symmetry-breaking is only seen at sufficiently large $u$, presumably because the increased DOS leads to stronger interaction effects \cite{Zhou2021_ABCmetals}.
The  $\gamma_2$ term then splits the band extrema into three shallow pockets related by $C_3$ rotations about $K/K^\prime$.
As shown in Fig.~\ref{fig:FS}(c), as the electron density is reduced below  neutrality, the topology of the Fermi surface within each valley first transitions from three $C_3$-related pockets to an annulus via a van-Hove singularity, and finally to a distorted disc via a Lifshitz transition.
For hole-dopings large enough such that $(vk_F)^3/\gamma_1^2 \gtrsim u,\gamma_2$, the DOS at the Fermi surface is low and interesting interaction effects disappear.

The interacting Hamiltonian $H_{\rm C}$ is given by:
\beq
H_{\rm C} = \frac{1}{2A} \sum_{\q} V_C(\q) :\rho(\q) \rho(-\q):
\label{eq:HC}
\eeq
where $A$ is the sample area, $V_C(\q) = e^2 \tanh{(q D)}/(2 \epsilon q)$ is the repulsive dual gate-screened Coulomb interaction with sample-gate distance $D$, and $\rho(\q) = \sum_{\k,\tau,s,\sigma} c^\dagger_{\tau,s,\k,\sigma} c_{\tau,s,\k+\q,\sigma}$ is the Fourier component of the electron density operator, with $|\k|$ and $|\q|$ being restricted to small values relative to the inverse lattice spacing $a^{-1}$.

The symmetries of $H$ include charge conservation U(1)$_c$, valley-charge conservation U(1)$_v$ generated by $\tau_z$, time-reversal $\cal{T}$, translations $T_{1,2}$, mirror reflection $M_x$,  and rotation $C_3$.
Note there is no inversion symmetry whenever $u \neq 0$, the case of interest, reducing the point group symmetry from $D_{3d}$ to $C_{3v}$ \cite{Latil,Cvetkovic}.
The absence of spin-orbit coupling allows us to define a spinless time-reversal $\tilde{\mathcal{T}} = \tau^x K$ which relates dispersions of the n$^{th}$ bands in the two valleys as $\varepsilon_{\tau,n}(\k) = \varepsilon_{-\tau,n}(-\k)$. 
However, trigonal warping splits the valleys locally in momentum space, so $\varepsilon_{\tau,n}(\k) \neq \varepsilon_{-\tau,n}(\k)$.
Finally, for the interaction defined by $H_C$ there is a separate spin-rotation symmetry in each valley, denoted by SU(2)$_+ \times$ SU(2)$_-$. In reality this symmetry is broken by lattice-scale effects such as optical phonons and inter-valley Coulomb scattering \cite{CBZ2020} to a global SU(2) spin rotation;  we will return later to the effect of this `Hund's' coupling $J_H$.

\subsection{Inter-valley coherent order}
\label{sec:IVC}

\emph{Isospin symmetry breaking-}
We begin by reviewing the experimental constraints on isospin symmetry breaking in the vicinity of SC1 \cite{Zhou2021_ABCmetals,Zhou_ABCSC}.
Upon approaching charge neutrality from  the  hole-doped  side,  a  series  of  phase  transitions  is  observed. 
The phase transitions are accompanied by Fermi surface reconstruction, visible in quantum oscillations.
The first transition is from a fully symmetric phase with fourfold-degenerate annular Fermi surfaces (corresponding to the four isospin degrees of freedom), to a symmetry-broken metallic phase (the PIP phase) with two large and two small Fermi surfaces.
The critical density is displacement field (e.g. $u$) dependent, and within our model at $u = 30$meV, it occurs in the general vicinity of $n_e \sim \SI{-1.4e12}{cm^{-2}}$.
The boundary between the two phases is insensitive to an in-plane magnetic field, indicating that the PIP phase is not spin-polarized (this is in contrast to other regions of parameter space, where such dependence is clearly visible). 
Furthermore, the PIP phase does not have an observable anomalous Hall effect \cite{AndreaPrivate}, which suggests it is time-reversal symmetric.
In other regions of the phase diagram, the system is valley-polarized, which produces an experimentally observed anomalous Hall effect due to the valleys' opposing Berry curvature \cite{Zhou2021_ABCmetals, SM}.

The absence of spin and valley polarization suggests that the PIP phase instead has broken U(1)$_v$ symmetry, i.e, it is inter-valley coherent.
An alternate possibility would be a spin-valley locked state (SVL) with spins polarized in each valley, but oppositely aligned between the valleys.
While such a state is compatible with experiment, we note that it would be disfavored by a ferromagnetic Hund's coupling. As mentioned above, the presence of nearby spin-polarized, valley-unpolarized phases suggests that the Hund's coupling is ferromagnetic. We shall assume that this is the case, and will not discuss the possibility of a SVL phase further.

In the absence of symmetry breaking, the band dispersion of the two valleys $\varepsilon_{\pm}(\mathbf{k})$ cross at certain high symmetry points related by $C_3$ and $M_x$. 
The IVC order hybridizes the valleys, gapping out the band crossings and deforming the $\mathcal{T}$-related annular Fermi surfaces of the two valleys into a small and large annulus, see Fig.~\ref{fig:FS}(a,d).
We identify this as the ``PIP'' phase in which quantum oscillations give evidence for a spin-unpolarized state featuring multiple Fermi surfaces with different areas; SC1 lies adjacent to this phase.

To verify that an IVC metal can be energetically favorable, we conduct self-consistent Hartree-Fock (HF) calculations within the  six-band model \cite{Jung2013}.
In these calculations we phenomenologically account for screening from the itinerant electrons by modifying $V_C$ within the Thomas-Fermi approximation, with screening wavevector $q_{\mathrm{TF}}$ based on the non-interacting density of states (for details, refer to SM \cite{SM}).
The resulting phase diagram as a function of hole-doping and displacement field is presented in Fig.~\ref{fig:IVC}(b), and a line cut at a fixed displacement field is shown in Fig~\ref{fig:IVC}(a).
Over significant regions of hole-doping and displacement fields of $20 - \SI{40}{\meV}$, a spin-unpolarized IVC metal is energetically competitive with the isospin polarized phase (without a Hund's coupling $J_H$, different patterns of isospin polarization, e.g. full spin vs full valley polarization, are degenerate within HF). 
The precise energetic ordering of the phases depends on details such as $u$ and $q_{\textrm{TF}}$.
Nevertheless, we note that the broad features of our phase diagram (Fig.~\ref{fig:IVC}(b)), such as interaction-induced symmetry breaking at large displacement fields, and the phase boundary between the spin-unpolarized IVC metal and the fully symmetric metal, are consistent with experiments.

\emph{Physical description of IVC states-}
In the absence of $J_H$ the set of IVC ground states form a degenerate U(2) manifold related by the action of SU(2)$_+ \times$ SU(2)$_-$ spin-rotations \cite{HalperinRice,Aleiner2007}.
Out of this manifold, inter-valley Hund's coupling, as we will elaborate on later, selects either a spin-singlet or triplet IVC. 
These states have simple real-space structures, as shown in Fig.~\ref{fig:FS}(b).
The spin-singlet IVC is a $\T$-symmetric CDW at momentum $\K - \Kp$, tripling the unit cell.
Unlike  monolayer and bilayer graphene, where the active sublattices form a honeycomb lattice, in RTG the active sublattices $A_1/B_3$ are stacked vertically, forming a single \emph{triangular} lattice (see Fig.~\ref{fig:FS}(a) inset). 
We define the $A_1/B_3$-projected density operator about $\K - \K^\prime$ momentum transfer 
\beq
n^{\rm IV}_{\rm S}(\q) =  \sum_{\R} e^{- i(\Kp - \K + \q)\cdot \R} \, \rho(\R),
\eeq
where $\R$ is the two-dimensional position vector for $A_1/B_3$ sublattices, and $\rho(\R) = \sum_{\sigma = A_1/B_3} \rho_\sigma(\R)$ is the total electron density summed over the two sublattices at position $\R$.
Thus, we conclude that $n^{\rm IV}_{\rm S}(\q=0)$ serves as a complex order parameter for the singlet/CDW IVC.
In fact, HF calculations show that the valley off-diagonal part of the self-consistent HF Hamiltonian $H_{\textrm{HF}}$ is very well-approximated by the operator $\Delta_{\textrm{IVC}} \, n^{\rm IV}_{\rm S}(0) + h.c.$, where $\Delta_{\textrm{IVC}}$ is the amplitude of the IVC order parameter (see SM \cite{SM}, Fig.~3 for a quantitative comparison). 
Under $C_3$ about an $A_1/B_3$ site, $n^{\rm IV}_{\rm S}(\q) \to n^{\rm IV}_{\rm S}(C_3 \q)$.
Therefore, the IVC order preserves $A_1/B_3$-site centered $C_3$.
While a unit-cell tripling would generically be described by a $\mathbb{Z}_3$ order parameter, corresponding to pinning of the U(1)$_v$ phase of the IVC order parameter to one of three distinct values, quartic interactions do not allow for Umklapp terms that break U(1)$_v \to \mathbb{Z}_3$, such terms appear only at the sextic level \cite{Aleiner2007}. 

The spin-triplet IVC can be obtained from the singlet one using the $\rm{SU}(2)_+\times \rm{SU}(2)_-$ symmetry, by applying a spin rotation of $\pi$ on one valley relative to the other around an arbitrary axis. The triplet IVC is a collinear SDW at momenta $\K - \K^\prime$.
In analogy with the singlet IVC, we define the $A_1/B_3$ projected spin-density operator $\s(\R) = \sum_{\sigma = A_1/B_3} \s_\sigma(\R)$ about $\K - \K^\prime$ momentum transfer:
\beq
\n^{\rm IV}_{\rm T}(\q) = \sum_{\R} e^{- i(\Kp - \K + \q)\cdot \R} \, \s(\R)
\eeq
The spin-triplet IVC parameter is $\n^{\rm IV}_{\rm T}(\q = 0)$. 
Thus, the SDW IVC breaks both valley U(1)$_v$ and global SO(3)$_s$ spin-rotation symmetry. 
Note that a change of the order parameter phase by U(1)$_v$ rotations can be offset by a global spin-rotation about $\hat{\n}$, leading to an order parameter manifold of U(1)$_v\times$SO(3)/U(1)$_{v+s} \cong$ SO(3) \cite{Mukerjee2006,LS2021,Eyal}.
Thus, such a state formally has no long-range or algebraic order at finite temperature \cite{MW,Chaikin}, although it may appear to order at low-enough temperature in finite size systems due to an exponentially diverging correlation length. 
We also note that within Landau theory, symmetry-allowed couplings between a SDW with momenta $\Q$ and a CDW with momenta $2\Q$ can nucleate such a CDW in presence of long-range SDW order \cite{Zachar1998}. 
Thus, the triplet IVC can induce a CDW at $\K - \K^\prime$, which is precisely the singlet IVC order parameter.
As such, the strict symmetry distinction between the triplet and singlet IVC is the lack of magnetic order for singlet. 

An alternative characterization of the IVC order parameters, useful for studying IVC energetics as well as superconductivity mediated by IVC fluctuations, may be obtained in momentum space.
To do so, we use the band-basis, defined via $c^\dagger_{\tau,s,\k,\sigma} = \sum_{n} u^*_{n,\tau,s,\k}(\sigma) \psi^\dagger_{n,\tau,s,\k}$, where $u^*_{n,\tau,s,\k}(\sigma)$ are the Bloch wave-functions and $n$ labels the band index.
We define a valence-band projected operator $n^{\rm IV}_{s s^\prime}(\q) = \sum_{\k} \lambda^{+-}_{\q}(\k) \psi^\dagger_{+,s,\k} \psi_{-,s^\prime,\k + \q}$, where $\lambda^{+-}_{\q}(\k) = \braket{ u_{+,s,\k} }{ u_{-,s,\k + \q} }$ is the inter-valley form factor that captures overlap of wavefunctions from opposite valleys in the valence band, and $U_{s s^\prime}$ is any unitary matrix in spin-space.
In this formulation, it is evident that IVC order parameter $n^{\rm IV}_{s s^\prime}(\q = 0)$ lies in the U(2) manifold. 
This degeneracy is broken by the inter-valley Hund's coupling, which either picks the spin-singlet CDW with $n^{\rm IV}_{s s^\prime} \propto \delta_{s s^\prime}$ or the spin-triplet SDW with $n^{\rm IV}_{s s^\prime} \propto (\hat{\n} \cdot \s)_{s s^\prime}$ with an arbitrary unit-vector $\hat{\n}$.

\emph{Energetics of IVC-}
We now turn to the energetics of the IVC phase. 
The IVC order parameter necessarily involves overlap of Bloch states from opposite valleys, and therefore has non-trivial winding originating from opposite chirality of threefold Dirac cones around $K$ and $K^\prime$ points at $u=0$. 
The winding of the IVC order parameter in momentum space contributes an additional energy cost relative to an isospin polarized phase \cite{SM}.
This additional energy cost is responsible for stabilizing an isospin polarized state relative to an IVC state in certain insulators with non-trivial band topology, such as magic angle graphene at certain odd integer filling of flat bands \cite{Goldhaber,Serlin2019,BCZ2020,ZMS2019,NickPRX}.
This raises an important question: why, then, is the IVC state energetically favored over a valley-polarized state?

This puzzle can be resolved by noting that an IVC metal can reduce its kinetic energy cost by local valley-polarization \cite{AdrianTBG,EslamNC}.
To visualize this, it is convenient to think of the IVC order at each $\k$ point as a vector in the x-y plane on the Bloch sphere corresponding to the valley isospin. 
The trigonal-warping induced kinetic energy mismatch between the valence bands in the $K/K^\prime$ valleys, given by $\varepsilon_{+}(\k) - \varepsilon_{-}(\k) = \varepsilon_{+}(\k) - \varepsilon_{+}(-\k)$, results in a local valley-Zeeman field $B_{VZ}(\k)$. 
The IVC state can thus benefit energetically by canting the valley isospin vector towards $B_{VZ}(\k)$ (much like an antiferromagnet gains energy by canting towards an applied magnetic field), without carrying any net valley-polarization as $B_{VZ}(\k)$ averages to zero.
We explicitly illustrate this energy gain in the supplement \cite{SM}, under the approximations of weak IVC order and linearized dispersion close to the Fermi surface. 
Consistent with this intuition, the self-consistent IVC order parameter obtained from HF also shows local valley-polarization in the vicinity of the Fermi-surface (see Fig.~\ref{fig:IVC}(c)).
On the other hand, a valley-polarized phase (corresponding to a vector polarized along $\hat{z}$ on the Bloch sphere) cannot benefit from this local valley-Zeeman field without losing significant interaction energy.
This is again in accordance with our HF results, where the valley-polarized phase shows no local canting in the parameter regime where it is energetically favorable.

Experimentally, as the hole-density is further reduced towards neutrality there is another sequence of transitions, first to a spin-polarized and valley-unpolarized `half-metal' (with zero spontaneous Hall resistance, $R_{xy} = 0$), subsequently to a second PIP phase, and finally to a spin and valley polarized `quarter metal' (where $R_{xy} \neq 0$) \cite{Zhou2021_ABCmetals}.
While the Hall response of the intervening PIP phase is unknown, a reasonable candidate for this phase, which borders SC2, is a spin-polarized IVC metal, which HF calculations also show is competitive in this density region (see Supplementary Fig.~2 \cite{SM}).
Starting with spin-polarized Fermi surfaces, the same interplay of kinetic energy benefit and interaction energy penalty can favor IVC over a valley-polarized state. 
Further reduction of hole-doping can suppress this kinetic energy gain, and tilt the energetic balance towards the observed spin-valley polarized `quarter metal'.

\subsection{Hund's coupling}
\label{sec:Hunds}
As alluded to previously, the inter-valley Hund's coupling plays a crucial role in determining the nature of iso-spin symmetry breaking.
We derive this term for an arbitrary translationally invariant interaction potential matrix $U_{\sigma \sigma^\prime}(\q)$ in the SM \cite{SM}, where $\sigma,\sigma^\prime$ refer to $A_1/B_3$ sublattice indices within each unit cell. 
However, to illustrate the physical effect, we focus on a simple limit $U_{\sigma \sigma^\prime}(\q) = U$, i.e,  a local interaction $U$ that acts only within the unit cell and is independent of the sublattice index. In this limit, the Hund's coupling takes the form:
\beq
H_{\rm Hund's} = - \frac{J_H}{A} \sum_{\q} \s_{+-}(\q) \cdot \s^\dagger_{+-}(\q)
\label{eq:HHunds}
\eeq
where  $\s_{+-}(\q) = \sum_{\k} \lambda^{+-}_{\q}(\k) \psi^\dagger_{+,s,\k} \s_{s s^\prime} \psi_{-,s^\prime,\k + \q}$ is the inter-valley spin-density projected to the valence bands, and $J_H = U$.
The Hund's coupling breaks the SU(2)$_+ \times$ SU(2)$_-$ symmetry down to the physical spin SU(2)$_s$ symmetry. 
While the short-range component of the Coulomb interaction is thus expected to give $J_H >0$, other lattice-scale effects, such interactions between electrons and optical phonons, may also contribute: so we treat $J_H$ as a phenomenological parameter to be constrained by experiments.

For $J_H > 0$, the Hund's coupling term favors a triplet IVC, as $\s_{+-}$ is nothing but the triplet IVC order parameter $\n_{\rm T}$.  
This can be understood by noting that a local repulsive interaction would disfavor excess accumulation of charge that characterizes a CDW such as the singlet IVC.
On the other hand, an attractive $U < 0$ favors the singlet IVC. 

We note that $H_{\rm Hund's}$ differs from another symmetry-allowed Hund's term $\tilde{H}_{\rm Hund's} = - \frac{\tilde{J}_H}{A} \sum_{\q} \s_{+}(\q) \cdot \s_{-}(-\q)$, 
where $\mathbf{s}_\tau$ is the spin-density in valley $\tau$.
While $H_{\rm Hund's}$ and $\tilde{H}_{\rm Hund's}$ are related by a Fierz transformation at the lattice scale, after projection into the valence band they are not,  giving rise to different physical effects. 
While  a ferromagnetic $H_{\rm Hund's}$ favors a triplet IVC state at the Hartree level as discussed above,  $\tilde{H}_{\rm Hund's}$  prefers either a spin-polarized or spin-valley locked state for $\tilde{J}_H > 0$  or $\tilde{J}_H < 0$ respectively.
The difference between these two distinct Hund's terms is rooted in the opposite Berry curvature of the two valleys.
Specifically, $\tilde{H}_{\rm Hund's}$ contains only valley-diagonal form-factors $\lambda^{\tau,\tau}_{\mathbf{q}}(\mathbf{k})$, while the $H_{\rm Hund's}$ derived microscopically from short-range $U$ has only valley-off-diagonal ones $\lambda^{\tau,-\tau}_{\mathbf{q}}(\mathbf{k})$.
This is  distinct from the SU$(4)$ quantum Hall physics in monolayer graphene, where the Landau level wave-functions in both valleys have identical Berry curvature, in which case the two kinds of Hund's terms are related by Fierz identities \cite{Kharitonov}.
However, for small momenta $\mathbf{k}$, the wavefunctions are nearly sublattice polarized, in which case the Berry curvature vanishes and the form factors become trivial,  $\lambda^{\tau,-\tau},  \lambda^{\tau,\tau}  \approx 1$.
In this part of the BZ, the two types of Hund's terms \emph{are} related by exchange symmetry. 
Therefore, at lower hole-doping, one might expect a  lack of competition from kinetic energy and a small ferromagnetic $H_{\rm Hund's}$ will tilt the balance in favor of spin-polarization.
Indeed, a spin-polarized, valley-unpolarized `half metal' phase is observed at hole dopings slightly lower than the spin-unpolarized PIP phase.

\subsection{IVC fluctuation mediated superconductivity}
\label{sec:SC}
\emph{Superconducting instabilities-}
Motivated by the likely presence of IVC order in the vicinity of superconductivity, we study superconducting instabilities mediated by near-critical fluctuations of the IVC order parameter.
While the transition to the IVC state appears to be first order in the HF phase diagram of Fig.~\ref{fig:IVC}(a), we find that the precise nature of this transition depends on details such as screening by the itinerant electrons; for example, small adjustments to $q_{\textrm{TF}}$ can render it continuous. 
Experimentally, there is no evidence of a first order phase transition (such as a negative compressibility spike) between the symmetric metal and the IVC metal, indicating that this transition is second order or weakly first order.  
To microscopically justify that IVC fluctuations are nearly gapless close to the transition, we compute the IVC correlation length $\xi_{\rm IVC}$ within Hartree Fock (see SM \cite{SM} for details), and find that $\xi_{\rm IVC}/a \approx 10^2$, i.e., $\xi_{\rm IVC}$ becomes much larger than the microscopic lattice spacing $a$ near the transition.
Therefore, we start in the symmetric metallic state with no long range IVC order, but with IVC correlations peaked at $\q = 0$. We assume that fluctuations of the IVC are described by phenomenological propagator of the form $g_{\q} = g/(\q^2 + \xi_{\rm IVC}^{-2})$ at $\omega = 0$ (we provide an estimate of $g$ in the SM \cite{SM}).
In the spirit of spin-fermion models \cite{MonthouxLonzarich,RoussevMillis,AbanovAIP2003}, we then integrate out the fluctuating IVC fields to obtain an effective inter-electron interaction.
We first focus on the SU(2)$_+ \times$ SU(2)$_-$ symmetric case, where the effective interaction takes the form ($\Tr$ stands for tracing spin-indices): 
\beq
H^{\rm eff}_{\rm IVC} = - \frac{1}{A} \sum_{\q} g_{\q} \, \Tr\left[ n^{\rm IV}(\q) [n^{\rm IV}(\q)]^\dagger \right]
\label{eq:HeffIVC}
\eeq
We use the above effective Hamiltonian as the pairing-interaction, in conjunction with the single-particle band structure projected to the valence band, to numerically solve a linearized BCS gap equation (see \cite{SM} for justification of projection, and further numerical details).
We restrict attention to inter-valley pairing of the general form
\begin{align}
F_{s s'}(\mathbf{k}) \equiv \langle \psi_{-, s, -\mathbf{k}} \psi_{+, s', \mathbf{k}} \rangle.
\end{align}
Intra-valley pairing occurs at finite center of mass momentum, and is expected to be energetically unfavorable. 

\begin{figure}
    \centering
    \includegraphics[width=0.48\textwidth]{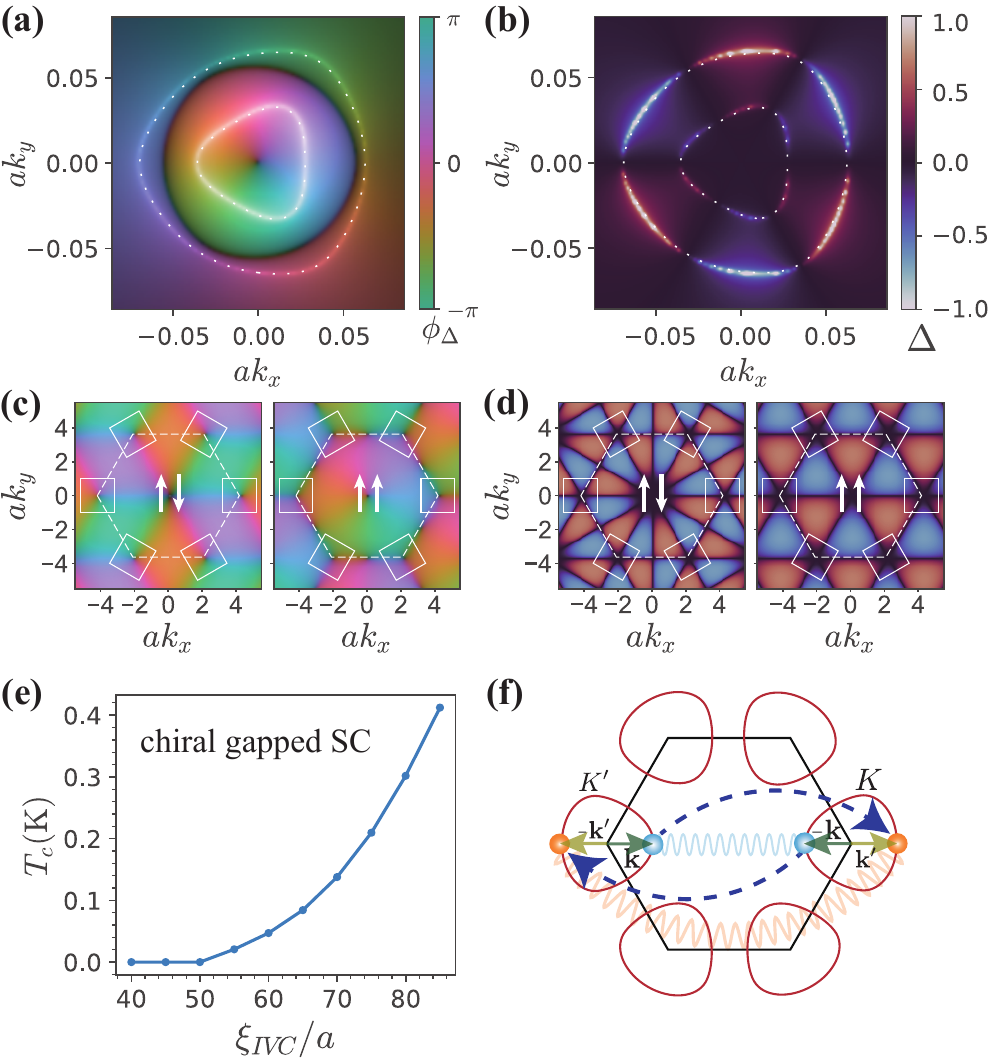}
    \caption{Unconventional superconductivity from IVC fluctuations (a) Complex pair wave-function of the gapped chiral superconductor showing magnitude (intensity) and phase (hue). Coulomb repulsion favors a sign change between the interior and exterior Fermi surfaces. (dotted white lines)
    (b) Real pair wave-function of nodal non-chiral superconductor with 6-fold oscillations around the annular Fermi surfaces.
    (c), (d) Schematic depiction of the favored superconducting pairings extended from momenta patches around $K/K^\prime$ points (white boxes) to the entire hexagonal BZ (dotted white lines), for spin-singlets (left) and triplets (right). 
    (e) $T_c$ of the gapped chiral superconductor (e.g. $d+id$ spin-singlet) within the self-consistent BCS calculations as a function of $\xi_{\rm IVC}$. Calculations at density $n_e = \SI{-1.7e12}{cm^{-2}}, u = \SI{30}{meV}$, including the effect of Coulomb repulsion with screening $q_{\textrm{TF}} = \frac{e^2}{\epsilon} \chi_0 $ and IVC fluctuations (Eq.\eqref{eq:HIVC_TvS}) of strength $g = n_e / \chi_0 \approx 6$meV, where $\chi_0$ is the DOS at the Fermi energy.
    (f) Electron-scattering between valleys by IVC fluctuations (indicated by blue arrows), showing how an attractive interaction in the IVC channel is converted to a repulsive interaction between inter-valley Cooper pairs. Only one Fermi surface in each valley is shown. The Fermi surfaces are enlarged relative to the BZ for clarity.}
    \label{fig:SC}
\end{figure}

Our numerical results are shown in Fig.~\ref{fig:SC}.
Remarkably, the leading superconducting instability is always towards a superconductor in which $F_{s s'}(\mathbf{k}) \approx -F_{s s'}(-\mathbf{k})$. 
It is tempting to call this `odd-parity', but due to the valley degree of freedom the parity depends on whether the spin structure is singlet vs. triplet (recall that $\mathbf{k}$ is measured relative to the $K$ or $K'$ point).
The precise pairing channel is sensitive to the correlation length $\xi_{\rm IVC}$.
For large $\xi_{\rm IVC}$, pairing occurs first in the chiral $k_x \pm i k_y$ channels, leading to a fully gapped superconductor (at the mean-field level) with orbital angular momentum $L_z = \pm 1$ about the $K,K^\prime$ points (Fig.~\ref{fig:SC}(a)).
The simplest extension of such an order parameter to the entire Brillouin Zone (BZ), consistent with fermionic anticommutation, is $d + i d$ for spin-singlet, and $p + i p$ for spin-triplet (Fig.~\ref{fig:SC}(c)) \cite{Annica_review,Nandkishore,AnnicaDoniach,Raghu,Lin}.
In contrast, a smaller $\xi_{\rm IVC}$ leads to a non-chiral nodal superconductor with a gap-function $\sim k_y(3k_x^2 - k_y^2) = \text{Im}[(k_x + i k_y)^3]$ about the $K,K^\prime$ points (Fig.~\ref{fig:SC}(b)).
We note that $C_3$ symmetry about the K point does not distinguish this nodal state from a trivial s-wave state ($L_z = 0$). 
Rather, such a gap function is odd under the combination of mirror $M_x$ and spinless time-reversal $\tilde{\mathcal{T}}$, leading to nodes at $k_y = 0$ and all $C_3$ related points: while an s-wave state is even under $M_x \tilde{\mathcal{T}}$ and non-nodal.
The simplest extension of the nodal pairing function to the entire BZ involve a twelve-fold oscillation about the $\Gamma$ point ($i$-wave) for the spin-singlet, and a six-fold oscillation ($f$-wave) for the spin-triplet (Fig.~\ref{fig:SC}(d)).

These results can be understood by analyzing the IVC fluctuation-mediated interaction in Eq.~\eqref{eq:HeffIVC}.
Decoupling  $H^{\rm eff}_{\rm IVC}$ in the Cooper channel, 
\beq
\langle H^{\rm eff}_{\rm IVC} \rangle = \frac{1}{A} \sum_{\k, \k^\prime} V_{\k \k^\prime} \Tr[F^\dagger(\k) F(\k^\prime)] 
\label{eq:HIVCCooper}
\eeq
where the effective interaction potential is $V_{\k \k^\prime} = g_{\q = -\k - \k^\prime}   |\lambda^{+-}_{\q = -\k - \k^\prime}(\k)|^2$. 
When $\xi_\textrm{IVC}$ becomes large, $V_{\k \k^\prime}$ is peaked at $\q = 0$. Thus, in contrast to the Coulomb interaction,  IVC-induced scattering is strongest between Cooper pairs with  \emph{opposite} momenta $\k = - \k'$.
An intuition for the resulting pairing channel is then gleaned from  the $\q = 0$ limit of Eq.~\eqref{eq:HIVCCooper}.
Due to the SU(2)$_+ \times $SU(2)$_-$ symmetry,  spin-singlet superconductivity with $F(\k) = i s^y f_{\k} $ and unitary spin-triplet superconductivity with $F(\k) = (i s^y) (\hat{\d} \cdot \s) f_\k$ are degenerate.
Inserting these ansatz into the $q \to 0$ limit, 
\beq \langle H^{\rm eff}_{\rm IVC} \rangle \approx \frac{2}{A} \sum_{\k} g_{\bm 0}|\lambda^{+-}_{\q=\bm 0}(\k)|^2 f^*_\k f_{-\k}
\eeq
Evidently, $\langle H^{\rm eff}_{\rm IVC} \rangle$ is minimized when $f^*_\k = - f_{-\k}$, corresponding to unconventional pairing, as found in our numerical calculations.
This result is reminiscent of Cooper-pairing due to spin fluctuations in $C_4$ symmetric systems, such as high-$T_c$ cuprates, where a repulsive interaction leads to sign-change of the pairing order parameter between points on the Fermi surface connected by the wavevector where the spin flutuations are strongest, resulting in a $d$-wave superconductivity \cite{Scalapino95}.
In $C_3$ symmetric RTG, inter-valley scattering by IVC fluctuations mediates an analogous repulsive interaction between inter-valley Cooper pairs \cite{YV2019}, and leads to sign-change in $f_{\k}$ across the Fermi surface within each valley (see Fig.~\ref{fig:SC}(f) for a schematic depiction).

Next, we turn to the $\xi_{\rm IVC}$-induced transition between chiral gapped and non-chiral nodal superconductivity.
When $\xi_{\textrm{IVC}}$ is large, the effective interaction strength $g_{\q}$ becomes increasingly singular at small $|\q|$. 
In this regime, the fully gapped $f_\k \sim k_x \pm i k_y$ is most energetically favorable, since it has a uniform magnitude of the gap on the Fermi surface, and gains the most from the singular part of the interaction. 
Further, the pairing amplitude is typically stronger on the inner Fermi surface (see Fig.~\ref{fig:SC}(a)), which hosts a larger density of states.
In contrast, when  $\xi_{\textrm{IVC}}$ is small, $g_{\q = \bm{0}} \approx g \, \xi_{\rm IVC}^2$ and $V_{\k \k'}$ is determined by the inter-valley form factor $|\lambda^{+-}_{\q=\bm{0}}(\k)|^2$. 
The form-factor has a six-fold oscillating structure across the Fermi surface, which induces an corresponding oscillating structure in $f_\k$, leading to the nodal superconductor observed numerically. 
In this case, pairing is much stronger on the outer Fermi surface which is at larger momenta, as opposed to the inner Fermi surface where the layer polarization term dominates and $|\lambda^{+-}_{\q = \bm{0}}(\k)|^2$ is approximately constant (see Fig.~\ref{fig:SC}(b)).
These considerations explain the $\xi_{\rm IVC}$-induced transition between preferred superconducting channels. 

Fig.~\ref{fig:SC}(e) shows the mean-field $T_c$ as a function of the correlation length $\xi_{\textrm{IVC}}$ for the chiral superconducting state, including the effect of long-range Coulomb repulsion (see SM \cite{SM} for further details of this calculation). We find that $T_c$ is a strongly increasing function of $\xi_{\rm{IVC}}$, and as a result $T_c$ is appreciable only in the regime where the fully-gapped chiral state dominates.
We therefore expect that this state, which is $d + id$ ($p + ip$) for spin-singlet (spin-triplet), is the one realized in the experiments. 
We note that in this calculation, we have ignored the frequency dependence of the interaction, and the damping of the electrons by bosonic IVC fluctuations.
Both effects are known to become important close to the critical point, and we defer a detailed study of these effects to future work \cite{abanov2001coherent}.

\emph{Effect of Hund's coupling-}
The inter-valley Hund's coupling splits the degeneracy between spin-singlet and spin-triplet superconductors, by amplifying SDW IVC fluctuations over CDW IVC fluctuations or vice versa, depending on the sign of $J_H$.
To see this, we use the Fierz identity $2 \delta_{\alpha \nu} \delta_{\beta \mu } = \s_{\alpha \beta} \cdot \s_{\mu \nu } + \delta_{\alpha \beta} \delta_{\mu \nu}$ to 
decompose the effective Hamiltonian for IVC fluctuations into of spin-singlet and spin-triplet IVC channels:
\beq
H^{\rm eff}_{\rm IVC} &=& - \frac{1}{2A} \sum_\q \Big( g^{\rm T}_\q \, \n^{\rm IV}_{\rm T}(\q) \cdot [\n^{\rm IV}_{\rm T}(\q)]^\dagger \nn && ~~~~~~~~~~~~ + g^{\rm S}_\q \, n^{\rm IV}_{\rm S}(\q) [n^{\rm IV}_{\rm S}(\q)]^\dagger \Big)
\label{eq:HIVC_TvS}
\eeq
In the SU(2)$_+ \times$ SU(2)$_-$ symmetric limit, the susceptibilities $g^{\rm S}_\q= g^{\rm T}_\q (= g_\q)$ for the singlet and triplet IVC states are identical.
However, including Hund's coupling breaks this symmetry and amplifies one susceptibility at the expense of the other, so more generally $g^{\rm S}_\q \neq g^{\rm T}_\q$, and we have:
\beq
\langle H^{\rm eff}_{\rm IVC} \rangle \approx \begin{cases} \displaystyle \frac{1}{A} \sum_{\k} (3g^{\rm T}_{\bm 0} - g^{\rm S}_{\bm 0})|\lambda^{+-}_{\bm q = \bm{0}}(\k)|^2 f^*_\k f_{-\k}, \text{singlet SC} \\ \displaystyle \frac{1}{A} \sum_{\k} (g^{\rm T}_{\bm 0} + g^{\rm S}_{\bm 0}) |\lambda^{+-}_{\bm q = \bm{0} }(\k)|^2 f^*_\k f_{-\k}, \text{ triplet SC} \end{cases}
\label{eq:HundsSplitSC} \nn
\eeq
From Eq.~\eqref{eq:HundsSplitSC}, we see that when triplet-IVC fluctuations are stronger, i.e, $g^{\rm T}_\q > g^{\rm S}_\q$, a spin-singlet superconductor becomes energetically favorable. 
Since a triplet IVC state is preferred by ferromagnetic Hund's coupling arising from short-range repulsive interactions ($J_H>0$), this leads to the surprising conclusion that such a Hund's coupling also prefers a spin-singlet superconductor.

Intuitively, this happens because ferromagnetic Hund's coupling promotes antiferromagnetic fluctuations that couple antipodal points on the Fermi surface, promoting singlet superconductivity with an order parameter that changes its phase between these points, in analogy to the cuprates \cite{Scalapino95} and magic angle twisted bilayer graphene \cite{YV2019,Isobe2018}.
In contrast, an antiferromagnetic Hund's term amplifies singlet-IVC fluctuations with $g^{\rm S}_\q > g^{\rm T}_\q$, and therefore leads to a spin-triplet p/f wave perturbatively away from the fully symmetric point. 
When it significantly enhances singlet-IVC fluctuations, the effective interaction $V_{\k,\k^\prime}$ turns attractive and a spin-singlet fully-gapped s-wave superconductor becomes the most favored pairing channel.

If we assume that the sign of the Hund's term does not change across the doping range studied in the experiment, we expect it to be ferromagnetic since it prefers spin-polarization at low doping. 
This leads to the interesting prediction that SC1 is a spin-singlet chiral $d + id$ superconductor. 
This conclusion is  consistent with fact that SC1 obeys the Pauli limit \cite{Zhou_ABCSC}.
Of course, as discussed previously, such a ferromagnetic Hund's term may also drive a transition to a spin-polarized IVC state, as possibly happens at lower doping. 
In this case, IVC fluctuations favor a spin-polarized (triplet) state, which we consider a candidate for SC2.

\emph{Effect of Coulomb repulsion-}
Finally, we comment on the effect of Coulomb interactions in our numerical solutions of the BCS gap equation.
Some intuition can be gained by analyzing $H_{\rm C}$ at a mean-field level, by decoupling the Coulomb interaction in the Cooper channel:
\beq
\langle H_{\rm C} \rangle = \frac{1}{A} \sum_{\k,\k^\prime} V^c_{\k, \k^\prime} \Tr[F^\dagger(\k) F(\k^\prime)]~~~~~
\label{eq:HCCooper}
\eeq
where $V^c_{\k, \k^\prime} = |\lambda^{++}_{\q = \k^\prime - \k}(\k)|^2 V_C(\q = \k^\prime - \k)$ is the effective repulsive potential.
The repulsion from Eq.~\eqref{eq:HCCooper} with static RPA screening was included in the BCS calculations for $T_c$ shown in Fig.~\ref{fig:SC}(c).

Noting that $V_C(\q)$ and $|\lambda^{++}_{\q}(\k)|^2$ are positive and peaked at $\q = 0$, the $\mathbf{k} \to \mathbf{k}'$ limit gives a large contribution to Eq.~\eqref{eq:HCCooper}. 
Since $\Tr[F^\dagger(\k) F(\k)]$ is always positive semi-definite, this leads to the expected conclusion that a repulsive Coulomb interaction disfavors superconductivity in all channels. 
However, for annular Fermi-surfaces, the superconductor can reduce the Coulomb penalty by flipping the sign of the pairing between the outer and inner Fermi surfaces, while leaving the  pairing symmetry unchanged.
This leads to an attractive contribution to Eq.~\eqref{eq:HCCooper} for wavevectors $\mathbf{q}$ which connect the inner and outer Fermi surfaces.
This sign change is indeed found in the solution to the linearized BCS equations shown in Fig.~\ref{fig:SC}(a). 
Furthermore, we find that the gapped chiral superconductor is quite  robust to Coulomb interactions, indicating that strong near-critical IVC fluctations can overcome repulsion between electrons and lead to Cooper-pairing.
In contrast,  the Coulomb interaction destabilizes the \textit{weaker} pairing in the nodal superconductor in favor of a metallic phase.

\section{Discussion}
\label{sec:Outlook}

In this paper, we showed that  IVC metallic phases, with and without net spin-polarization, are promising candidates for the symmetry broken phases adjacent to the SC2 and SC1 superconductors respectively.
Fluctuations in the IVC order parameter can provide the pairing glue for superconductivity in RTG, with $T_c$ comparable to experiments.
IVC fluctuations naturally favor gapped chiral superconductivity or non-chiral nodal superconductivity, depending on the correlation length $\xi_{\rm IVC}$.
In the SU(2)$_+ \times$ SU(2)$_-$-symmetric model, the spin-singlet and triplet channels are degenerate. 
The short-range Hund's coupling which breaks this symmetry then favors either (1) an IVC corresponding to a spin-singlet CDW, and triplet superconductivity or (2)  an IVC corresponding to a spin-triplet SDW, and singlet superconductivity. 
The latter  superconductor breaks only U(1)$_c$, and has a finite temperature BKT transition, and is Pauli limited, consistent the experimental observations for SC1. 

On the other hand, fully  spin-polarized IVC fluctuations at lower hole-densities can lead to a spin-polarized chiral or nodal superconductor, consistent with the Pauli limit violation observed for SC2. 
We note that such a superconductor has an order parameter manifold of SO(3) \cite{Mukerjee2006,LS2021,Eyal}, which would not have a finite temperature BKT transition in absence of a Zeeman field.
However, if the magnetic correlation length is large enough, we expect apparent superconducting behavior for low enough temperatures and finite-size systems.

\emph{Experimental probes-}
To experimentally verify the IVC metal in RTG, we note that it is either  a CDW, or a SDW with a small CDW component. 
Thus spin-polarized scanning tunneling microscopy (STM) \cite{SPSTM_RMP,SPSTM_ARMR} is the probe of choice, as it can directly access the spin and charge density distribution at the lattice scale.
However, since  symmetry considerations do imply that the SDW  will induce a weak CDW, a good first step is spin-unpolarized STM,  where a tripled unit cell should be observable in the site-resolved LDOS. 

Our theory predicts that the superconducting phases are unconventional in nature, in the sense that the average of the order parameter over the Fermi surface vanished. Such an order parameter is expected to be sensitive to small amounts of non-magnetic disorder \cite{Abrikosov,Larkin}. 
The chiral phase should produce  spontaneous edge currents \cite{Furusaki2001}, observable in scanning nano-SQUID experiments.
However, we carefully note that a chiral superconductor obtained from a parent metal with an annular Fermi surface is topologically trivial. 
To see this, we consider the BdG mean-field spectrum of the superconductor, where we first tune the chemical potential to empty all the bands, and subsequently tune the superconducting gap to zero. 
The chiral order parameter is gapless only at $K/K^\prime$ points, which never touch the annular Fermi surface as $\mu$ is tuned.
Thus, the bulk BdG gap never closes during this process, implying that the chiral superconductor is smoothly connected to the topologically trivial vacuum. 
Hence, we do not expect quantized edge modes, though the $\mathcal{T}$-breaking may still manifest in a bulk magnetization observable as edge currents.
Finally, current-noise spectroscopy using quantum impurity defects \cite{Agarwal2017} can efficiently distinguish between nodal and fully gapped chiral superconductors \cite{CD2021,DC2021}.

\emph{Alternative routes to superconductivity-}
Alternative mechanisms of superconductivity are possible, and deserve further investigation.
Ref.~\onlinecite{Chou2021} studies inter-electron attraction mediated by acoustic phonons as a possible pairing mechanism, and finds s-wave spin singlet/f-wave spin triplet superconductors to be favored. 
However, acoustic phonons do not choose between a singlet and a triplet superconductor, as the phonon-mediated interactions are fully SU(2)$_+\times$SU(2)$_-$ symmetric (optical phonons do not preserve this symmetry, but coupling of low-energy electrons to optical phonons is very weak in RTG under strong displacement fields \cite{DaChuan2022}).
Suppose we could characterize the phase diagram by a single Hund's coupling $J_H$.
Then, the presence of spin-polarized, valley-unpolarized phases in the phase diagram indicates that $J_H$ is ferromagnetic.
In such a scenario, a pairing mechanism based solely on acoustic phonons would predict a spin-triplet superconductor, in contradiction with the experimental observation for SC1. 
Our proposed scenario can explain both the presence of spin-polarized phases and spin-singlet superconductivity within a single, consistent picture.
Further, we note that the same acoustic phonons would act as an external bath for electrons, and lead to a strong linear in T resistivity in the metallic state above the Bloch–Grüneisen temperature, which has not been observed in RTG \cite{Zhou_ABCSC}.
While isospin fluctuations can also potentially increase the resistance above $T_c$, these fluctuations microscopically originate from the collective behavior of the electrons themselves.
Therefore, these result in electron-electron scattering that strongly affects single-particle lifetimes, but does not degrade the net momentum (in absence of umklapp scattering \cite{ZimanBook}). 
Thus, collective isospin fluctuations can only contribute to d.c. transport in the presence of disorder. 
We leave this interesting problem to future work.

On a different note, a two-dimensional annular Fermi surface allows for a Kohn-Luttinger mechanism for pairing \cite{KL1965,Baranov1992,Maiti2013,Raghu2011,Chubukov2017}. 
Similarly to the mechanism explored in this work, in the Kohn-Luttinger mechanism the pairing is driven by electronic fluctuations. However, no particular soft collective mode is assumed (i.e., the system is not assumed to be close to a continuous transition). Instead, all the particle-hole fluctuation channels contribute on the same footing. For RTG, this mechanism was recently found to lead to a chiral state \cite{Ghazaryan}, similar to the state predicted in this work in the vicinity of the critical point.

\emph{Outlook-}
Our study provides a starting point for further theoretical and experimental investigation of correlation effects and superconductivity in RTG in particular, and in \textit{non-moir\'e} few-layered graphene more generally.
It also shows that, somewhat contrary to usual belief, spin-singlet superconductors can be favored by ferromagnetic Hund's coupling when additional (valley) degrees of freedom are relevant. 
While our phenomenological treatment of coupling between electrons and soft-modes only allows us obtain an estimate of the superconducting critical temperature, our work motivates numerical explorations to determine $T_c$ accurately as a function of carrier density and electric field in RTG. 
Understanding the relevance of RTG physics to moir\'e graphene platforms, which also feature strong iso-spin fluctuations in topological flat bands \cite{Saito2021,Zondiner2020,NickPRX,Khalaf2020}, or to surface superconductivity in rhombohedral graphite \cite{Kopnin,Kopnin2} is left for future work.

\emph{Note added-} Recently, we became aware of another study of isospin fluctuation-mediated superconductivity in RTG \cite{Levitov}. Since this paper was submitted, several more studies of unconventional superconductivity in RTG have appeared \cite{Guinea,Roy,Yizhuang}.

\section*{Acknowledgements} 
We thank A. Black-Schaffer, A. Vishwanath, S. Whitsitt, Y. You and A. F. Young for helpful discussions.
S.C. and M.Z. were supported by the ARO through the MURI program (grant number W911NF-17-1-0323). S.C. also acknowledges support from the U.S. DOE, Office of Science, Office of Advanced Scientific Computing Research, under the Accelerated Research in Quantum Computing (ARQC) program via N.Y. Yao.
T.W. and M.Z. were supported by the U.S. DOE, Office of Science, Office of Basic Energy Sciences, Materials Sciences and Engineering Division, under Contract No. DE-AC02-05CH11231, within the van der Waals Heterostructures Program (KCWF16).
E.B. was supported by the European Research Council (ERC) under grant HQMAT (Grant Agreement No.~817799), by the Israel-USA Binational Science Foundation (BSF), and by a Research grant from Irving and Cherna Moskowitz.

\bibliography{arXiv3}

\newpage
\appendix
\onecolumngrid

\section{Model and symmetries}

\begin{figure}[htbp]
    \centering
    \includegraphics[width = 0.7\textwidth]{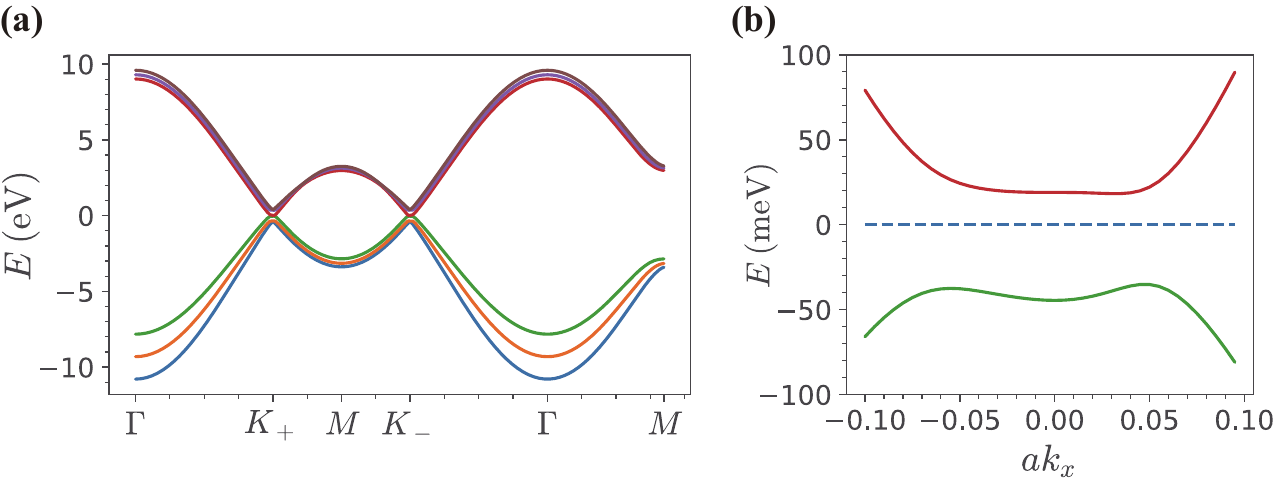}
    \caption{(a) Band structure along a high-symmetry line in the BZ for the 6-band model, and (b) Zoom-in near the $K$ point showing the flat spectrum and gapping out of cubic band touching at $\K$, due to the perpendicular electric field quantified by $u = \SI{30}{meV}$.}
    \label{fig:BandStrucSM}
\end{figure}

ABC trilayer graphene or RTG consists of three graphene monolayers, each displaced relative to the next by the same translation vector.
Each unit cell consists of six sites (two sublattice sites $A_i,B_i$ of the honeycomb lattice in each layer $i$), which we index by $(A_1,B_3,B_1,A_2,B_2,A_3)$.
Following Ref.~\onlinecite{Zhang2010,Zhou2021_ABCmetals}, we consider the following 6-band Hamiltonian (per valley, per spin) at low energy to describe the single-particle band structure. 
\beq
H_{6-band} = \begin{pmatrix} \Delta_1 + \Delta_2 + \delta  & \frac{\gamma_2}{2} & v_0 \Pi^\dagger & v_4 \Pi^\dagger & v_3 \Pi & 0 \\ \frac{\gamma_2}{2} & \Delta_2 - \Delta_1 + \delta & 0 & v_3 \Pi^\dagger & v_4 \Pi^\dagger & v_0 \Pi \\ v_0 \Pi & 0 & \Delta_1 + \Delta_2 & \gamma_1 & v_4 \Pi^\dagger & 0 \\ v_4 \Pi & v_3 \Pi & \gamma_1 & -2 \Delta_2 & v_0 \Pi^\dagger & v_4 \Pi^\dagger \\ v_3 \Pi^\dagger & v_4 \Pi^\dagger & v_4 \Pi & v_0 \Pi & -2 \Delta_2 & \gamma_1 \\ 0 & v_0 \Pi^\dagger & 0 & v_4 \Pi & \gamma_1 & \Delta_2 - \Delta_1 \end{pmatrix}
\eeq
where $\Pi = \tau k_x + i k_y$ ($\tau = \pm$ labels valleys), $\gamma_i$ are the bare hopping matrix elements, $v_i = \sqrt{3} a \gamma_i/2$ with $a = 0.246$ nm being the graphene lattice constant, $\delta$ is an on-site potential only present at the non-dimerized sites $A_1$ and $B_3$, and $\Delta_i$ account for the potential difference between the layers due to the perpendicular displacement field.
The values of all tight binding parameters of our study are chosen from Ref.~\onlinecite{Zhou2021_ABCmetals}.

The band structure for the 6-band model is shown in Supplementary Figure~\ref{fig:BandStrucSM}.
To analytically understand the symmetries of the problem, it is convenient to project the Hamiltonian onto the two \textit{active} low-energy bands per valley per spin, assuming a small enough density of doped holes or electrons near charge neutrality so that these bands are separated from the remote band by a gap of order $\gamma_1$.
In other words, most of the spectral weight of the active bands lie on the non-dimerized $A_1/B_3$ sites, which constitute an effecive sublattice space $\sigma$, and the remote bands have spectral weight concentrated on the dimerized sites $B_1/A_2$ and $B_2/A_3$.  
In the $\sigma$ pseudospin basis, the Hamiltonian can be approximately written as (N = total number of unit cells):
\beq
H &=& \sum_{\tau,s,\k} c^\dagger_{\tau,s,\k, \sigma} \left([h_{\tau}(\k)]_{\sigma, \sigma^\prime} - \mu \, \delta_{\sigma \sigma^\prime} \right) c_{\tau,s, \k, \sigma^\prime} + H_{\rm C}, \text{ with } [h_{\tau}(\k)]_{\sigma, \sigma^\prime} = \begin{pmatrix} - u & \frac{v^3}{\gamma_1^2}(\tau_z k_x + i k_y)^3 + \frac{\gamma_2}{2} \\ \frac{v^3}{\gamma_1^2}(\tau_z k_x - i k_y)^3 + \frac{\gamma_2}{2} & u 
\end{pmatrix}_{\sigma \sigma^\prime}, \nn 
 H_C &=& \frac{1}{2A} \sum_{\q} V_C(\q) :\rho(\q) \rho(-\q): \text{ and } \rho(\q) = \sum_{\tau,s,\k} c^\dagger_{\tau,s,\k} c_{\tau,s,\k+\q} 
\label{eq:HSM}
\eeq
In Eq.~\eqref{eq:HSM}, $c^\dagger_{\tau,s,\k, \sigma}$ denotes the electron creation operator at momenta $\k$ for valley/spin/sublattice indices $\tau/s/\sigma$ respectively. 
$u \sim \Delta_1$ is the difference in electrostatic potential in the top and bottom layers due to the perpendicular electric field.
$\rho(\q)$ is the slowly-varying component of the electron density operator, involving only `intra-valley' terms. 
Inter-valley scattering terms, which modulate on the lattice scale, and are suppressed at low densities near charge neutrality for long-range Coulomb interactions, will be discussed later.

For analytical arguments, we will often restrict our attention to the relevant `active' band (for a given valley and spin index) which is crossed by the chemical potential.
Hence, it would be convenient to move to the band-basis, and recast the Hamiltonian in terms of the Bloch-eigenstates of the single-particle Hamiltonian, denoted by $u_{n,\tau, \k, s}(\sigma)$. 
To do so, we write:  
\beq
c^\dagger_{\tau,s,\k, \sigma} = \sum_{n} u^*_{n,\tau, \k, s}(\sigma) \psi^\dagger_{n,\tau,s,\k},
\eeq 
where $n$ is the band-index and $\psi^\dagger_{n,\tau,s,\k}$ is the corresponding electron creation operator.
In this basis, the free Hamiltonian takes the simple form $\sum_{\k,n,s,\tau} \varepsilon_{n,\tau}(\k) \psi^\dagger_{n,\tau,s,\k} \psi_{n,\tau,s,\k}$, where $\varepsilon_{n,\tau}(\k)$ is the dispersion of the n$^{th}$ band in valley $\tau = \pm$, obtained by diagonalizing the matrix $h_{\tau}(\k)$.
To write the interaction term in the band-basis, it is convenient to define form-factors $[\lambda^{\tau \tau^\prime}_{\q}(\k)]^{n,n^\prime} = \bra{u_{n,\k,\tau,s}} \ket{u_{n^\prime, \tau^\prime, \k + \q, s}}$, which characterize the overlap of Bloch-wavefunctions.
Since we are interested in the slowly-varying part of electron density, $\rho(\q)$ involves only `intra-valley' form factors and takes the following form:
\beq
\rho(\q) = \sum_{\tau, s, \k} [\lambda^{\tau \tau}_{\q}(\k)]^{n,n^\prime} \psi^\dagger_{n,\tau, s, \k} \psi_{n^\prime,\tau, s, \k + \q}
\eeq
We note that in this work, we are mainly concerned with physics in the valence band in each valley, corresponding to small hole-doping near charge neutrality.
So unless otherwise mentioned in our analytical studies we will fix $n = n^\prime = $ `valence', and ignore the band index.
In this limit, $H_C$ consists of gate-screened Coulomb interaction $V_C(\q)$ projected onto the active valence bands.
The numerical studies are carried out in the full 6-band (per spin, per valley) basis without resorting to band-projection.

Let us now consider the symmetries of $H$. It conserves both total electric charge and electron number in each valley, and thus possesses U(1)$_c \times$ U(1)$_v$ symmetry.
The spatial symmetries include translations by the Bravais lattice vectors $\bm{a}_{1/2} = a(1, \pm \sqrt{3})/2$ corresponding to the honeycomb lattice $T_{1/2}$ ($a = 0.246$ nm is the graphene lattice constant), a mirror $M_x$ and $C_3$. 
At zero perpendicular displacement field, there is an additional inversion symmetry, forming the space group P$\bar{3}$m1 \cite{Latil,Cvetkovic}, but inversion is broken for $u \neq 0$, which is required for observing correlated physics. 
The symmetry actions are given by:
\beq
\text{U(1)}_c: c_{\tau,s,\k,\sigma} \to e^{i \theta_c} c_{\tau,s,\k,\sigma}, && ~~~ \textrm{U(1)}_v: c_{\tau,s,\k,\sigma} \to e^{i \tau \theta_v} c_{\tau,s,\k,\sigma} \nn
T_{j}: (x,y) \to (x,y) + \bm{a}_{j}, && ~~~ c_{\tau,s,\k,\sigma} \to e^{i \tau \K \cdot \bm{a}_{j}} c_{\tau,s,\k,\sigma} , \text{ where } j = 1 \text{ or } 2\nn
M_x: (x,y) \to (-x,y), && ~~~ c_{\tau,s,\k,\sigma} \to c_{-\tau,s,M_x(\k),\sigma}, \text{ where } M_x(k_x,k_y) = (-k_x, k_y) \nn
C_3: x + i y \to e^{2\pi i/3}(x + iy) , && ~~~ c_{\tau,s,\k,\sigma} \to c_{\tau,s,C_3[\k],\sigma} 
\label{eq:SymSM} 
\eeq
In Eq.~\eqref{eq:SymSM}, translations act as internal symmetries on the field operators $c_{\tau,s,\k,\sigma}$, $M_x$ preserves sublattice but flips valley and the x-component of momenta, and $C_3$, which preserves both valley and sublattice, is taken to be centered on the $A_1/B_3$ sites so that its only action on $c_{\tau,s,\k,\sigma}$ is to rotate the momenta. Note that any other choice of rotation center for $C_3$ will an an overall phase for the spinor in the sublattice space.
In particular, it does not act differently on the two active sublattices, that lie directly on top of each other, for any consistent choice of rotation center.
The internal symmetries include anti-unitary time-reversal $\cal{T}$ and global spin-rotation SU(2)$_s$. 
However, neglecting lattice-scale effects leads to an enhanced SU(2)$_+\times$ SU(2)$_-$ symmetry, corresponding to individual spin-rotations in the valleys.
\beq
\mathcal{T}: c_{\tau,s,\k,\sigma} \to (is_y)_{s s^\prime} c_{-\tau,s^\prime,-\k,\sigma}, ~~~ i \to -i \nn
\text{SU(2)}_{\pm}: c_{\pm,s,\k,\sigma} \to [e^{i \theta_\pm \hat{\n}_\pm \cdot \s}]_{s s^\prime} c_{\pm,s^\prime,\k,\sigma}
\label{eq:Sym2SM} 
\eeq
In Eq.~\eqref{eq:Sym2SM}, $\s = (s_x, s_y, s_z)$ denote the Pauli matrices in spin-space, and $(\theta, \hat{\n})_\pm$ correspond to the angle and axis of spin-rotations in each valley labeled $\tau = \pm$.
Including lattice scale effects such as inter-valley electron scattering will reduce the spin-rotation to a global SU(2)$_s$, corresponding to a single choice of $(\theta, \hat{\n})$ for both valleys. 
The absence of spin-orbit coupling in graphene allows us to define a anti-unitary spinless time-reversal $\tilde{\cal{T}}$ that preserves spin, but flips valley and momentum, i.e, acts as $\tau_x K$ and takes $\k \to -\k$. 
$\tilde{\cal{T}}$ relates the band-dispersion in the two valleys, setting $\varepsilon_{n,\tau}(\k) = \varepsilon_{n,-\tau}(-\k)$ (note that the dispersion is independent of spin).
Further, demanding that the Bloch wave-functions in opposite valleys are related by time-reversal implies:
\beq 
u_{n,\tau,s,\k} = u^*_{n,-\tau,s,-\k}, ~~~~  [\lambda^{\tau, \tau^\prime}_\q(\k)]^* = \lambda^{-\tau, -\tau^\prime}_{-\q}(-\k) 
\label{eq:STRSM}
\eeq
Eq.~\eqref{eq:STRSM} relates the `intra-valley' form factors from opposite valleys, and also `inter-valley' form factors at different momenta, and will prove useful later when we look at the IVC and superconducting phases.

\section{Details of self-consistent Hartree-Fock calculations}

In the Hartree-Fock calculation, we solve the self-consistent equations for Slater determinant states characterized by the one-electron covariance matrix $P_{\tau, \tau^{\prime}}^{s s^{\prime}}(\mathbf{k})=\left\langle\psi_{\tau, s, \mathbf{k}}^{\dagger} \psi_{\tau^{\prime}, s^{\prime}, \mathbf{k}}\right\rangle$ using the formulation described in Ref.~\onlinecite{NickPRX}. In the following, we will only consider kinetic energy and intra-valley Coulomb scattering in the HF calculation such that the enhanced SU(2)$_+ \times$ SU(2)$_-$ spin-rotation symmetry remains. The absence of spin-orbit coupling further decouples spin from all other degrees of freedom, which allows us to first block diagonalize the covariance matrix in the spin space such that $P_{\tau, \tau^{\prime}}^{s s^{\prime}}(\mathbf{k}) = P_{\tau, \tau^{\prime}}^{s}(\mathbf{k}) \delta_{s,s'}$. 
Then we use both the `ODA' and `EDIIS' algorithms to solve the self-consistency equation \cite{ODA,EDIIS}.

HF typically over-estimates the exchange energy-gain in a metallic state since it neglects screening of the interaction by mobile electrons. To capture this screening effect, we consider a random phase approximation (RPA) correction by itinerant fermions \cite{coleman_2015}:
\beq
H_C = \frac{1}{2A} \sum_\q V_C(\q) :\rho(\q) \rho(-\q):  \xrightarrow{\text{RPA}} \frac{1}{2A} \sum_\q V_{\rm RPA}(\q) :\rho(\q) \rho(-\q): ~, \text{ where }  V_{\rm RPA}(\q) = \frac{V_C(\q)}{1 + \chi_{\rho \rho}(\q) V_C(\q)} ~~
\label{eq:HRPASM}
\eeq
where $V_C(\q) = e^2 \tanh{(q D)}/(2 \epsilon q)$ is the repulsive dual gate-screened Coulomb interaction, and $\chi_{\rho \rho}(\q)$ is the static Lindhard response function. We have neglected the frequency dependence of the screening, and $\chi_{\rho \rho}(\q) = \chi_0(1 - c \q^2/k_F^2 + \cdots)$ for small $q/k_F$, where $-\chi_0 = -\partial_\mu n_e$ is simply the density of states at the Fermi surface at $T = 0$, and $c$ is an O(1) constant that depends on Fermi surface details \cite{coleman_2015}. In the limit $q \gg D^{-1}$, we can define a Thomas-Fermi screening wavevector $q_{\textrm{TF}} = -\frac{e^2}{\epsilon} \chi_0$. For the purpose of the HF calculation, we only keep $\chi_0$, and neglect further $\q$ dependence of $\chi_{\rho \rho}(\q)$. To ensure convergence of the self-consistent calculation, we use the non-interacting density of states $\chi_0$ with fourfold isospin degeneracy.

The numerical results shown in the paper are obtained with $\epsilon = 4.4$, gate distance $D=\SI{50}{nm}$, using the projected valance band per spin per valley on a $71 \times 71$ momentum grid, with UV momentum cutoff $0.085 a^{-1}$. We take $u = \SI{30}{meV}$ in main text Fig.~2(a) showing the full competition among all candidate states. We also explored the phase diagram at $u = 20 - \SI{40}{meV}$ and confirmed the robustness of all our observations. In main text Fig.~2(b), we show the phase diagram at $u = 0 - \SI{30}{\meV}$ only concerning the competition between the IVC state and the fully symmetric state.
We take $\chi_0 = \SI{0.16}{eV^{-1}}$ per unit cell for all figures except for Supplementary Figure~\ref{fig:HFDOS8} where we explored the effect of changing $\chi_0$. 

Depending on which symmetries are explicitly enforced, we find several self-consistent solutions that can be grouped into four categories: (i) a `half-metal', including a spin polarized (SP) state that breaks the global SU(2)$_s$ symmetry, a valley polarized (VP) state that breaks the spinless time-reversal $\tilde{\mathcal{T}}$, and a spin-valley locked (SVL) state that breaks both global SU(2)$_s$ and $\tilde{\mathcal{T}}$ but preserves their combination, (ii) a spin-singlet/triplet IVC ``half-metal'' that breaks U(1)$_v$ but preserves $\tilde{\mathcal{T}}$ and global SU(2)$_s$, (iii) a metal that breaks both global SU(2)$_s$ and $\tilde{\mathcal{T}}$, including a spin-valley polarized ``quarter metal'' (SVP) and a partially spin and valley-polarized (SP-v) state, (iv) a metallic IVC state that breaks both global SU(2)$_s$ and U(1)$_v$, including a spin polarized IVC ``quarter metal'' (SP-IVC) and a partially spin-polarized IVC (IVC-s) state. States within the latter two groups cannot be distinguished by symmetry, but they appear at very different hole doping. Close to SC1, the competitive candidates are SP, VP, SVL, spin-singlet/triplet IVC, SP-v, and IVC-s; and close to SC2, only SVP and SP-IVC are energetically competitive. Due to the enlarged SU(2)$_+ \times$ SU(2)$_-$ symmetry of the Hamiltonian, states within the first two groups are degenerate, so we only plot one example within each group. We note that SP, VP, and SVL are not always fully polarized for weaker interaction strength, and will adjust population between two spin/valley sectors to minimize energy. 

\begin{figure}[htbp]
    \centering
    \includegraphics[width = 0.6\textwidth]{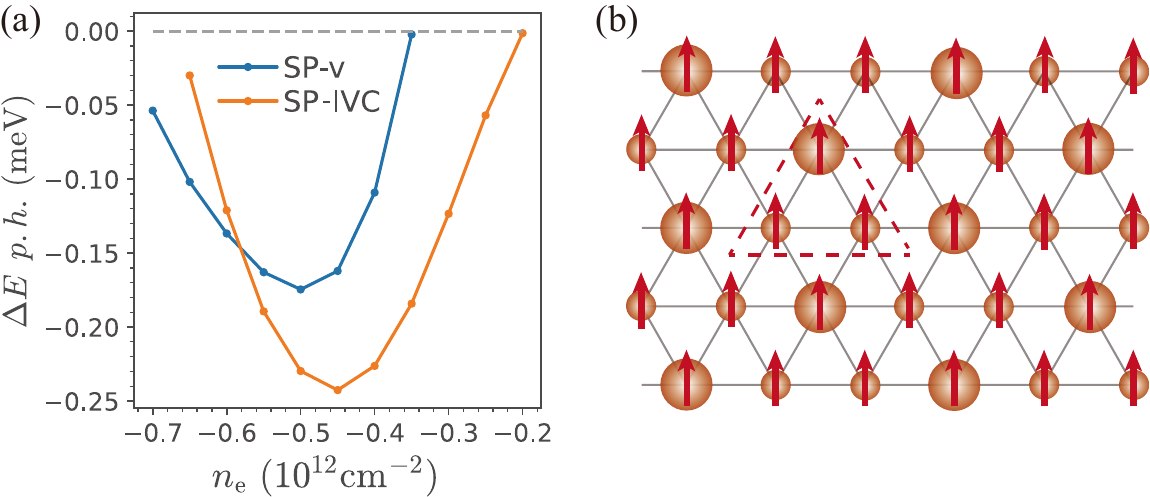}
    \caption{(a) Hartree-Fock energetics assuming full spin-polarization at lower hole-doping, showing the competition between partially valley-polarized (SP-v) state and spin-polarized inter-valley coherent (SP-IVC) states, for parameters mentioned in the text. (b) Real space cartoon of the SP-IVC state, which is a ferromagnetic charge-density wave with a tripled unit cell (shown with dotted red lines).}
    \label{fig:SC2}
\end{figure}

As shown in Fig.~2 in the main text, the precise energetic ordering of the phases and the reconstructed Fermi surface topology are sensitive to the interaction strength, which is mainly controlled by the density of states $\chi_0$. However, across a wide parameter regime, SP/VP/SVL and spin-singlet/triplet IVC are close in energy and favored over a fully symmetric metal close to SC1, which is expected from the Fock energy gain. SP-v and IVC-s are even lower in energy close to the transition to fully symmetric metal. However, these two states are not observed in the experiment due to the Hund's coupling that is not included in the HF calculation. We will discuss the role Hund's coupling plays in this competition later in the SM. We also note that close to SC2, the SP-IVC phase (a ferromagnetic CDW in real space) can be energetically competitive with the SP-v phase (Supplementary Figure~\ref{fig:SC2}).

\begin{figure}[htbp]
    \centering
    \includegraphics[width = 0.35\textwidth]{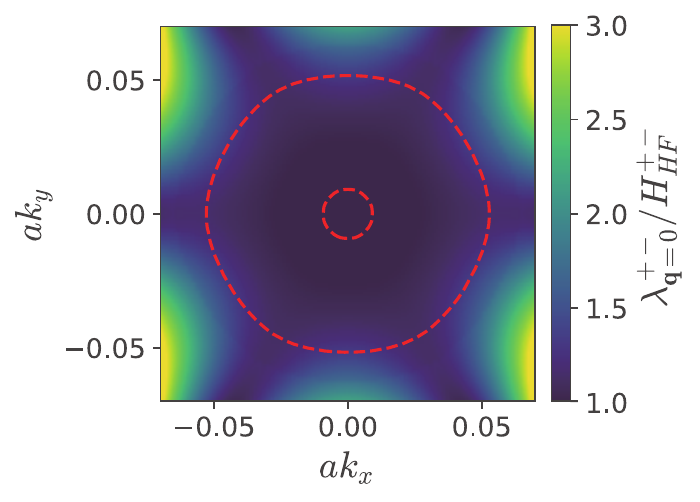}
    \caption{The valley off-diagonal part of $H_{HF}$ compared to the IVC density operator $n_S^{\mathrm{IV}}(\q)$ at $\q = 0$ in magnitude with proper normalization at $n_e = \SI{-1.05E12}{cm^{-2}}$, which is almost uniform within the annular HF Fermi surface (indicated by dotted red lines).} 
    \label{fig:HFIVC}
\end{figure}

Now we turn to analyze the structure of the self-consistent HF Hamiltonian $H_{HF}$ deep in the IVC phase. In fact, the valley off-diagonal part of $H_{HF}$ is very well approximated by the operator $n_{\rm S}^{\mathrm{IV}}(\q)$ at $\q = \mathbf{0}$. In the momentum space, $n_{\rm S}^{\mathrm{IV}}(\q)$ takes the form $\sum_{\k,s} \lambda^{+-}_{\q}(\k) \psi^\dagger_{+,s,\k} \psi_{-,s,\k + \q}$. We compared $\lambda^{+-}_{\q = \mathbf{0}}(\k)$ with the valley off-diagonal part of HF Hamiltonian $H^{+-}_{HF}$ at each $\k$, with proper normalization in Supplementary Figure~\ref{fig:HFIVC}. The fact that $\lambda^{+-}_{\q}/H^{+-}_{HF}$ is mostly uniform within the Fermi sea suggests that $H^{+-}_{HF}$ captures a purely local $n_{\rm S}^{\mathrm{IV}}$ perturbation.

\begin{figure}[htbp]
    \centering
    \includegraphics[width = 0.43\textwidth]{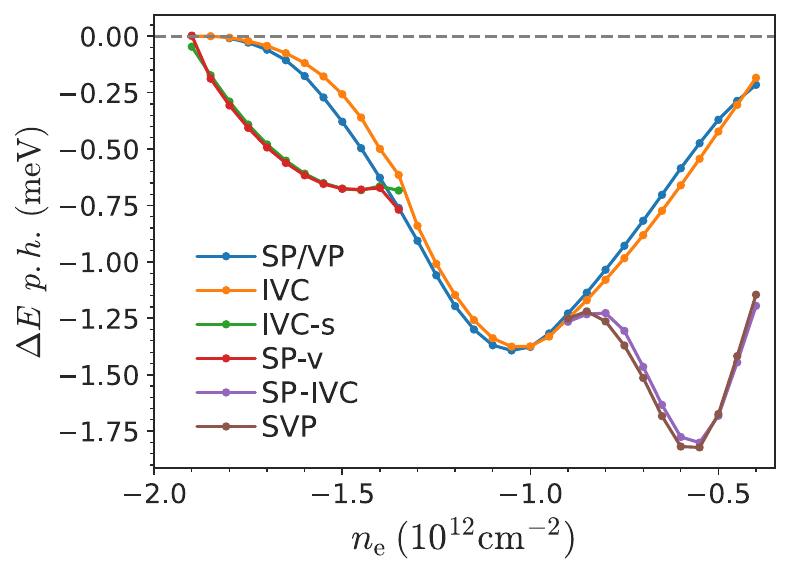}
    \caption{Self-consistent HF energetics of isospin symmetry-broken states, including SP/VP/SVL, spin singlet/triplet IVC, SP-v, IVC-s, SVP, and SP-IVC using $\chi_0 = \SI{0.08}{eV^{-1}}$ per unit cell, at $u = 30$ meV and $\epsilon = 4.4$. SC1 is observed approximately around $n_e = \SI{-1.9E12}{cm^{-2}}$ and SC2 is observed around $n_e = \SI{-0.9E12 }{cm^{-2}}$.}
    \label{fig:HFDOS8}
\end{figure}

Finally, we comment on the effect of non-interacting density of states $\chi_0$ used in the RPA screening. A smaller $\chi_0$ can enhance the interaction and thus change the precise energetic ordering of the phases. When $\chi_0 = \SI{0.08}{eV^{-1}}$, the SP/VP/SVL state and the spin singlet/triplet IVC state are much closer in energy and therefore their competition is almost entirely determined by the Hund's couple. In addition, we see that the phase transition toward an IVC phase becomes a second-order transition, which also leads to a divergent correlation length $\xi_{\rm IVC}$.

\section{Perturbative Hartree-Fock analysis of IVC energetics}

To analyze the energetics of various isospin symmetry broken states analytically, we evaluate the energy of a general Slater determinant state characterized by a covariance matrix $P_{\tau, \tau^{\prime}}^{s s^{\prime}}(\mathbf{k})=\left\langle\psi_{\tau, s, \mathbf{k}}^{\dagger} \psi_{\tau^{\prime}, s^{\prime}, \mathbf{k}}\right\rangle$. In this appendix, we will focus on two types of strong candidates close to SC1: (1) spin polarized (SP) state, valley polarized (VP) state, and spin-valley locked (SVL) state, and (2) spin-singlet/triplet IVC state.
In the absence of Hund's coupling $J_H$, states within each group are degenerate within our Hartree-Fock analysis. Thus, we will only analyze the VP state and the spin-singlet IVC state, both of which are spin-singlet such that $P_{\tau, \tau^{\prime}}^{s s^{\prime}}(\mathbf{k}) = P_{\tau, \tau^{\prime}}(\mathbf{k}) \delta_{s,s'}$. Similar perturbative Hartree-Fock analysis can also apply to states close to SC2, which includes the spin valley polarized (SVP) state and the spin-polarized IVC state.

The general Slater determinant state can be viewed as the ground state of a mean-field Hamiltonian $H_{MF}$,
\begin{equation} \label{eq:MF}
    H_{MF}=\sum_{\tau, \boldsymbol{k}} \psi_{\tau, \boldsymbol{k}}^{\dagger} h_{\tau, \tau^{\prime}}^{M F}(\boldsymbol{k}) \psi_{\tau^{\prime}, \boldsymbol{k}}, \quad \text{with } h_{\tau, \tau^{\prime}}^{M F}(\boldsymbol{k})=\left(\begin{array}{cc}\varepsilon_{\tau}(\boldsymbol{k})-\mu+\Delta_{V P}(\k) & \Delta_{I V C}^{*}(\boldsymbol{k}) \\ \Delta_{I V C}(\boldsymbol{k}) & \varepsilon_{-\tau}(\boldsymbol{k})-\mu-\Delta_{V P}(\k)\end{array}\right)
\end{equation}
where $\Delta_{VP}(\k)$ is the valley polarization, and $\Delta_{IVC}(\k) = |\Delta_{IVC}(\k)|e^{i \phi_\k}$ is the IVC order parameter. 
Then the covariance matrix takes the form
\begin{gather}
    P_{\tau,\tau^\prime} = \left[ \frac{1}{2}\left( 1 + \frac{\bm{\Delta_\k}^* \cdot \bm{\tau}}{|\bm{\Delta_\k|}} \right) n_F(E_{\k,+}) + \frac{1}{2}\left( 1 - \frac{\bm{\Delta_\k}^* \cdot \bm{\tau}}{|\bm{\Delta_\k|}} \right) n_F(E_{\k,-}) \right], \nonumber\\
    \text{where }
    \bm{\Delta_\k} \equiv \left( |\Delta_{IVC}(\k)| \cos(\phi_\k), |\Delta_{IVC}(\k)| \sin(\phi_\k),  \xi_a (\k) + \Delta_{V P}(\k) \right), \quad E_{\k, \pm} = \xi_s(\k) \pm |\bm{\Delta_\k}|.
\label{eq:PmatrixSM}
\end{gather}
In Eq.~\eqref{eq:PmatrixSM}, $\xi_s(\k) = (\varepsilon_{+,\k} + \varepsilon_{-,\k})/2 - \mu $ and $\xi_a (\k) = (\varepsilon_{+,\k} - \varepsilon_{-,\k})/2$ denote the valley-symmetric and valley-antisymmetric components of the dispersion respectively. 
Then we can evaluate the mean-field energy per spin using Wick's theorem,
\begin{equation}
    \langle H\rangle_{MF} =\sum_{\boldsymbol{k}, \tau} \xi_{\tau}(\boldsymbol{k}) P_{\tau \tau}(\boldsymbol{k})+\frac{V_C(\mathbf{0}) N^{2}}{A}-\frac{1}{2A} \sum_{\k, \boldsymbol{q}} V_C(\boldsymbol{q})\lambda^{\tau \tau}_{ \boldsymbol{q}}(\boldsymbol{k}) [\lambda^{\tau^{\prime} \tau^{\prime}}_{ \boldsymbol{q}}(\boldsymbol{k}) ]^{*}P_{\tau \tau^{\prime}}(\boldsymbol{k}) P_{\tau^{\prime} \tau}(\boldsymbol{k}+\boldsymbol{q})
\end{equation}
The first term is the kinetic term, the second is the Hartree term which simply counts the total number of electrons per spin species $N=\sum_{\tau, k} P_{\tau \tau}(\boldsymbol{k})$ and the last term is the Fock term. Since the Hartree term does not distinguish different isospin symmetry broken states, we will neglect it from now on and consider only the other two
terms. In the following, we will compare the Hartree-Fock energy of the VP state and the IVC state in two different limits.

\begin{figure}
    \centering
    \includegraphics[width=0.5\textwidth]{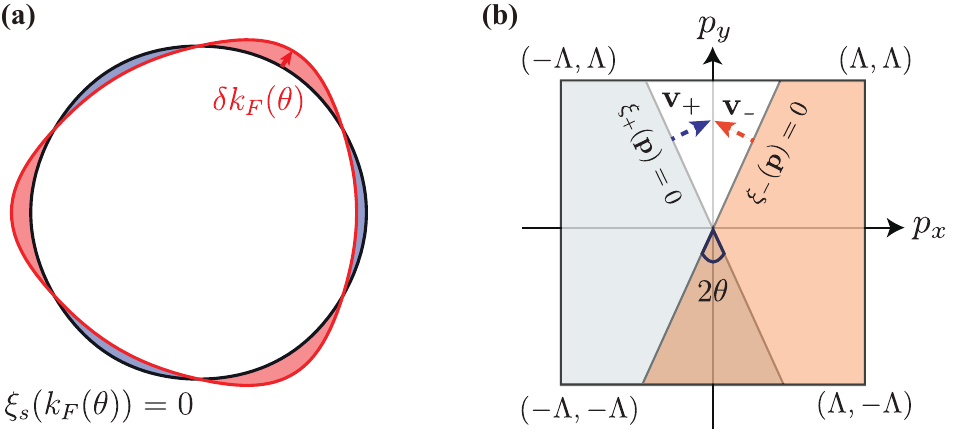}
    \caption{(a) The black curve is the reference Fermi surface defined by $\xi_s(k_F(\theta)) = 0$. The red curve is the deformed Fermi surface $k_F (\theta) \to k_F (\theta) + \delta k_F (\theta)$ for a particular isospin symmetry broken state. (b) Intersecting Fermi surfaces of the two valleys at an angle $2\theta$ (zoomed-in view with linearized dispersions). Filled regions are indicated by solid colors. $\v_\pm$ denote the Fermi velocities of the two valleys near the band-crossing point, and $\Lambda$ is an upper momentum cutoff.}
    \label{fig:FSCrossing}
\end{figure}

\subsection{Deep in the IVC phase}

When the hole doping is low, only the lower mean field band is filled, then we can take $n_F(E_{\k,+}) = 0$ for all $\k$. The kinetic energy becomes 
\begin{equation} \label{eq:kinetic}
    \langle H_{kin} \rangle = \sum_{\tau, \k} \xi_\tau(\k) P_{\tau \tau}(\k) =  \sum_{o.c.c} \xi_s(\k) - \sum_{o.c.c} \frac{\Delta^*_{z}(\k)}{|\boldsymbol{\Delta}_{\k}|} \xi_a(\k)
\end{equation}
We first define a reference state with dispersion $\xi_s(\k)$ and work with a reference Fermi surface defined by $\xi_s(k_F(\theta)) = 0$ (Supplementary Figure~\ref{fig:FSCrossing}(a)). In the following, we will fill this reference Fermi pocket for all isospin symmetry broken states instead of the Fermi pocket defined by the mean-field Hamiltonian in Eq.\eqref{eq:MF}. We will add back Fermi surface deformation $\delta k_F$ from the reference Fermi surface perturbatively later. If we neglect Fermi surface deformation, the first term is the kinetic energy of the reference state and does not care about isospin symmetry breaking, so we will focus on the second term.

We start with two analytical limits: VP state with $\boldsymbol{\Delta}(\k) = \Delta_z(\k) \hat{z}$ (with $\Delta_z(\k)>0$) and IVC state with $\boldsymbol{\Delta}(\k) = \Delta_x(\k) \hat{x} + \Delta_y(\k) \hat{y}$. In these two limits, neither of them benefits from the kinetic energy term. For the IVC state, $\Delta^*_{z}(\k) = 0$ for all $\k$. For the VP state, the summation of $\xi_a(\k)$ over occupied states gives zero due to time reversal symmetry.
Now we discuss small perturbations to these two limits. For the VP state, if we introduce a small in-plane component $\boldsymbol{\Delta}(\k) = \left ( \delta_x(\k), \delta_y(\k), \Delta_z(\k)\right)$, leading order correction to the kinetic energy is to the second order in $\delta/\Delta$ since $\Delta/\sqrt{\Delta^2 + \delta^2} \approx 1 - \delta^2/(2 \Delta^2)$. However, a small local valley-polarization in the IVC state $\boldsymbol{\Delta}(\k) = \left ( \Delta_x(\k) , \Delta_y(\k),  \delta_z(\k) \right)$ results in a kinetic energy change which is linear in $\delta/\Delta$,
\begin{equation}
    \Delta E_{kin}^{IVC} = - \sum_{|\k|<k_F(\theta)} \frac{\delta_z(\k)}{|\Delta_{\k}|} \xi_a(\k)
\end{equation}
which is negative if $\delta_z(\k)$ follows the sign of the local valley-Zeeman field $B_{VZ}(\k) = 2 \xi_a(\k)$. 
Since $B_{VZ}(-\k) = - B_{VZ}(\k)$, we can consistently choose $\delta_z(-\k) = - \delta_z(\k)$, resulting in no net valley-polarization.
This kinetic energy gain from valley-isospin vector canting is the major reason why the IVC state can be energetically competitive in spite of the interaction energy cost due to the opposite chirality of two valleys. It may seem that the kinetic energy gain is unbounded and the IVC state will automatically flow to the a fully symmetric metal such that $\mathbf{\Delta}(\k) = \xi_a(\k) \hat{z}$. This does not happen because there is an interaction energy cost associated with local canting as we will show later.

We also note that the VP state gains kinetic energy through Fermi surface deformation $\delta k_F(\theta)$ (Supplementary Figure~\ref{fig:FSCrossing}(a)). 
\begin{align}
    \Delta E_{kin}^{VP} = \sum_{|\k| < k_F(
    \theta) +\delta k_F(
    \theta)} \xi_{-,\k} - \sum_{|\k| < k_F(
    \theta)} \xi_{-,\k}\approx - \left \{ \sum_{|\k| < k_F(
    \theta) +\delta k_F(
    \theta)} \xi_{a,\k} - \sum_{|\k| < k_F(
    \theta)} \xi_{a,\k} \right \}
\end{align}
where we take $\delta k_F(\theta + \pi) \approx - \delta k_F(\theta)$ to ensure that the area within the Fermi surface remains the same, and group terms at $\k$ and $-\k$ in the second equality. The final result corresponds a summation over shaded areas in Supplementary Figure~\ref{fig:FSCrossing}(a), which contributes positively (negatively) in the red (blue) shaded regime. $\Delta E_{kin}^{VP}$ is negative if $\delta k_F(\theta)$ follows the sign of the local valley-Zeeman field $B_{VZ}(\k)$. This can be intuitively understood as the kinetic energy gain by filling within the Fermi surface of the mean-field Hamiltonian defined by $\xi_{-}(k_F^{-}(\theta)) = 0$ instead of the reference Fermi surface. The actual Fermi surface would not deform all the way toward $k_F^{-}(\theta)$ since Fermi surface deformation has an associated interaction energy cost. However, it sets a hard upper limit on how much kinetic energy Fermi surface deformation can gain, which is achieved when the Fermi surface becomes the Fermi surface of the mean-field Hamiltonian, $k_F(\theta)   +\delta k_F(\theta) = k_F^{-}(\theta)$. This is why when the interaction is sufficiently weak, the IVC state gains much more energy than the VP state of which the Fermi surface deformation has already saturated.

Now we turn to discuss the interaction (Fock) energy in this two limits. For the VP state, the covariance matrix takes a simple form
\begin{equation}
    P_{\tau \tau'}(\boldsymbol{k}) = \begin{pmatrix}
        0 & 0\\
        0 & 1
    \end{pmatrix}
\end{equation}
in the $\{+,-\}$ basis. Then the Fock energy becomes
\begin{equation}
    H_{Fock}^{VP} = -\frac{1}{2 A} \sum_{\substack{\mathbf{k}, \mathbf{q}\\o.c.c.}} V_C(\boldsymbol{q}) |\lambda^{--}_{\boldsymbol{q}}(\boldsymbol{k})|^2
\end{equation}
where the summation over $\k$ and $\q$ is performed with $\k$ and $\k + \q$ both inside the reference Fermi surface. We note that this is lower than what one would expect from a fully symmetric metal characterized by $P(\k) = \tau^0/2$,
\begin{equation}
    \Delta E_{Fock}^{
    Stoner} = -\frac{1}{4 A} \sum_{\substack{\mathbf{k}, \mathbf{q}\\o.c.c.}} V_C(\boldsymbol{q}) |\lambda^{--}_{\boldsymbol{q}}(\boldsymbol{k})|^2
\end{equation}
which is the driving force of isospin polarization. The IVC state has a slightly more complicated covariance matrix structure,
\begin{equation}
    P_{\tau \tau'}(\boldsymbol{k}) = \frac{1}{4} \begin{pmatrix}
        1 - \frac{ \delta_z(\k)}{\left | \Delta_{IVC}(\k) \right |}  & e^{i \phi_{\mathbf{k}}} \left ( 1 - \frac{ \delta_z^2(\k)}{2\left | \Delta_{IVC}(\k) \right |^2} \right )\\
        e^{-i \phi_{\mathbf{k}}} \left ( 1 - \frac{ \delta_z^2(\k)}{2\left | \Delta_{IVC}(\k) \right |^2} \right ) & 1 + \frac{ \delta_z(\k)}{\left | \Delta_{IVC}(\k) \right |}
    \end{pmatrix}
\end{equation}
up to second order in $\delta/\Delta$. Then the Fock energy becomes
\begin{align}
    H_{Fock}^{IVC} = &-\frac{1}{4 A} \sum_{\substack{\mathbf{k}, \mathbf{q}\\o.c.c.}} V_C(\boldsymbol{q}) \bigg \{ |\lambda^{++}_{\boldsymbol{q}}(\boldsymbol{k})|^2 \left ( 1 + \frac{ \delta_z(\k) \delta_z(\k+\q)}{\left | \Delta_{IVC}(\k) \right |^2} \right ) \nonumber\\
    &\phantom{=} - \left ( |\lambda^{++}_{\boldsymbol{q}}(\boldsymbol{k})|^2 - |\lambda^{--}_{\boldsymbol{q}}(\boldsymbol{k})|^2\right ) \frac{ \delta_z(\k)+ \delta_z(\k+\q)}{\left | \Delta_{IVC}(\k) \right |} \nonumber \\
    & \phantom{=} + |\lambda^{++}_{\boldsymbol{q}}(\boldsymbol{k})| |\lambda^{--}_{\boldsymbol{q}}(\boldsymbol{k})| \left ( 1 - \frac{ \delta_z^2(\k) + \delta_z^2(\k+\q)}{2\left | \Delta_{IVC}(\k) \right |^2} \right ) \cos \left(\phi_{\boldsymbol{k}}-\phi_{\boldsymbol{k}+\boldsymbol{q}}+\boldsymbol{q} \cdot\left(\boldsymbol{A}^+_{\boldsymbol{k}}-\boldsymbol{A}^-_{\boldsymbol{k}}\right)\right) \bigg \}\\
    &\approx -\frac{1}{2 A} \sum_{\substack{\mathbf{k}, \mathbf{q}\\o.c.c.}} V_C(\boldsymbol{q}) |\lambda^{++}_{\boldsymbol{q}}(\boldsymbol{k})|^2 \Bigg \{ \left ( 1 - \frac{ q^2 \left | \boldsymbol{\nabla}_{\boldsymbol{k}} \delta_z(\k) \right |^2 }{4\left | \Delta_{IVC}(\k) \right |^2} \right ) - \frac{q^2}8 \left(\boldsymbol{\nabla}_{\boldsymbol{k}} \phi_{\boldsymbol{k}}-\left(\boldsymbol{A}^+_{\boldsymbol{k}}-\boldsymbol{A}^-_{\boldsymbol{k}}\right)\right)^{2} \Bigg \}
\end{align}
where we used $\lambda^{\tau \tau}_{\boldsymbol{q}}(\boldsymbol{k}) \approx |\lambda^{\tau \tau}_{\boldsymbol{q}}(\boldsymbol{k})| e^{i \boldsymbol{q} \cdot \boldsymbol{A}^\tau_{\boldsymbol{k}}}$ in the first step (for small $|\q|$), where $\A^\tau(\k) = - i \langle u_{\tau,s,\k} | \mathbf{\nabla}_\k u_{\tau,s,\k} \rangle$ is the Berry connection in valley $\tau = \pm$ \cite{Khalaf2020}. In the second line, we further approximated $|\lambda^{++}_{\boldsymbol{q}}(\boldsymbol{k})| \approx |\lambda^{--}_{\boldsymbol{q}}(\boldsymbol{k})|$, Taylor expanded around small $\mathbf{q}$ and averaged over the dot product.

There are two additional energy cost compared to the VP state. One is the energy cost associated with the phase winding of the IVC order parameter due to opposite chirality in two valleys \cite{Goldhaber,Serlin2019,BCZ2020,ZMS2019,NickPRX}, 
\begin{equation}
    \Delta E_{Fock}^{IVC,wind} = \frac{1}{16 A} \sum_{\substack{\mathbf{k}, \mathbf{q}\\o.c.c.}} q^2 V_C(\boldsymbol{q}) |\lambda^{++}_{\boldsymbol{q}}(\boldsymbol{k})|^2  \left(\boldsymbol{\nabla}_{\boldsymbol{k}} \phi_{\boldsymbol{k}}-\left(\boldsymbol{A}^+_{\boldsymbol{k}}-\boldsymbol{A}^-_{\boldsymbol{k}}\right)\right)^{2}
\end{equation}
The other term is associated with canting of the valley isospin vector,
\begin{equation}
    \Delta E_{Fock}^{IVC,cant} = \frac{1}{8 A } \sum_{\substack{\mathbf{k}, \mathbf{q}\\o.c.c.}} V_C(\boldsymbol{q}) q^2 |\lambda^{++}_{\boldsymbol{q}}(\boldsymbol{k})|^2 \frac{\left | \boldsymbol{\nabla}_{\boldsymbol{k}} \delta_{z}(\k) \right |^2}{\left |\Delta^2_{IVC}(\k)\right|^2}
\end{equation}
Along with the kinetic energy gain $\Delta E_{kin}^{IVC}$, it determines the local valley polarization $\delta_{z}(\k)$ in the IVC state. Thus, the IVC state is energetically more favorable when the Coulomb repulsion is weak since the winding energy cost reduces and the kinetic energy gain increases due to a larger local valley polarization. This is precisely the case in RTG since the large density of states near the Fermi surface at low hole-doping can strongly screen the Coulomb repulsion.

\subsection{Close to the onset of IVC phase}
In this section, we carry out a complimentary analysis of the IVC kinetic energy using linearized band dispersions, by focusing on the crossing points of the valence bands from the two valleys that are gapped out by the development of the IVC order. 
This applies when the order parameter magnitude is small.
It has the advantage of being amenable to explicit analytical evaluation of the IVC kinetic energy including Fermi surface deformation effects, at the expense of introducing a cutoff momentum $\Lambda$ away from  beyond which the IVC order parameter vanishes.

To set up the problem, we consider the intersection points of the Fermi surfaces from the two valleys at an angle $2\theta$, as shown in Supplementary Figure~\ref{fig:FSCrossing}(b). 
All symmetry-related crossings will have  identical contributions to the energy, so it is sufficient to just focus on one crossing.
We introduce a local coordinate system $(p_x,p_y)$ centered at the crossing point, and linearize the band dispersions about this point:
\beq
\xi_\pm(\p) = \v_\pm \cdot \p = v(p_y \sin\theta \pm p_x \cos\theta)
\label{eq:lindisp}
\eeq 
We first establish that at a given filling, the chemical potential remains unchanged across the transition within this linear approximation. 
To do so, we need the mean-field spectrum of the symmetry-broken band structure, which in the most general case ($\Delta_{VP} \neq 0$ and $\Delta_{IVC} \neq 0$) is given by:
\beq
E_{\pm}(\p) &=& \xi_{s}(\p) \pm \sqrt{(\xi_{a}(\p) + \Delta_{VP})^2+ |\Delta_{IVC}(\p)|^2}. 
\eeq 
Within the linearized dispersion, we have:
\beq 
\xi_{s}(\p) &=& \frac{\xi_{+,\p} + \xi_{-,\p}}{2} = v p_y \sin\theta, \text{ and }  \xi_{a}(\p) = \frac{\xi_{+,\p} - \xi_{-,\p}}{2} = v p_x \cos \theta, \nn \text{ and }
E_{\pm}(\p) & = & v p_y \sin \theta \pm \sqrt{(v p_x \cos\theta + \Delta_{VP})^2 + |\Delta_{IVC}|^2}
\eeq
where we have also assumed that we can neglect the $\p$ dependence of the IVC gap $|\Delta_{IVC}|$ near the crossing point. 
We note that $\xi_s(p_x,-p_y) = - \xi_s(p_x, p_y)$ and $\xi_a(p_x,-p_y) = \xi_a(p_x,p_y)$, such that $E_{+}(p_x,p_y) = -E_-(p_x,-p_y)$, indicating that the size of the hole Fermi pocket for the lower band ($E_-$) is the same as the size of the electron Fermi pocket of the upper band ($E_+$). 
More rigorously, we have (within a patch of size $2\Lambda \times 2\Lambda$ centered at the crossing point $\p = 0$):
\beq
\sum_{\p} \Theta(- E_-(\p)) + \Theta(- E_+(\p))  &=& \sum_{\p} \Theta(- E_-(p_x,p_y)) + \Theta(- E_+(p_x, -p_y))
=  \sum_{\p} \Theta(- E_-(\p)) + \Theta(E_-(\p))
=\Lambda^2 \nn
 &=& \sum_{\p} \Theta(- \xi_-(\p)) + \Theta(- \xi_+(\p)) 
\eeq
Thus, we have shown that the occupancy remains unchanged if we retain the same chemical potential, indicating that the chemical potential remains unchanged when $|\Delta_{IVC}| \neq 0$ and/or $\Delta_{VP} \neq 0$ (this continues to hold for $\Delta_{VP}(\p) \propto \xi_a(\p)$ too, i.e, when there is local canting, within the linearized dispersion approximation).

Let us now take specialize to the IVC phase with no local or global valley polarization ($\Delta_{VP} = 0$). 
The covariance matrix for this phase is given by:
\beq
P_{\tau \tau}(\p) &=& \xi_s(\p) (\Theta_{\p,+} + \Theta_{\p,-}) + \frac{\tau \xi_a(\p)}{\sqrt{\xi_{a}^2(\p) + |\Delta_{IVC}(\p)|^2}} (\Theta_{\p,+} - \Theta_{\p,-}), ~~~ \tau = \pm \nn
\langle H_{kin} \rangle_{IVC} &=& \sum_{\p,\tau} \xi_{\tau}(\p) P_{\tau \tau}(\p) =  \sum_{\k} \xi_s(\p) (\Theta_{\p,+} + \Theta_{\p,-}) + \frac{\xi^2_a(\p)}{\sqrt{\xi_{a}^2(\p) + |\Delta_{IVC}(\p)|^2}} (\Theta_{\k,+} - \Theta_{\k,-}) \nn
  && \text{ where } E_{\pm}(\k) = \xi_{s}(\k) \pm \sqrt{(\xi_{a}^2(\k)+ |\Delta_{IVC}(\k)|^2}
\label{eq:HkinIVC}
\eeq
It is instructive to compare the kinetic energy of the IVC state with the kinetic energy $\langle H_{kin} \rangle_0$ of the fully symmetric metal ($\Delta_{IVC}(\k) = 0 = \Delta_{VP}$), and that of the valley polarized metallic phase (VP) where $\Delta_{IVC}(\k) = 0$ but $\Delta_{VP} \neq 0$ in the basis of states with eigen-energies $E_{\pm}$ (i.e, we are no longer labeling the states by valley index $\tau$, even when valley is a good quantum number). 
\beq
\langle H_{kin} \rangle_0 &=& \sum_{\p} \xi_s(\p) (\Theta_{\p,+} + \Theta_{\p,-})  + |\xi_a(\p)|  (\Theta_{\p,+} - \Theta_{\k,-}), ~ E_\pm(\p) = \xi_s(\p) \pm |\xi_a(\p)|  \nn
\langle H_{kin} \rangle_{VP} &=&  \sum_{\p} \xi_s(\p) (\Theta_{\p,+} + \Theta_{\p,-}) + \frac{\xi_a(\p) (\xi_a(\p)+ \Delta_{VP})}{|\xi_a(\p)+ \Delta_{VP}|} (\Theta_{\p,+} - \Theta_{\p,-}), ~ E_\pm(\p) = \xi_s(\p) \pm |\xi_a(\p)+ \Delta_{VP}| ~~~~~~
\label{eq:HkinVP}
\eeq
From Eq.~\eqref{eq:HkinVP}, we note that the VP phase has a higher kinetic energy than the symmetry-preserving metal, as $(\Theta_{\p,+} - \Theta_{\p,-}) < 0$ is always true, and $|\xi_a(\p)| > 0$ while $\xi_a(\p) \text{sign}(\xi_a(\p)+\Delta_{VP})$ takes both positive and negative values. 
Further, comparing Eqs.~\eqref{eq:HkinIVC} and \eqref{eq:HkinVP}, we note that the IVC can have a higher kinetic energy than the symmetry-preserving metal, so long as we neglect the difference in deformation of Fermi surfaces,  and $|\xi_a(\p)| \leq \sqrt{\xi_{a}^2(\p) + |\Delta_{IVC}(\p)|^2}$. 
However, it is incorrect to neglect Fermi surface deformation, and it turns out that within the linearized dispersion approximation in Eq.~\eqref{eq:lindisp}, we can evaluate the kinetic energy $\langle H_{kin} \rangle$ of each of these states analytically. 
These results are presented below for different states (for a single patch, for the whole BZ we have to multiply by the appropriate number of symmetry related crossing points), along series expansions to lowest non-trivial order in $\Delta_{VP/IVC}^2$.

\beq
\langle H_{kin} \rangle_0 &=& - 2 v \Lambda^3 \left( \frac{\sin^2\theta}{3\cos\theta} + \cos\theta \right) \nn
\langle H_{kin} \rangle_{VP} & = & - 2 v \Lambda^3 \left( \frac{\sin^2\theta}{3\cos\theta} + \cos\theta \right) + \frac{2 \Lambda \Delta^2_{VP}}{v \cos\theta} =  \langle H_{kin} \rangle_0 + \frac{2 \Lambda \Delta^2_{VP}}{v \cos\theta}  \nn
\langle H_{kin} \rangle_{IVC} & =&  -   \frac{2 v \sin^2\theta}{3\cos\theta} \left( \Lambda^2 - \frac{\Delta^2_{IVC}}{v^2 \sin^2 \theta} \right)^{3/2} - 2 v \cos\theta \Lambda^2 \left( \Lambda^2 + \frac{\Delta^2_{IVC}}{v^2 \cos^2 \theta}  \right)^{1/2} \nn
&& +  \frac{2 \Delta_{VC}^2}{v \cos\theta} \left[ \left( \Lambda^2 - \frac{\Delta^2_{IVC}}{v^2 \sin^2 \theta}  \right)^{1/2} -  \Lambda \coth^{-1}\left( \frac{\Lambda}{\sqrt{\Lambda^2 - \frac{\Delta^2_{IVC}}{v^2 \sin^2 \theta} }} \right) + \Lambda \tanh^{-1} \left( \frac{\Lambda}{\sqrt{ \Lambda^2 + \frac{\Delta^2_{IVC}}{v^2 \cos^2 \theta}} } \right) \right] \nn
& \approx & - 2 v \Lambda^3 \left( \frac{\sin^2\theta}{3\cos\theta} + \cos\theta \right) + \frac{2 \Lambda \Delta_{IVC}^2}{v \cos\theta} \left(1 + \ln(\cot\theta) \right) \nn
& = &  \langle H_{kin} \rangle_0  +  \frac{2 \Lambda \Delta_{IVC}^2}{v \cos\theta} \left(1 + \ln(\cot\theta) \right) 
\label{eq:Hkin}
\eeq
We note that the IVC state tends pay more kinetic energy penalty as $\theta \to 0$ or $\theta \to \pi/2$, corresponding to $2\theta = 0$ or $\pi$, i.e, perfect nesting of the Fermi surfaces from the two valleys in the patch considered. 
Further, generally for same magnitude of gap, $\langle H_{kin} \rangle_{IVC}  > \langle H_{kin} \rangle_{VP}$, although this should not be taken very seriously as the UV cutoff imposed is somewhat arbitrary and the IVC order parameter $\Delta_{IVC}$ does depend on $\p$ (it is only approximately constant close to the band-crossing points). 

However, the kinetic energy term in the IVC state stands to gain when we introduce a momentum dependent valley polarization $\Delta_{VP}(\p)  \propto \xi_a(\p)$ (valley-antisymmetric part of dispersion). 
While this doesn't lead to an overall valley polarization as time-reversal symmetry of the single-particle band structure implies that $\langle \Delta_{VP}(\p) \rangle = 0$ averaged over a Fermi pocket, it aids the kinetic term. 
This can be seen explicitly by using a mean-field ansatz with $\Delta_{VP}(\p) = (\alpha - 1) \xi_a(\k)$, where $\alpha$ is a control parameter that tunes the degree of canting ($\alpha = 1$ implies no canting). 
For such a state, we note that $E_{\pm}(\p) = \xi_s(\p) \pm \sqrt{(\alpha \xi_a(\p))^2 + |\Delta_{IVC}(\p)|^2}$. Consequently, 
\beq
P_{\tau \tau}(\k) &=& \xi_s(\p) (\Theta_{\p,+} + \Theta_{\p,-}) + \frac{\tau \alpha \xi_a(\p)}{\sqrt{( \alpha \xi_{a}(\p))^2 + |\Delta_{IVC}(\p)|^2}} (\Theta_{\p,+} - \Theta_{\p,-}), ~~~ \tau = \pm \nn
\langle H_{kin} \rangle &=& \sum_{\p,\tau} \xi_{\tau}(\p) P_{\tau \tau}(\p) =  \sum_{\p} \xi_s(\p) (\Theta_{\p,+} + \Theta_{\p,-}) + \frac{\alpha \xi^2_a(\p)}{\sqrt{ ( \alpha \xi_{a}(\p))^2 + |\Delta_{IVC}(\p)|^2}} (\Theta_{\p,+} - \Theta_{\p,-}) 
\eeq
We again approximate $\Delta_{IVC}(\p) \approx \Delta_{IVC}$ near the band-crossing points, and linearize the band dispersions. 
By our previous arguments, the chemical potential remains unchanged for such an ansatz, and the kinetic energy can be calculated explicitly.
\beq
\langle H_{kin} \rangle  &=&  \left( \Lambda^2 - \frac{\Delta^2_{IVC}}{v^2 \sin^2 \theta} \right)^{3/2}  \left[ - \frac{4 v \sin^2\theta}{3 \alpha \cos\theta}+ \frac{2 v \sin^2\theta}{3 \alpha^2 \cos\theta} \right] - 2 v \cos\theta \Lambda^2 \left( \Lambda^2 + \frac{\Delta^2_{IVC}}{\alpha^2 v^2 \cos^2 \theta}  \right)^{1/2} \nn
&+&  \frac{2 \Delta_{IVC}^2}{\alpha^2 v \cos\theta} \left[ \left( \Lambda^2 - \frac{\Delta^2_{IVC}}{v^2 \sin^2 \theta}  \right)^{1/2} -  \Lambda \coth^{-1}\left( \frac{\Lambda}{\sqrt{\Lambda^2 - \frac{\Delta^2_{IVC}}{v^2 \sin^2 \theta} }} \right) + \Lambda \tanh^{-1} \left( \frac{\Lambda}{\sqrt{ \Lambda^2 + \frac{\Delta^2_{IVC}}{\alpha^2 v^2 \cos^2 \theta}} } \right) \right] \nn
& \approx & -\frac{2 v \Lambda^3}{3 v \cos\theta} \left[  \left( \frac{2}{\alpha} - \frac{1}{\alpha^2} \right) \sin^2\theta + 3 \cos^2\theta \right] +  \frac{2 \Delta_{IVC}^2}{\alpha^2 v \cos\theta} \left[ \frac{1}{\alpha} + \frac{1}{\alpha^2}\ln(\alpha \cot\theta) \right]
\eeq
First, we note that the the $\alpha = 1$ limit recovers the IVC kinetic energy in Eq.~\eqref{eq:Hkin}, which acts as a sanity check. 
Next, we note that if we write $\alpha = 1 + \delta_Z$ and expand in small $\delta_Z$, there are corrections to both linear and quadratic orders in $\delta_Z$:
\beq
\langle H_{kin} \rangle = \langle H_{kin} \rangle_{IVC} + \frac{2v\Lambda^3}{3 \cos \theta} \delta_Z^2 +  \frac{2 \Delta_{IVC}^2}{v \cos\theta} \left[ -2 \ln(\cot \theta) \delta_Z + 3\left( \ln(\cot \theta) - \frac{1}{2} \right)\delta_Z^2  \right]
\eeq
Roughly speaking, the linear correction arises from the term involving the valley-symmetric part of the dispersion $\sum_\p \xi_s(\p) (\Theta_{\p,+} + \Theta_{\p,-})$. While $\xi_s(\p)$ itself does not change, the Fermi surface occupancies change, resulting in a linear correction in $\delta_Z$ to the lowest order. 
The quadratic correction arises from the next higher order correction to the symmetric part, as well as to leading order from the valley-antisymmetric part of dispersion, as $\Delta_{VP}(\p)$ directly adds on to $\xi_a(\p)$. 
This implies that the kinetic energy for a given non-zero $\Delta_{IVC}$ cannot have a minima at $\delta_Z = 0$, and thus such a minima will always be shifted to $\delta_Z \neq 0$. 
Thus, the kinetic energy of the IVC state can always be lowered by canting, i.e, by having some $\Delta_{VP}(\p) \propto \xi_a(\p)$ near the Fermi surface. 
This is exactly what is observed in Fig.~2(c) in the main text.

\section{Derivation of inter-valley Hund's coupling}
The inter-valley Hund's coupling plays a crucial role in determining the phases of ABC graphene in presence of a displacement field, as discussed above. 
Therefore, in this section, we work out the inter-valley Hund's coupling term for RTG microscopically by including an arbitrary translation invariant density-density interaction $U(\r,\rp) = U(\r - \rp)$ in real space. 
The following interaction Hamiltonian is our starting point ($:~:$ denotes normal ordering):
\beq
H_I = \frac{1}{2} \sum_{\r,\rp} U(\r,\rp) :\rho(\r) \rho(\rp): 
\label{eq:ReSpU}
\eeq
Each lattice site $\r$ can be labeled by a Bravais lattice position $\R$ of the two-dimensional unit cell, together with a basis position index $a \in \{A_1,B_1,A_2,B_2,A_3,B_3\}$ which determines both sublattice and layer ($\sigma$ in previous/later sections denote a subset of these indices relevant for active bands near charge neutrality). 
Now, we note that the local electron operator can be written in terms of the monolayer low-energy Dirac fermions $c_{\tau,s,\k,a}$ as ($N = $ number of unit cells):
\beq
c_{\R,a,s} = \frac{1}{\sqrt{N}} \sum_{\k, \tau} e^{i (\k + \tau \K) \cdot \R_a} c_{\tau, s, \k, a}
\eeq
The summation over $\k$ is implicitly restricted to $|\k| \leq \Lambda_{UV}$, where $\Lambda_{UV}$ is some ultra-violet cutoff on the scale of the lattice spacing, that also satisfies $\Lambda_{UV} \ll |\K|$.
This implies that the density operator takes the following form:
\beq
\rho(\r) = \rho(\R,a) = \sum_{s} c^\dagger_{\R,a,s} c_{\R,a,s} = \frac{1}{N}\sum_{\k, \q, \tau, \tilde{\tau}, s} e^{i [\q + (\tilde{\tau} - \tau) \K] \cdot \R_a} c^\dagger_{\tau, s, \k, a} c_{\tilde{\tau}, s, \k + \q, a} \equiv \frac{1}{N}\sum_{\q} e^{i \q \cdot \R_a} \rho_a(\q)
\eeq
where by the last equality we have have defined the electron density $\rho_a(\q)$ in momentum space.
Note that it has a slowly varying component which modulates at momenta $|\q|$ (when $\tilde{\tau} = \tau$), and a fast-varying component that modulates at momenta $\q \pm 2\K$ for $\tilde{\tau} =-\tau$.
While only the slowly varying component of density was considered in the interaction term for our Hartree-Fock numerics, here we keep both terms.
Since we want to study the effect of this interaction projected to the `active' low-energy bands, we re-write the valence-band projected density operator in the Bloch or band-basis of RTG:
\beq
\rho_{a}(\q)  =  \sum_{\k,s} \sum_{\tau, \tilde{\tau}} \lambda_{a,\q}^{\tau, \tilde{\tau}}(\k) e^{i [\q + (\tilde{\tau} - \tau) \K] \cdot \R_a}  \psi^\dagger_{\tau, s, \k} \psi_{\tilde{\tau}, s, \k + \q}
\eeq
where have restricted to the valence band above and neglected the $n$ index.
$\lambda_{a,\q}^{\tau, \tilde{\tau}}(\k)$ is the sublattice and layer projected form-factor, given by:
\beq
\lambda_{a,\q}^{\tau, \tilde{\tau}}(\k)  = \bra{u_{\tau,\k}} P_{a} \ket{ u_{\tilde{\tau},\k + \q}}
\eeq
For a translation invariant potential $U(\r - \rp)$ with $\r = (\R, a)$ and $\rp = (\R^\prime, b)$, we can define a Fourier transform as:
\beq 
U(\r,\rp)  = U(\r - \rp) = \frac{1}{A} \sum_{\p} U_{ab}(\p) e^{-i \p \cdot (\R_a - \R^\prime_b)}
\eeq
Plugging this into Eq.~\eqref{eq:ReSpU} and carrying out the summations over the lattice positions, we find that overall momentum conservation leads to net valley-charge conservation (in absence of Umklapp processes which doesn't appear at this order \cite{Aleiner2007}).
Specifically, the interaction term takes the form:
\beq
H_I = \frac{1}{2A} \sum_{\k,s, \k^\prime, s^\prime} \sum_{\tau, \tilde{\tau}} U_{ab}(\p)  \lambda_{a,\q}^{\tau, \tilde{\tau}}(\k) \lambda_{b,-\q}^{\tau^\prime, \tilde{\tau^\prime}}(\k^\prime) : \psi^\dagger_{\tau, s, \k} \psi_{\tilde{\tau}, s, \k + \q} \psi^\dagger_{\tau^\prime, s, \k^\prime} \psi_{\tilde{\tau^\prime}, s, \k^\prime - \q} : \nn 
\times \left( \frac{1}{N} \sum_{\R_a} e^{i [-\p + \q + (\tilde{\tau} - \tau) \K] \cdot \R_a} \right) \left( \frac{1}{N} \sum_{\R_b^\prime}  e^{i [\p - \q + (\tilde{\tau^\prime} - \tau^\prime) \K]\cdot \R_b^\prime} \right)
\eeq
where the summation over $\R_a$ and $\R_b^\prime$ gives:
\beq
\p = \q + (\tilde{\tau} - \tau)\K = \q - (\tilde{\tau^\prime} - \tau^\prime)\K \implies 
(\tau - \tau^\prime + \tilde{\tau} - \tilde{\tau^\prime})\K = 0 \text{ mod } \G
\eeq
where $\G$ is any reciprocal lattice vector. 
Since $3 \K$ satisfies this condition but not $2 \K$, we are only left with two options: (i) $\tau = \tau^\prime,  \tilde{\tau} = \tilde{\tau^\prime}$ which corresponds to intra-valley scattering terms with small momentum transfer $|\p|$, such terms respect SU(2)$_+\times$SU(2)$_-$, and (ii) $\tau = -\tau^\prime = - \tilde{\tau} = \tilde{\tau^\prime}$ which correspond to terms which scatter between valleys, such terms break SU(2)$_+\times$SU(2)$_-$ to a global SU(2)$_s$ and give rise to Hund's. 
Type (i) terms allow for small momenta scattering, and are more important for long-range Coulomb interactions which decay at large momenta as $1/q$.  
Type (ii) terms necessarily involve momentum transfer of $|\p| \approx 2|\K|$, and therefore are more important for short-range electron-electron interactions (such as on-site Hubbard U).
Thus, we arrive at the following form for the inter-valley interaction:
\beq
H_{\rm inter-valley} = \frac{1}{2A} \sum_{\k, \k^\prime, \q, \tau} \sum_{s,s^\prime} \sum_{a,b} \left[ U_{ab}(\q + 2\tau \K) \lambda_{a,\q}^{\tau, -\tau}(\k) \lambda_{b,-\q}^{-\tau, \tau}(\k^\prime) \right] : \psi^\dagger_{-\tau,s,\k} \psi_{\tau, s, \k + \q} \psi^\dagger_{\tau, s^\prime, \k^\prime} \psi_{-\tau, s^\prime, \k^\prime - \q}  : 
\label{eq:HinterSM}
\eeq

To derive the Hund's coupling, we use the Fierz identity $2 \delta_{\alpha \nu} \delta_{\beta \mu } = \s_{\alpha \beta} \cdot \s_{\mu \nu } + \delta_{\alpha \beta} \delta_{\mu \nu}$ on the inter-valley scattering term derived in Eq.~\eqref{eq:HinterSM}.
\beq
\sum_{s,s^\prime} : \psi^\dagger_{-\tau,s,\k} \psi_{\tau, s, \k + \q} \psi^\dagger_{\tau, s^\prime, \k^\prime} \psi_{-\tau, s^\prime, \k^\prime - \q} : ~~ 
= \sum_{\alpha, \beta, \mu, \nu} : \psi^\dagger_{-\tau,\alpha,\k} \psi_{\tau, \beta, \k + \q} \psi^\dagger_{\tau, \mu, \k^\prime} \psi_{-\tau, \nu, \k^\prime - \q} :~ \delta_{\alpha \beta} \delta_{\mu \nu} \nn
= - \sum_{\alpha,
\beta,\mu,\nu} :( \psi^\dagger_{-\tau,\alpha,\k} \s_{\alpha  \beta} \psi_{\tau, \beta, \k + \q} ) \cdot (\psi^\dagger_{\tau, \mu, \k^\prime} \s_{\mu \nu} \psi_{-\tau, \nu, \k^\prime - \q}): + 2 \sum_{\alpha,\beta} : \psi^\dagger_{-\tau,\alpha,\k} \psi_{\tau, \beta, \k + \q} \psi^\dagger_{\tau, \beta, \k^\prime} \psi_{-\tau, \alpha, \k^\prime - \q}:
\label{eq:Fierz1}
\eeq
Note that the second term in Eq.~\eqref{eq:Fierz1} is SU(2)$_+\times$SU(2)$_-$ symmetric, while the first term has only global SU(2) symmetry. Defining a gauge-invariant (inter-valley) site-projected spin-operator as:
\beq
\s_{+-,a}(\q) =  \sum_\k \lambda_{a,\q}^{+ -}(\k) \psi^\dagger_{+,\alpha,\k} \s_{\alpha  \beta} \psi_{-, \beta, \k + \q} 
\eeq
we see that the interaction Hamiltonian can be re-written in a particularly simple manner as:
\beq
H_{\rm Hund's} = -\frac{1}{2A} \sum_{\q,a,b} U_{ab}(\q) \left[ \s_{+-,a}(\q) \cdot \s_{-+,b}(-\q) +  \s_{-+,a}(\q) \cdot \s_{+-,b}(-\q) \right]
\eeq
This takes a particularly simple form when $U_{\R_a, \R^\prime_b} = \tilde{U} \delta_{\R, \R^\prime}$, i.e, there is short-range interaction which is equal on all sublattice sites within the unit cell. 
While this is not physically accurate because of interlayer separation being much larger than intralayer separation, it is nevertheless useful for illustrating the basic physics. 
In this limit, we have $U_{ab}(\q) = (\sqrt{3}a^2/2)\tilde{U} \equiv U$, and therefore:
\beq
H_{\rm Hund's} = -\frac{U}{A} \sum_{\q}\s_{+-}(\q) \cdot \s_{-+}(-\q) = -\frac{U}{A} \sum_{\q}\s_{+-}(\q) \cdot \s_{+-}^\dagger(\q), \text{ where } \s_{\tau,-\tau}(\q) = \sum_a \s_{\tau,-\tau,a}(\q) 
\label{eq:HHundsSM}
\eeq
which is Eq.~(5) in the main text.
Note that the magnitude of $J_H$ is suppressed for long-range Coulomb interactions by a factor of $k_F/|\K| \ll 1$, $k_F$ being the typical Fermi momentum which is small near charge neutrality.
However, it is not necessarily small for short-range e-e scattering, and can thus contribute significantly to the choice of preferred ground state when iso-spin symmetry is broken.

We comment that form of coupling in Eq.~\eqref{eq:HHundsSM} is quite different from the usual Hund's coupling (taken to be of the form $\s_+ \cdot \s_-$), familiar in the context of quantum hall ferromagnets and phenomenologically introduced in Ref.~\onlinecite{Zhou2021_ABCmetals} to explain the flavor polarization observed in experiments.
As we argued in the main text, the data in Refs.~\onlinecite{Zhou2021_ABCmetals,Zhou_ABCSC} can also be explained using the $H_{\rm Hund's}$ microscopically derived in Eq.~\eqref{eq:HHundsSM}. 
Because of the valley-crossed form-factors, after projection to the low-energy bands there is no Fierz transformation which can convert it to the form:
\beq
\tilde{H}_{\rm Hund's} = - \frac{\tilde{J}_H}{A} \sum_{\q} \s_+(\q) \cdot \s_-(-\q), \text{ where } \s_{\tau}(\q) =  \sum_\k \lambda_{\q}^{\tau, \tau}(\k) \psi^\dagger_{\tau,\alpha,\k} \s_{\alpha  \beta} \psi_{\tau, \beta, \k + \q} 
\label{eq:HHunds2SM}
\eeq
Thus, while the Hund's coupling in Eq.~\eqref{eq:HHunds2SM} is symmetry-allowed, any inter-valley electron-electron scattering, whether mediated by local interactions or by phonons, would not give rise to such a term (at least within a hopping model which neglects the orbital overlaps on different sites).
For general $\q$ and $\k$, the opposite Berry-curvature of the two valleys imply that $\lambda_{\q}^{\tau, \tau}(\k) \neq \lambda_{\q}^{\tau, -\tau}(\k)$.
However, when the wave-functions are polarized to a single sublattice (say $A_1$), the distinction between $ \lambda_{\q}^{\tau, \tau}(\k)$ and $ \lambda_{\q}^{\tau, -\tau}(\k)$ vanishes, as these are both equal to one.

To summarize, there are two kinds of symmetry-allowed Hund's coupling, which can be written as follows by allowing for more general non-local interactions.
\beq
H_{\rm Hund's} &=& - \frac{1}{A} \sum_{\q} J_H(\q) ~ \s_{+-}(\q) \cdot \s_{-+}(-\q), \\
\tilde{H}_{\rm Hund's} &=& - \frac{1}{A} \sum_{\q} \tilde{J}_H(\q) ~ \s_+(\q) \cdot \s_-(-\q), 
\eeq
$H_{\rm Hund's}$ and $\tilde{H}_{\rm Hund's} $ are not related by Fierz transformations for Pauli matrices, unlike in a $\nu = 0$ graphene QH system \cite{Kharitonov}, because of the presence of different form-factors arising from valley-projection.
The only exception is the sublattice polarized limit (large displacement field or small kinetic energy), when the form-factors become trivial ($\lambda^{\tau,\tau^\prime}_\q(\k) \approx 1$) and and the two Hamiltonians are equivalent upto SU(2)$_+\times$SU(2)$_-$ symmetric terms.

\section{Effect of Hunds coupling on iso-spin symmetry broken states}
In this section, we consider the perturbative effect of Hund's coupling on the different isospin symmetry-broken states.
In particular, we derive the Hund's coupling induced splitting of the U(2) IVC manifold, relevant to the PIP phase near SC1.
Simultaneously, we also discuss its effect on the spin-polarized IVC state, a possible candidate for the PIP phase near SC2. 
Finally, we also consider how inter-valley scattering can break the degeneracy between spin and valley polarization.

\subsection{IVC states}
First, we consider the Hund's coupling $H_{\rm Hund's}$ derived from microscopic considerations in Eq.~\eqref{eq:HHundsSM}.
We first evaluate its expectation value in a general mean-field state, which is a Slater determinant characterized by a covariance matrix $P_{\tau,\tau^\prime}^{s s^\prime}(\k) = \langle \psi^\dagger_{\tau,s,\k} \psi_{\tau^\prime, s^\prime, \k} \rangle$, and then apply it to the different IVC states on interest (traces indicate tracing over spin indices, sum on index $i=x,y,z$ for Pauli matrices $s^i$ is implicit):
\beq
\langle H_{\rm Hund's} \rangle &=& - \frac{1}{A} \sum_{\q, \k, \k^\prime} J_H(\q) \lambda^{+-}_\q(\k) \lambda^{-+}_{-\q}(\k^\prime) s^i_{\alpha \beta} s^i_{\mu \nu} \langle \psi^\dagger_{+,\alpha,\k} \psi_{-,\beta, \k + \q} \psi^\dagger_{-,\mu,\k^\prime} \psi_{+,\nu,\k^\prime - \q} \rangle_{\rm IVC} \nn
& =& - \frac{1}{A} \sum_{\k, \k^\prime} J_H(\mathbf{0}) \lambda^{+-}_{\q = \mathbf{0}}(\k) \lambda^{-+}_{\q = \mathbf{0}}(\k^\prime)    (\Tr[P_{+-}(\k) (s^i)^T]) (\Tr[P_{-+}(\k^\prime) (s^i)^T])  \nn 
&& + \frac{1}{A} \sum_{\q, \k} J_H(\q) \lambda^{+-}_{\q}(\k) \lambda^{-+}_{-\q}(\k+\q) \Tr[P_{++}(\k)(s^i)^T P_{--}(\k+\q) (s^i)^T ]\nn
& = & - \frac{J_H(\mathbf{0})}{A}  \left( \sum_\k \lambda^{+-}_{\q = \mathbf{0}}(\k) \Tr[P_{+-}(\k) (s^i)^T]  \right) \left( \sum_{\k^\prime} \lambda^{-+}_{\q = \mathbf{0}}(\k^\prime) \Tr[P_{-+}(\k^\prime) (s^i)^T] \right)  \nn
& &  + \frac{1}{A} \sum_{\q, \k} J_H(\q) |\lambda^{+-}_{\q}(\k)|^2 \Tr[P_{++}(\k)(s^i)^T P_{--}(\k+\q) (s^i)^T ]
\label{eq:HHund'sIVC}
\eeq
(i) Spin-singlet CDW IVC: For this state, the U(2) matrix $U_{s s^\prime} = \delta_{s,s^\prime}$, upto an overall phase.
Using this, we find that the covariance matrix takes the form:
\beq
P_{\tau,\tau^\prime}^{s,s^\prime}(\k) &=&  \left[ \frac{1}{2}\left( 1 + \frac{\bm{\Delta_\k}^* \cdot \bm{\tau}}{|\bm{\Delta_\k|}} \right) n_F(E_{\k,+}) + \frac{1}{2}\left( 1 - \frac{\bm{\Delta_\k}^* \cdot \bm{\tau}}{|\bm{\Delta_\k|}} \right) n_F(E_{\k,-}) \right]\delta_{s,s^\prime}, \text{ where } \nn
\bm{\Delta_\k} &\equiv& \left( |\Delta_{IVC}(\k)| \cos(\phi_\k), |\Delta_{IVC}(\k)| \sin(\phi_\k),  \xi_a(\k) \right), \text{ and } E_{\k, \pm} = \xi_s(\k) \pm |\bm{\Delta_\k}|
\eeq
In this case, the first term in Eq.~\eqref{eq:HHund'sIVC} vanishes, and we find:
\beq
\langle H_{\rm Hund's} \rangle_{\rm CDW ~IVC} = \frac{6}{A} \sum_{\q, \k} J_H(\q) |\lambda^{+-}_{\q}(\k)|^2 P_{++}(\k) P_{--}(\k+\q)
\eeq
We note that there is an energy penalty for the CDW IVC for local ferromagnetic Hund's ($J_H > 0$) which is proportional to the overlap of Fermi surfaces, at least in the limit where $J_H$ is local, i.e, nearly independent of $\q$. 
This is expected, as a local repulsive interaction that gives rise to ferromagnetic Hund's also penalizes accumulation of excess charge density. 
The converse is true for antiferromagnetic Hund's $J_H < 0$, which arises from a local attractive interaction and favors accumulation of excess charge density. 

(ii) Spin-triplet SDW IVC: For this state, the U(2) matrix $U_{s s^\prime} = (\hat{\n}\cdot \s)_{s s^\prime}$, upto an overall phase.
Using this, the covariance matrix takes the form ($s^0_{s,s^\prime} = \delta_{s,s^\prime}$): 
\beq 
P^{s,s^\prime}_{\tau,\tau^\prime} &=& \frac{1}{2}\left( 1 + \frac{\left[\bm{\Delta}_{\k}\right]_{s s^\prime}^* \cdot \bm{\tau}}{|\bm{\Delta_\k|}} \right) n_F(E_{\k,+}) + \frac{1}{2}\left( 1 - \frac{\left[\bm{\Delta}_{\k}\right]_{s s^\prime}^* \cdot \bm{\tau}}{|\bm{\Delta_\k|}} \right) n_F(E_{\k,-}) \text{ where } \nn
\left[\bm{\Delta}_{\k}\right]_{s s^\prime} &\equiv& \left( (\hat{\n} \cdot \s) |\Delta_{IVC}(\k)| \cos(\phi_\k) , (\hat{\n} \cdot \s) |\Delta_{IVC}(\k)| \sin(\phi_\k),  s^0 \xi_a(\k) \right)_{s s^\prime}
\eeq
In this case, both terms in Eq.~\eqref{eq:HHund'sIVC} contribute, and we find:
\beq
\langle H_{\rm Hund's} \rangle_{\rm SDW ~IVC} = - \frac{4 J_H(\mathbf{0})}{A}  \bigg| \sum_\k \lambda^{+-}_{\q = \mathbf{0}}(\k)  P_{+-}(\k) \bigg|^2  + \frac{6}{A} \sum_{\q, \k} J_H(\q) |\lambda^{+-}_{\q}(\k)|^2 P_{++}(\k) P_{--}(\k+\q) ~~~~~~
\eeq
Since the mean-field band structures are identical for both CDW and SDW IVC, we see that the first (Hartree) term gives a contribution which is local in real space, and does not depend on the overlap of Fermi surfaces, while the second (Fock) term will depend on such an overlap. 
Accordingly, we see that a ferromagnetic Hund's will strongly favor a SDW IVC, while AF Hund's will disfavor it --- consistent with our previous arguments. 
In fact, this can be directly seen by re-writing $H_{\rm Hund's}$ (for short range Hund's) in terms of the triplet IVC order parameter $\n^{\rm IV}_{\rm T}(\q)$, which is nothing but $\s_{+-}(\q)$:
\beq
H_{\rm Hund's}  = - \frac{J_H}{A} \sum_{\q}  \n^{\rm IV}_{\rm T}(\q) \cdot [\n^{\rm IV}_{\rm T}(\q)]^\dagger, \text{ where } \n^{\rm IV}_{\rm T}(\q) \equiv \sum_\k \lambda^{+-}_\q(\k) \psi^\dagger_{+,s,\k} (\hat{\n} \cdot \s)_{s,s^\prime} \psi_{-,s^\prime, \k + \q}  = \s_{+-}(\q) ~~~~~~~
\eeq

(iii) Spin-polarized IVC: For this state which is relevant at lower doping near SC2, the covariance matrix takes the form (taking spin-polarization axes to be $\hat{z}$):
\beq
P_{\tau,\tau^\prime}^{s,s^\prime}(\k) &=&  \left[ \frac{1}{2}\left( 1 + \frac{\bm{\Delta_\k}^* \cdot \bm{\tau}}{|\bm{\Delta_\k|}} \right) n_F(E_{\k,+}) + \frac{1}{2}\left( 1 - \frac{\bm{\Delta_\k}^* \cdot \bm{\tau}}{|\bm{\Delta_\k|}} \right) n_F(E_{\k,-}) \right]\delta_{s,\ua} \delta_{s^\prime,\ua},
\eeq
For the spin-polarized (SP) IVC as well, both terms in Eq.~\eqref{eq:HHund'sIVC} contribute, and we find:
\beq
\langle H_{\rm Hund's} \rangle_{\rm SP ~IVC} = - \frac{J_H(\mathbf{0})}{A}  \bigg| \sum_\k \lambda^{+-}_{\q = \mathbf{0}}(\k)  P_{+-}(\k) \bigg|^2  + \frac{1}{A} \sum_{\q, \k} J_H(\q) |\lambda^{+-}_{\q}(\k)|^2 P_{++}(\k) P_{--}(\k+\q)
\eeq
We note that a ferromagnetic Hund's term prefers spin-polarization over a singlet IVC, but it prefers SDW IVC over spin-polarized ferromagnetic IVC.
Intuitively this happens because the spin-polarized IVC is still a CDW in one-spin species, so although some exchange energy is gained from spin-polarization it is not enough to offset the energy penalty from non-uniform distribution of charge density from a local repulsion that gives rise to ferromagnetic Hund's coupling. 

Having discussed in detail the miscroscopically derived Hund's term, we now consider the other symmetry-allowed Hund's coupling, as detailed in Eq.~\eqref{eq:HHunds2SM}, i.e, $\tilde{H}_{\rm Hund's} = - \frac{\tilde{J}_H}{N} \sum_{\q} \s_{+}(\q) \cdot \s_{-}(-\q)$. 
Note that the above Hamiltonian is Hermitian only if $\tilde{J}_H(\q)= \tilde{J}_H(-\q)$, which we will implicitly assume in what follows. The expectation value of this term in a mean-field state is given by:
\beq
\langle \tilde{H}_{\rm Hund's} \rangle &=& - \frac{1}{A} \sum_{\q, \k, \k^\prime} \tilde{J}_H(\q) \lambda^{++}_\q(\k) \lambda^{--}_{-\q}(\k^\prime) s^i_{\alpha \beta} s^i_{\mu \nu} \langle \psi^\dagger_{+,\alpha,\k} \psi_{+,\beta, \k + \q} \psi^\dagger_{-,\mu,\k^\prime} \psi_{-,\nu,\k^\prime - \q} \rangle_{\rm IVC} \nn
& =& - \frac{\tilde{J}_H(\mathbf{0})}{A} \sum_{\k, \k^\prime} \lambda^{++}_{\q = \mathbf{0}}(\k) \lambda^{--}_{\q = \mathbf{0}}(\k^\prime)   (\Tr[P_{++}(\k)(s^i)^T] ) (\Tr[P_{--}(\k^\prime) (s^i)^T] ) \nn 
&& + \frac{1}{N} \sum_{\q, \k} \tilde{J}_H(\q) \lambda^{++}_{\q}(\k) \lambda^{--}_{-\q}(\k+\q) \Tr[(s^i)^T M (s^i)^T M] \nn
& = & -\frac{\tilde{J}_H(\mathbf{0})}{A} \left( \sum_{\k}  \Tr[P_{++}(\k)(s^i)^T] \right) \left( \sum_{\k^\prime} \Tr[P_{--}(\k^\prime)(s^i)^T] \right)  \nn
& & + \frac{1}{A} \sum_{\q, \k} \tilde{J}_H(\q) \lambda^{++}_{\q}(\k) [\lambda^{--}_{\q}(\k)]^* \Tr[(s^i)^T P_{+-}(\k) (s^i)^T P_{-+}(\k + \q)] 
\eeq
where $P^{s,s^\prime}_{\tau,\tau^\prime}$ is the projector onto the HF mean-field IVC ground state, the trace is over spin degrees of freedom, and we have used $\lambda^{\tau \tau}_{\q = 0}(\k) = 1$ by virtue of normalization of Bloch-wavefunctions.
Note that for unitary IVC, the first term vanishes as the valley-diagonal projectors are proportional to $\delta_{s,s^\prime}$. 
Further, the second term always features non-trivial winding, as discussed earlier in the context of IVC energetics.
In particular, we note that $P_{+-}(\k) \sim e^{-i \phi_\k}$, and $\lambda^{++}_{\q}(\k) [\lambda^{--}_{\q}(\k)]^* \sim e^{- i \q \cdot (\bm{A}_+ - \bm{A}_-)}$ where $\bm{A}_\pm(\k)$ denote the Berry-connection in the $\tau = \pm$ valleys (see previous discussion on IVC energetics for a detailed description).
So the sign of this term is not uniform and its effect will be quite small if $\tilde{J}_H(\q)$ is local, i.e, approximately independent of $\q$. 
Therefore, even if it is present, we generally expect the effect of this term to be quite small for unitary IVCs. 
More explicitly, we have the following contributions for the three kinds of IVC states.

(i) Spin-singlet CDW IVC is weakly favored by antiferromagnetic coupling  ($\tilde{J}_H < 0$), as:
\beq
\langle \tilde{H}_{\rm Hund's} \rangle_{\rm CDW ~IVC} = \frac{6}{A} \sum_{\q, \k} \tilde{J}_H(\q) \lambda^{++}_{\q}(\k) [\lambda^{--}_{\q}(\k)]^* P_{+-}(\k) P_{-+}(\k + \q) 
\eeq

(ii) Spin-triplet SDW IVC is weakly favored by ferromagnetic coupling ($\tilde{J}_H > 0$), as:  
\beq
\langle \tilde{H}_{\rm Hund's} \rangle_{\rm SDW ~IVC} = -\frac{2}{A} \sum_{\q, \k} \tilde{J}_H(\q) \lambda^{++}_{\q}(\k) [\lambda^{--}_{\q}(\k)]^* P_{+-}(\k) P_{-+}(\k + \q) 
\eeq

(iii) Spin-polarized IVC: For such a state, the expectation value of $H_{\rm Hund's}$ includes both a \textit{Hartree} and \textit{Fock} contribution, unlike the previous two cases where the \textit{Hartree} contribution was zero due to lack of net spin-polarization. 
Taking the spin-quantization axis to be $\hat{\n} = \hat{z}$, we have (using $\lambda^{\tau \tau}_{\q = 0}(\k) = 1$):
\beq
\langle \tilde{H}_{\rm Hund's} \rangle_{\rm SP ~ IVC} &=& - \frac{\tilde{J}_H(\mathbf{0})}{A} \left( \sum_{\k} P_{++}(\k) \right)  \left(\sum_{\k^\prime} P_{--}(\k^\prime) \right)  + \frac{1}{A} \sum_{\q, \k} \tilde{J}_H(\q) \lambda^{++}_{\q}(\k) [\lambda^{--}_{\q}(\k)]^* P_{+-}(\k) P_{-+}(\k + \q) ~~~~~~~~~ 
\eeq
The first term involves a sum over $\k$, and is therefore completely local. 
In fact, since the total number of dopants is given by $\sum_{\k,\tau} P_{\tau \tau}(\k) = N_h$, if we assume that only one band is filled then this term just gives the net alignment energy of all the spins $- J_H N_h^2/4A$. 
The second term can be thought of as a \textit{Fock} contribution, which depends on the overlap of Fermi surfaces of the mean-field bands when displaced by $\q$, and the product of the form factors which decay with $\q$. 
Therefore, it decays fast a function of $\q$, and also contains non-trivial winding which further decreases the overall magnitude of this contribution. 
Thus, the \textit{Hartree} contribution dominates, and aids spin-polarization for FM Hund's $\tilde{J}_H > 0$ (opposes it or penalizes a spin-polarized IVC for AFM Hund's $\tilde{J}_H  < 0$). 

To summarize, the microscopically derived Hund's coupling $J_H$ prefers the SDW IVC when it originates from spatially local repulsive interaction and is ferromagnetic, and it prefers the CDW IVC when it originates from local attractive interactions and is antiferromagnetic. 
This follows from the observation that local attractive interactions favor a CDW, while local repulsive interactions prefer equal charge density on all sites. 
The spin-polarized IVC state (which is also a CDW) is only weakly favored by a ferromagnetic Hund's term, as the exchange energy gained by aligning spins competes with the energy penalty from excess charge density accumulation. 
In contrast, even though a Hund's coupling with $\tilde{J}_H$ is symmetry-allowed, it does not directly arise within our microscopic calculation, and is expected to be quite small. 
This form of Hund's term (which is, for example, discussed in Ref.~\onlinecite{Zhou2021_ABCmetals}) has a weak effect on the IVC states, and prefers spin-polarization instead when it is ferromagnetic. 
When antiferromagnetic, it prefers the spin-singlet CDW IVC, but only weakly due to the winding of form-factors.

\subsection{Isospin polarized states without IVC}
Since the experiment \cite{Zhou_ABCSC,Zhou2021_ABCmetals} finds spin-polarization at smaller hole-doping, it is natural to ask if our proposed Hund's term can lead to spin-polarization while valley remains a good quantum number. 
Here we show that ferromagnetic $J_H$ can indeed favor such spin-polarized states over a spin-valley locked state (which has oppositely aligned spins for the valleys), and over a valley-polarized phase. 
For this purpose, we will evaluate $\langle H_{\rm Hund's} \rangle$ for (a) a spin-polarized ferromagnetic state, and (b) a spin-valley locked state, with spins pointing in opposite directions in the two valleys. 
\beq
\langle H_{\rm Hund's} \rangle_{\rm SP/SVL}  &=& - \frac{1}{A} \sum_{\q, \k, \k^\prime} J_H(\q) \lambda^{+-}_{\q}(\k) \lambda^{-+}_{-\q}(\k^\prime) s^i_{\alpha \beta} s^i_{\mu \nu} \left( P^{\alpha \beta}_{+-}(\k)  P^{\mu \nu}_{-+}(\k^\prime) \delta_{\q,\mathbf{0}} - P^{\alpha \nu}_{++}(\k) P^{\mu \beta}_{--}(\k^\prime) \delta_{\k^\prime, \k + \q} \right) \nn
&=& \frac{1}{A} \sum_{\q, \k} J_H(\q) |\lambda^{+-}_{\q}(\k)|^2  \Tr[P_{++}(\k) (s^i)^T P_{--}(\k + \q) (s^i)^T]
\eeq 
where we have noted that in both cases, $P^{s,s^\prime}_{\tau,\tau^\prime} \propto \delta_{s,s^\prime} \delta_{\tau, \tau^\prime}$ so the Hartree term does not contribute.

(a) Spin-polarized ferromagnet: The covariance matrix is given by $P^{s,s^\prime}_{\tau,\tau^\prime}(\k) = P_{\tau,\tau^\prime}(\k) \left( \frac{1 + \hat{\n} \cdot \s}{2} \right)_{s,s^\prime}$. Choosing $\hat{\n} = \hat{z}$ for simplicity (although our answer does not depend on this choice), we find that:
\beq
\langle H_{\rm Hund's} \rangle_{\rm SP} = \frac{1}{A} \sum_{\q,\k} \tilde{J}_H(\q) |\lambda^{+-}_{\q}(\k)|^2 P_{++}(\k) P_{--}(\k + \q)
\eeq

(b) Spin-valley locked state: This state has spins pointing in opposite directions $\pm \hat{\n}$ in the two valleys, with a covariance matrix given by $P^{s,s^\prime}_{\tau,\tau^\prime}(\k) = P_{\tau,\tau^\prime}(\k) \left( \frac{1 + \tau \hat{\n} \cdot \s}{2} \right)_{s,s^\prime}$. 
Once again, choosing $\hat{\n} = \hat{z}$ for simplicity, we have:
\beq
\langle H_{\rm Hund's} \rangle_{\rm SVL} = \frac{2}{A} \sum_{\q,\k} \tilde{J}_H(\q) |\lambda^{+-}_{\q}(\k)|^2 P_{++}(\k) P_{--}(\k + \q)
\eeq
Thus, this Hund's term has a larger penalty for the spin-valley locked state compared to the spin-polarized state, and thus favors the spin-polarized state. Note that this is exactly the kind of behavior that would also be a consequence of the other kind of Hund's term $\tilde{H}_{\rm Hund's}$ in Eq.~\eqref{eq:HHunds2SM}, which will directly favor the spin-polarized phase over the spin-valley locked phase at the Hartree level. The surprising feature is that even $H_{\rm Hund's}$ in Eq.~(5) in the main text chooses the same term, and microscopically the origin of this lies at noting that a locally repulsive interaction that gives rise to ferromagnetic $J_H$ will naturally favor a spatially anti-symmetric wavefunction to minimize local replusion, leading to spin polarization.

Since $\langle H_{\rm Hund's} \rangle = 0$ for a valley-polarized state, one might be tempted to conclude that it favors valley-polarization over spin-polarization.
However, this conclusion is incorrect.
The reason is that the SU(2)$_+ \times$SU(2)$_-$ symmetric interaction terms that we discarded following the application of the Fierz identity in Eq.~\eqref{eq:Fierz1} play an important role in determining the energy difference between spin and valley polarized states, as these terms break the putative SU(4) symmetry that allows rotating between spin and valley degrees of freedom.
Therefore, we should start directly with Eq.~\eqref{eq:HinterSM} to see that this is not the case, and spin-polarization is favored over valley-polarization, as we do explicitly next.
For simplicity, we assume $U(\r - \rp)$ is short-range, and does not depend on sublattice index $a$, i.e, $ U_{ab} (\q + 2 \tau \K) = U$. 
Then, we have:
\beq
H_{\rm inter-valley} = \frac{U}{2A} \sum_{\k, \k^\prime, \q, \tau} \lambda^{-\tau,\tau}_{\q}(\k) \lambda^{\tau,-\tau}_{-\q}(\k^\prime) :\psi^\dagger_{-\tau,s,\k} \psi_{\tau, s, \k + \q} \psi^\dagger_{\tau, s^\prime, \k^\prime} \psi_{-\tau, s^\prime, \k^\prime - \q} :
\eeq
For any-valley and spin-diagonal ansatz, i.e, $P^{s,s^\prime}_{\tau,\tau^\prime}(\k) \neq 0$ only if $\tau = \tau^\prime$ and $s = s^\prime$ we note that only the Fock term contributes. Therefore:
\beq
\langle H_{\rm inter-valley} \rangle =  -\frac{U}{2A} \sum_{\k, \q} |\lambda^{\tau,-\tau}_\q(\k)|^2 \Tr\left[ P_{\tau,\tau}(\k) P_{-\tau,-\tau}(\k + \q) \right]
\eeq
Using this, we see that:
\beq
\langle H_{\rm inter-valley} \rangle = \begin{cases}  - \frac{U}{2A} \sum_{\k, \q, \tau} |\lambda^{\tau,-\tau}_\q(\k)|^2 P_{\tau,\tau}(\k) P_{-\tau, -\tau}(\k + \q), \text{ spin-polarized} \\ 0, \text{ spin-valley locked} \\ 0, \text{ valley-polarized}   \end{cases}
\eeq
Thus, a locally repulsive interaction ($U > 0$) prefers the spin-polarized (ferromagnetic) state over the spin-valley locked state (spins oppositely aligned in the two valleys) and valley-polarized state. 
Note, however, that the latter two are not affected by the inter-valley coupling to lowest order in perturbation theory --- a conclusion that relies on perturbatively studying the full inter-valley scattering term (and not just the sign of inter-valley Hund's coupling). 

\section{Superconductivity}

\subsection{Symmetries of pairing correlations}
In this section, we elaborate on the symmetries of the pairing correlation function.
The action of the symmetry operators on the $c$ fermions are given in Eq.~\eqref{eq:SymSM}, the same action carries over to the $\psi$ fermions since these are related by a change of basis.
In the most general scenario, the pair-correlation function $4 \times 4$ matrix in spin and valley space: $F^{\tau, \tau^\prime}_{s,s^\prime}(\k) = \langle \psi_{\tau,s,-\k} \psi_{\tau^\prime,s^\prime,\k} \rangle$.
Anticommutation of the fermion field operators implies that:
\beq
F^{\tau, \tau^\prime}_{s,s^\prime}(\k) = - F^{\tau^\prime,\tau}_{s^\prime,s}(-\k)
\label{eq:Fantisym}
\eeq
We focus on inter-valley pairing so that $\tau^\prime = -\tau$, so that  we can write $F^{\tau, \tau^\prime}_{s,s^\prime}(\k) = \tau^\alpha F_{s,s^\prime}(\k)$, with $\alpha = x$ or $y$, and $F_{s,s^\prime}(\k) = \langle \psi_{-,s,-\k} \psi_{+,s^\prime,\k} \rangle$ is $2 \times 2$ matrix in spin-space.

We first classify $F_{s,s^\prime}(\k)$ by the action of global spin-rotation symmetry SU(2)$_s$. 
Spin-singlet pair-correlations are invariant under  SU(2)$_s$ and can be written as $F_{s,s^\prime}(\k) =  i s^y f_\k$, while (unitary) spin-triplet correlations transform as a vector under SU(2)$_s$ and can be written as $F_{s,s^\prime}(\k) = i s^y (\hat{\d}\cdot \s) f_\k$, where $\hat{\d}$ is a real unit-vector. 
However, we note that $f_\k$ can be either odd or even under $\k \to -\k$ for both singlets and triplets, depending on the choice of $\tau^y$ or $\tau^x$ for the inter-valley pairing, such that $F^{\tau, \tau^\prime}_{s,s^\prime}(\k)$ is appropriately antisymmetric (see Eq.~\eqref{eq:Fantisym}) . 

Next we discuss spatial symmetries. 
Since we consider inter-valley pairing, $f_\k$ is invariant under translation symmetry. 
We may also classify by $C_3$ rotations about the $K/K^\prime$ points, since $\k$ is measured relative to these points in the BZ.
Since $C_3$ is a symmetry about $K/K^\prime$ points \cite{Cvetkovic}, we have $f_\k \to f_{C_3 \k} = e^{2 \pi i L_z/3} f_\k$, where $L_z = 0,1,2$ are distinct. 
Note that $L_z = 3$ transforms trivially under $C_3$, and cannot be directly used to distinguish order parameters of the form $f_\k = \text{Im}[(k_x + i k_y)^3]$ from $f_\k$ independent of $\k$ ($L_z = 0$).
However, there is an additional mirror symmetry $M_x$, as we discussed previously (see Eq.~\eqref{eq:SymSM}). 
Consider the spin-singlet nodal order parameter $F(\k) = \tau^y s^y f_\k$, with $f_\k = \text{Im}[(k_x + i k_y)^3] = k_y(3 k_x^2 - k_y^2)$. 
Then, under $M_x \tilde{\mathcal{T}}$, where $\tilde{\mathcal{T}}$ is spinless time-reversal, $f_\k \to f^*_{M_y(\k)} = -f_{\k}$, where $M_y(\k) = (k_x, - k_y)$. 
So the spin-singlet nodal superconductor is odd under $M_x \tilde{\mathcal{T}}$, with nodes at $k_y = 0$ and $C_3$-related points in the Brillouin Zone.
In contrast, the fully-gapped spin-singlet s-wave with $F(\k) = \tau^x s^y$ is even under $M_x \tilde{\mathcal{T}}$.
Thus, the nodal superconductor is distinguished from the gapped s-wave superconductor by $M_x \tilde{\mathcal{T}}$. 
Finally, we comment that $\text{Im}[(k_x + i k_y)^3]$ and $\text{Re}[(k_x + i k_y)^3]$ transform as distinct irreps of the symmetry group, and therefore are not energetically degenerate, as can be seen from the fact that $\text{Re}[(k_x + i k_y)^3]$ is even under $M_x \tilde{\mathcal{T}}$. This is unlike $\text{Re}(k_x + i k_y)$ and $\text{Im}(k_x + i k_y)$ which belong to a single two-dimensional irrep, and are degenerate.

\begin{figure}
    \centering
    \includegraphics[width=\textwidth]{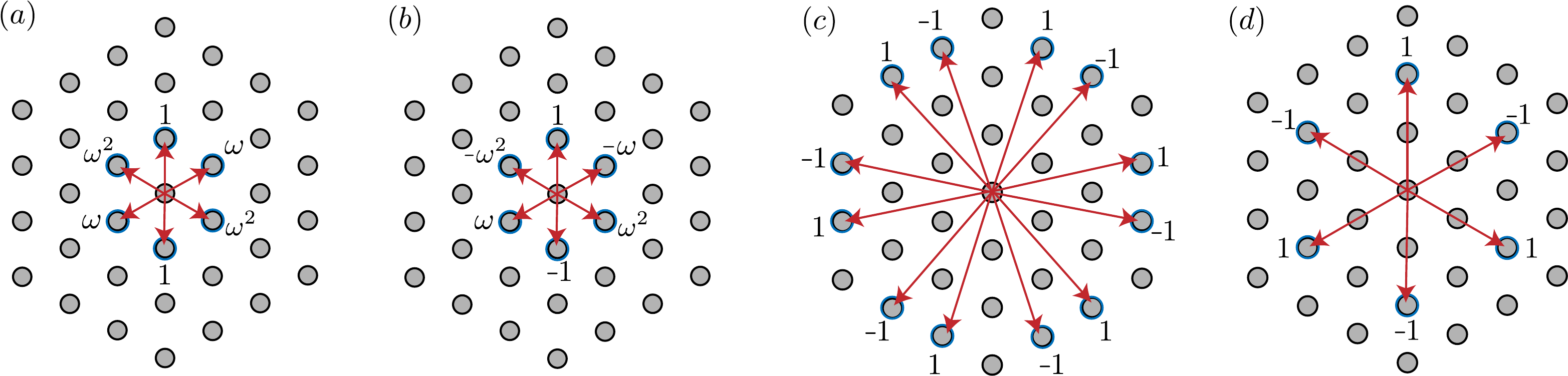}
    \caption{Recipe for extension of pair-correlation functions to the full BZ for (a) chiral spin-singlet, $d + id$, (b) chiral spin-triplet, $p +ip$, (c) nodal spin-singlet, $i$-wave, and (d) nodal spin-triplet, $f$-wave. The momentum-dependent part in each case is given by $\Delta(\q)\sim \sum_{i} c_i e^{i \q \cdot \mathbf{a}_i}$, where the required lattice translation vectors $\mathbf{a}_i$ corresponding to the real space triangular lattice formed by $A_1/B_3$ are shown by red arrows, and the corresponding weights $c_i$ (when non-zero) are noted beside each site.}
    \label{fig:BZpairings}
\end{figure}

It is also instructive to consider the extensions of the pairing wave-functions from small momenta patches around the $K/K^\prime$ points, to the entire BZ.
Fermionic anticommutation relations constrain that for a spin-singlet (spin-triplet) superconductor, such a pair-correlation function is necessarily even (odd) in the momentum $\q$ measured relative to the $\Gamma$-point in the BZ, i.e, have definite transformation under $C_2: \q \to - \q$. 
To construct the simplest extensions (with least number of nodes) of the pair-correlations that are compatible with their behavior around $K,K^\prime$ points and periodic in the BZ (i.e, obey lattice translation symmetry), we consider functions of the form $\Delta(\q) \sim \sum_{i} c_i e^{i \q \cdot \mathbf{a}_i}$, where $c_i$ are complex weights that are chosen to be consistent with the behavior under $C_3$ (about $\Gamma$) and $M_x$, and $\mathbf{a}_i$ form the minimal set of lattice translation vectors (on the triangular lattice formed by $A_1/B_3$) that allow us to impose the required odd/even behavior under $M_x$ and $C_2$.
These choices are shown pictorially in Supplementary Figure~\ref{fig:BZpairings}.
In particular, note that simplest extension compatible with the behavior of the chiral gapped superconductor ($L_z = 1$ about $K$) is $d_{x^2 - y^2} + i d_{xy}$ for a spin-singlet, and $p_x + i p_y$ for a spin-triplet.
For the nodal superconductor, the structure is more complicated. 
For the spin-triplet, it is f-wave (angular momentum 3 about the $\Gamma $ point), while for the spin-singlet, it is i-wave (angular momentum 6 about the $\Gamma $ point).

\subsection{Derivation of gap equations}
In this section, we derive superconducting gap equations, which we solve numerically, in different scenarios.
To this end, we will project all interactions onto the valence bands, which is a reasonable approximation as long as the typical interaction energy scales are less relative to the band-gap.
To check this, note that near the K (or K$^\prime$) point, the bands are quite flat and the displacement-field induced gap between the valence band and the conduction band is around 64 meV at $u = 30$ meV. 
The typical interaction scale, for unscreened Coulomb interaction, is given by $E_C = e^2/ 4\pi \epsilon \epsilon_0 \langle r \rangle$, where $\langle r \rangle \approx 1/\sqrt{n_h}$ is the average separation between the charge carriers. 
For the experimentally relevant carrier-density $n_h \sim 10^{12} $ cm$^{-2}$ and hBN dielectric constant $\epsilon = 4.4$, we find that $E_C \approx 33$ meV, which is smaller than the band-gap, albeit not by an order of magnitude.
Dual-gate screening reduces this estimate to $E_C \approx 28$ meV, and screening due to itinerant hole-carriers in the sample itself will further decrease $E_C$. 
In addition, the matrix elements coming from imperfect wavefunction overlaps between the valence and conduction bands would act to suppress interband scattering.
Therefore, it is reasonable to project the interaction onto the valence band to analyze fluctuation-mediated superconductivity.

We first consider IVC fluctuations in the SU(2)$_+\times$SU(2)$_-$ symmetric limit where spin-singlet and triplet superconductors are degenerate, and then turn to their splitting due to relative amplification of CDW vs SDW fluctuations due to Hund's coupling.
Simultaneously, we also consider spin-polarized IVC fluctuations which may be relevant for SC2.
Finally, we will also consider the effect of Coulomb interactions on pairing at a mean-field level.

In the SU(2)$_+\times$SU(2)$_-$ symmetric limit, a phenomenological Hamiltonian for IVC fluctuations is given by:
\beq
H^{\rm eff}_{\rm IVC} = - \frac{1}{A} \sum_{\q} g_{\q} \, \Tr\left[ n^{\rm IV}(\q) [n^{\rm IV}(\q)]^\dagger \right], , \text{ where } n^{\rm IV}_{s,s^\prime}(\q) = \sum_\k \lambda^{+-}_{\q}(\k) \psi^\dagger_{+,s,\k} \psi_{-, s^\prime \k + \q}, \text{ and } g_\q = \frac{g}{\q^2 + \xi_{IVC}^{-2}} ~~~
\eeq
We decouple this interaction in the BCS channel:
\beq
H^{\rm eff}_{\rm IVC} \xrightarrow{\q = - (\k + \k^\prime)} \frac{1}{A} \sum_{\k, \k^\prime} V_{\k, \k^\prime} \psi^\dagger_{+,s, \k} \psi^\dagger_{-,s^\prime, -\k} \psi_{-,s^\prime,-\k^\prime} \psi_{+, s, \k^\prime}, \text{ with } V_{\k, \k^\prime} = g_{\q = - \k - \k^\prime} |\lambda^{+-}_{\q= - (\k + \k^\prime)}(\k)|^2 
\label{eq:HIVCBCS}
\eeq
Note that the effective interaction $V_{\k, \k^\prime}$ in the valence band projected basis is symmetric under $\k \to \k^\prime$, and it is repulsive. Further, it is maximum at $\k^\prime = -\k$, stemming for the fact that both $g_{\q}$ is maximized at $\q = 0$, and and $| \lambda^{+-}_{\q}(\k)|$ is relatively featureless at small $|\q$ and $|\k|$ due to substantial sublattice polarization. 
We define the mean-field superconducting gap matrix (in spin-space):
\beq
\Delta_{ss^\prime}(\k) = \frac{1}{A} \sum_{\k} V_{\k, \k^\prime} \langle \psi_{-,s,-\k^\prime} \psi_{+,s^\prime, \k} \rangle = \frac{1}{A} \sum_{\k} V_{\k, \k^\prime} F_{ss^\prime}(\k^\prime)
\eeq
This implies that:
\beq
\Delta^*_{ss^\prime}(\k) =  \frac{1}{A} \sum_{\k} V_{\k, \k^\prime} \langle  \psi^\dagger_{+,s^\prime, \k} \psi^\dagger_{-,s,-\k^\prime} \rangle
\eeq
Now, we perform a mean-field decoupling of $H^{\rm eff}_{\rm IVC}$  as follows:
\beq
H_{\rm IVC} \xrightarrow{\rm mean ~ field} \sum_{\k,s,s^\prime} \Delta^*_{s^\prime s}(\k)  \psi_{-,s^\prime,-\k^\prime} \psi_{+,s, \k} + \Delta_{s^\prime s}(\k) \psi^\dagger_{+,s, \k} \psi^\dagger_{-,s^\prime,-\k^\prime}  - A \sum_{\k, \k^\prime, s, s^\prime}   \Delta^*_{s^\prime s}(\k) V^{-1}_{\k, \k^\prime} \Delta_{s^\prime s}(\k^\prime) ~~~~~~
\label{eq:MFIVCsym}
\eeq
where by $V^{-1}$ we have denoted the inverse of the matrix $V_{\k, \k^\prime}$ in momentum space. 
Therefore, the total mean-field Hamiltonian can be written in the Nambu (particle-hole) space as:
\beq
H_{MF} &=& H_{\rm BdG} - A \sum_{\k, \k^\prime, s, s^\prime}  \Delta^*_{ s^\prime s}(\k) V^{-1}_{\k, \k^\prime} \Delta_{s^\prime s}(\k^\prime), \text{ where }  \nn H_{\rm BdG} &=& \sum_{\k, s, s^\prime} \begin{pmatrix}
\psi^\dagger_{+, s, \k} & \psi_{-,s^\prime, -\k}
\end{pmatrix} \begin{pmatrix}
\xi_{+,\k} & \Delta_{s^\prime s}(\k) \\ \Delta^*_{s^\prime s}(\k) & - \xi_{-,-\k}
\end{pmatrix} \begin{pmatrix}
\psi_{+, s, \k} \\ \psi^\dagger_{-,s^\prime, -\k} 
\end{pmatrix}
\label{eq:MFHBDGsym}
\eeq
The free-energy of the system can be written as:
\beq
\mathcal{F} = -\frac{1}{\beta} \ln(Z)  - A \sum_{\k, \k^\prime, s, s^\prime}   \Delta^*_{s^\prime s}(\k) V^{-1}_{\k, \k^\prime} \Delta_{s^\prime s}(\k^\prime)  
\eeq
where $Z$ is the partition function of the system, given in terms of a sum over momenta and fermionic Matsubara frequencies $\omega_n = (2n+1)\pi/\beta$:
\beq
\frac{1}{\beta} \ln(Z)  = \frac{1}{\beta} \sum_{\k, i\omega_n, s, s^\prime} \ln\left[ (- i \omega_n + \xi_+(\k)) (- i\omega_n - \xi_-(-\k)) - |\Delta_{s^\prime s}(\k)|^2 \right]
\eeq
Now, we expand the logarithm in a power series in $\Delta_{s^\prime s}(\k)$, with the aim to derive a gap equation close to $T_c$ where the gap goes to zero. 
\beq
\frac{1}{\beta} \ln(Z) &\approx& \text{const.} - \sum_{\k, s, s^\prime} |\Delta_{s^\prime s}(\k)|^2 \left( \frac{1}{\beta} \sum_{\omega_n} \frac{1}{(- i \omega_n + \xi_+(\k)) (- i\omega_n - \xi_-(-\k))} \right) \nn
& = & \text{const.} - \sum_{\k, s, s^\prime} |\Delta_{s^\prime s}(\k)|^2 \left( \frac{1 - 2 n_F(\xi_+(\k))}{2 \xi_+(\k)} \right)
\eeq
where we have used time-reversal symmetry to set $ \xi_-(-\k) = \xi_+(\k)$, and $n_F(\xi) = (\exp(\beta \xi) +1)^{-1}$ is the Fermi-function. This leads to the following expression for the free energy:
\beq
\mathcal{F}  =  \text{const.} + \sum_{\k, s, s^\prime} |\Delta_{s^\prime s}(\k)|^2 \left( \frac{1 - 2 n_F(\xi_+(\k))}{2 \xi_+(\k)} \right)  - A \sum_{\k, \k^\prime, s, s^\prime}  \Delta^*_{s^\prime s}(\k) V^{-1}_{\k, \k^\prime} \Delta_{s^\prime s}(\k^\prime) 
\eeq
The gap-equation can be derived setting the variation the free energy with respect to the gap function to be zero.
\beq
\frac{\partial \mathcal{F}}{\partial ( \Delta^*_{s^\prime s}(\k))} &=& 0 \implies  \Delta_{s^\prime s}(\k) \left( \frac{1 - 2 n_F(\xi_+(\k))}{2 \xi_+(\k)} \right) - A \sum_{\k^\prime}V^{-1}_{\k, \k^\prime} \Delta_{s^\prime s}(\k^\prime) = 0 \nn
\implies \Delta_{s^\prime s}(\k) &=& - \frac{1}{A} \sum_{\k^\prime} V_{\k, \k^\prime} \left( \frac{\tanh(\beta \xi_+(\k^\prime)/2) }{2 \xi_+(\k^\prime)} \right) \Delta_{s^\prime s}(\k^\prime)
\eeq
For spin-singlet superconductors, we can write $\Delta(\k) = i s^y f_\k$, where $f_\k$ is a scalar. 
For spin-triplet superconductors, we can write  $\Delta(\k) = (\d \cdot \s) i s^y f_\k$, where $\d$ is a unit-vector.
In both cases, we find the following equation for the `orbital' part of the pair-wavefunction $f_\k$:
\beq
f_\k = -\frac{1}{A} \sum_{\k^\prime} V_{\k, \k^\prime} \left( \frac{\tanh(\beta \xi_+(\k^\prime)/2) }{2 \xi_+(\k^\prime)} \right) f_{\k^\prime}
\label{eq:fkIVCSM}
\eeq
Since $V_{\k, \k^\prime}$ is maximized near $\k + \k^\prime = 0$, we may focus on this parameter regime near the Fermi surface.
Since $V$ is repulsive, this means that the gap equation can be satisfied when $f_{\k^\prime = -\k} = - f_\k$, which indicates unconventional pairing.
The actual channel is determined by the IVC correlation length $\xi_{\rm IVC}$, for small $\xi_{\rm IVC}$ it is a nodal and non-chiral, while for large $\xi_{\rm IVC}$ it is gapped and chiral as explained in the main text.

As discussed in the main text, introducing a Hund's coupling will amplify either singlet and or triplet IVC fluctuations, and one important consequence of this is the splitting of the degeneracy between singlet and triplet superconductors. 
Therefore, we start with Eq.~(10) in the main text (reproduced below for convenience), where the effective IVC interaction Hamiltonian $H^{\rm eff}_{\rm IVC}$ was decomposed into singlet (CDW) and triplet (SDW) IVC fluctuations with distinct susceptibilities $g^{\rm S}_{\q}$ and $g^{\rm T}_{\q}$ respectively:
\beq
H^{\rm eff}_{\rm IVC} = - \frac{1}{2A} \sum_{\q} g^{\rm S}_{\q} \, \Tr\left[ n_{\rm S}^{\rm IV}(\q) [n_{\rm S}^{\rm IV}(\q)]^\dagger \right] - \frac{1}{2A} \sum_{\q} g^{\rm T}_{\q} \, \Tr\left[ n_{\rm T}^{\rm IV}(\q) [n_{\rm T}^{\rm IV}(\q)]^\dagger \right]
\label{eq:HIVC_TvSSM}
\eeq
Now, we derive the linearized gap equation for each of the two terms in Eq.~\eqref{eq:HIVC_TvSSM}, and then combine these to get the gap equation when both are present.
We follow the same procedure as described for the fully symmetric scenario, so we only outline the basic steps here.
For the singlet IVC channel, the BCS decoupling takes the following form:
\beq
-\frac{1}{2A} \sum_{\q} g^{\rm S}_{\q} \, \Tr\left[ n_{\rm S}^{\rm IV}(\q) [n_{\rm S}^{\rm IV}(\q)]^\dagger \right] \xrightarrow{\rm mean ~ field} && \sum_{\k,s,s^\prime} \Delta^*_{s^\prime s}(\k)  \psi_{-,s,-\k^\prime} \psi_{+,s^\prime, \k} + \Delta_{ss^\prime}(\k) \psi^\dagger_{+,s, \k} \psi^\dagger_{-,s^\prime,-\k^\prime} \nn 
&& - A \sum_{\k, \k^\prime, s, s^\prime}   \Delta^*_{s s^\prime}(\k) [V^{\rm S}_{\k, \k^\prime}]^{-1} \Delta_{s^\prime s}(\k^\prime), \nn
\text{ where } V^{\rm S}_{\k, \k^\prime} = \frac{1}{2} g^{\rm S}_{\q = - \k - \k^\prime} |\lambda^{+-}_{\q= - (\k + \k^\prime)}(\k)|^2 && \text{ and } \Delta_{ss^\prime}(\k) = \frac{1}{A} \sum_{\k} V^{\rm S}_{\k, \k^\prime} \langle \psi_{-,s,-\k^\prime} \psi_{+,s^\prime, \k} \rangle
\eeq
Therefore, the total mean-field Hamiltonian can be written as:
\beq
H_{MF} &=& H_{\rm BdG} - A \sum_{\k, \k^\prime, s, s^\prime}  \Delta^*_{ s^\prime s}(\k) [V^{\rm S}_{\k, \k^\prime}]^{-1} \Delta_{s s^\prime}(\k^\prime), \text{ where }  \nn H_{\rm BdG} &=& \sum_{\k, s, s^\prime} \begin{pmatrix}
\psi^\dagger_{+, s, \k} & \psi_{-,s^\prime, -\k}
\end{pmatrix} \begin{pmatrix}
\xi_{+,\k} & \Delta_{s s^\prime}(\k) \\ \Delta^*_{s s^\prime}(\k) & - \xi_-(-\k) 
\end{pmatrix} \begin{pmatrix}
\psi_{+, s, \k} \\ \psi^\dagger_{-,s^\prime, -\k} 
\end{pmatrix}
\eeq
Evaluating the free energy and minimizing it leads to:
\beq
\Delta_{s^\prime s}(\k) &=& - \frac{1}{A} \sum_{\k^\prime} V_{\k, \k^\prime} \left( \frac{\tanh(\beta \xi_+(\k^\prime)/2) }{2 \xi_+(\k^\prime)} \right) \Delta_{s s^\prime}(\k^\prime)
\eeq
In turn, this implies the following equation for the spatial wave-functions of spin-singlet and spin-triplet superconductors:
\beq
f_\k = \begin{cases} \frac{1}{A} \sum_{\k^\prime} V^{\rm S}_{\k, \k^\prime} \left( \frac{\tanh(\beta \xi_+(\k^\prime)/2) }{2 \xi_+(\k^\prime)} \right) f_{\k^\prime}, \text{ for spin-singlets: }\Delta(\k) = i s^y f_\k \\ -\frac{1}{A} \sum_{\k^\prime} V^{\rm S}_{\k, \k^\prime} \left( \frac{\tanh(\beta \xi_+(\k^\prime)/2) }{2 \xi_+(\k^\prime)} \right) f_{\k^\prime}, \text{ for spin-triplets: }\Delta(\k) =  (\d \cdot \s) i s^y f_\k \end{cases}
\eeq
Since $V^{\rm S}_{\k, \k^\prime}$ is always positive by definition, this means that for spin-singlets, a non-nodal s-wave solution (where $f_\k$ is a constant independent of $\k$) is allowed when only CDW fluctuations are present. 
One might expect that this solution corresponds to the highest $T_c$ for a given $\xi_{\rm IVC}$, as the order-parameter is not required to modulate in real-space - this is also seen in our numerical results.
For spin-triplets we can again focus on the limit near $\k + \k^\prime = 0$, since this is where $V^{\rm S}_{\k, \k^\prime}$ is maximized. 
Since $V^{\rm S}$ is repulsive in the triplet Cooper channel, this means that the gap equation can be satisfied when the gap-function changes sign, i.e, $f_{\k^\prime = -\k} = - f_\k$, indicating unconventional pairing.
However, these channels will typically have lower $T_c$ than the spin-singlet s-wave channel discussed above, as the order parameter modulates in real space.

For the triplet IVC channel, the BCS decoupling takes the following form:
\beq
 - \frac{1}{2A} \sum_{\q} g^{\rm T}_{\q} \, \Tr\left[ n_{\rm T}^{\rm IV}(\q) [n_{\rm T}^{\rm IV}(\q)]^\dagger \right] \xrightarrow{\rm mean ~ field} && \sum_{\k,s,s^\prime} \left[(s^i)^T\Delta^*(\k) s^i \right]_{s^\prime s}  \psi_{-,s,-\k^\prime} \psi_{+,s^\prime, \k} +\left[ s^i \Delta(\k) (s^i)^T \right]_{ss^\prime} \psi^\dagger_{+,s, \k} \psi^\dagger_{-,s^\prime,-\k^\prime} \nn
&&  - A \sum_{\k, \k^\prime} [V^{\rm T}_{\k, \k^\prime}]^{-1} \Tr\left[ (s^i)^T  \Delta^*(\k) s^i \Delta(\k^\prime) \right]  \nn
\text{ where } V^{\rm T}_{\k, \k^\prime} = \frac{1}{2} g^{\rm T}_{\q = - \k - \k^\prime} |\lambda^{+-}_{\q= - (\k + \k^\prime)}(\k)|^2 && \text{ and } \Delta_{ss^\prime}(\k) = \frac{1}{A} \sum_{\k} V^{\rm T}_{\k, \k^\prime} \langle \psi_{-,s,-\k^\prime} \psi_{+,s^\prime, \k} \rangle
\eeq
Therefore, the total mean-field Hamiltonian can be written as:
\beq
H_{MF} &=& H_{\rm BdG} -  A \sum_{\k, \k^\prime} [V^{\rm T}_{\k, \k^\prime}]^{-1} \Tr\left[ (s^i)^T  \Delta^*(\k) s^i \Delta(\k^\prime) \right], \text{ where }  \nn H_{\rm BdG} &=& \sum_{\k, s, s^\prime} \begin{pmatrix}
\psi^\dagger_{+, s, \k} & \psi_{-,s^\prime, -\k}
\end{pmatrix} \begin{pmatrix}
\xi_{+,\k} & \left[s^i  \Delta(\k)(s^i)^T \right]_{s s^\prime} \\ \left[(s^i)^T\Delta^*(\k) s^i \right]_{s s^\prime} & - \xi_-(-\k)
\end{pmatrix} \begin{pmatrix}
\psi_{+, s, \k} \\ \psi^\dagger_{-,s^\prime, -\k} 
\end{pmatrix}
\eeq
and summation on $i = x,y,z$ is implied.
Evaluating the free energy and minimizing it leads to:
\beq
\Delta_{s^\prime s}(\k) &=& - \frac{1}{A} \sum_{\k^\prime} V^{\rm T}_{\k, \k^\prime} \left( \frac{\tanh(\beta \xi_+(\k^\prime)/2) }{2 \xi_+(\k^\prime)} \right) \left[s^i  \Delta^{\rm T}(\k^\prime)(s^i)^T \right]_{s s^\prime}
\eeq
In turn, this implies the following equation for the spatial wave-functions of spin-singlet and spin-triplet superconductors:
\beq
f_\k = \begin{cases} -\frac{3}{A} \sum_{\k^\prime} V^{\rm T}_{\k, \k^\prime} \left( \frac{\tanh(\beta \xi_+(\k^\prime)/2) }{2 \xi_+(\k^\prime)} \right) f_{\k^\prime}, \text{ for spin-singlets: }\Delta(\k) = i s^y f_\k \\ -\frac{1}{A} \sum_{\k^\prime} V^{\rm T}_{\k, \k^\prime} \left( \frac{\tanh(\beta \xi_+(\k^\prime)/2) }{2 \xi_+(\k^\prime)} \right) f_{\k^\prime}, \text{ for spin-triplets: }\Delta(\k) =  (\d \cdot \s) i s^y f_\k \end{cases}
\eeq
We see that the interaction $V^{\rm T}$ is repulsive both in the spin-singlet and spin-triplet channels, hence a non-nodal s-wave superconductor is not favored.
However, pair-correlation with $f_{\k^\prime = -\k} = - f_\k$, i.e, unconventional superconductivity will be favored in both cases, but because of the additional factor of $3$ coming from the spin-trace, $T_c$ will be larger for the spin-singlets. 
Hence, the preferred channel due to SDW fluctuations is either a gapped chiral, or a nodal non-chiral, spin-singlet superconductor.

Combining the results from the analysis of CDW and SDW fluctuations, we find that when both are present, $f_\k$ is given by the self-consistent solution of the following equations.
\beq
f_\k = \begin{cases} -\frac{1}{2A} \sum_{\k^\prime} (3 g^{\rm T}_{\q = - \k - \k^\prime} - g^{\rm S}_{\q = - \k - \k^\prime} ) |\lambda^{+-}_{\q = - \k - \k^\prime}(\k)|^2 \left( \frac{\tanh(\beta \xi_+(\k^\prime)/2) }{2 \xi_+(\k^\prime)} \right) f_{\k^\prime}, \text{ for spin-singlets: }\Delta(\k) = i s^y f_\k \\ -\frac{1}{2A} \sum_{\k^\prime} (g^{\rm T}_{\q = - \k - \k^\prime} + g^{\rm S}_{\q = - \k - \k^\prime} ) |\lambda^{+-}_{\q = - \k - \k^\prime}(\k)|^2 \left( \frac{\tanh(\beta \xi_+(\k^\prime)/2) }{2 \xi_+(\k^\prime)} \right) f_{\k^\prime}, \text{ for spin-triplets: }\Delta(\k) =  (\d \cdot \s) i s^y f_\k \end{cases} ~~~
\label{eq:fkSC}
\eeq
Eq.~\eqref{eq:fkSC} constitute the generalizations of the mean-field analysis in Eq.~(11) in the main text, and identical conclusions regarding the preferred superconducting channels are reached from both sets of equations.

To end this section, we carry out an identical analysis for the gate-screened SU(2)$_+ \times$ SU(2)$_-$-symmetric Coulomb repulsion $V_C(\q)$, which we recall below.
\beq
H_C = \frac{1}{2A} \sum V_{C}(|\q|) : \rho(\q)  ~\rho(-\q) :  
\eeq
The BCS decoupling takes the following form:
\beq
H_{C} \xrightarrow{\q = \k - \k^\prime} \frac{1}{A} \sum_{\k, \k^\prime} V^c_{\k, \k^\prime} \psi^\dagger_{+,s, \k} \psi^\dagger_{-,s^\prime, -\k} \psi_{-,s^\prime,-\k^\prime} \psi_{+, s, \k^\prime}, \text{ with } V^c_{\k, \k^\prime} = |\lambda^{++}_{\q= \k^\prime - \k)}(\k)|^2 V_C(|\q = \k^\prime - \k|) ~~~~~~
\label{eq:HCBCS}
\eeq
The structure of Eq.~\eqref{eq:HCBCS} is identical to that of Eq.~\eqref{eq:HIVCBCS}, with $V_{\k,\k^\prime} \to V^c_{\k, \k^\prime}$.
Therefore, the mean-field decoupling amd BdG Hamiltonian also follow Eqs.~\eqref{eq:MFIVCsym} and \eqref{eq:MFHBDGsym} respectively, with the same replacement.
In turn, the gap equation leads to the following condition for the spatial wave-functions for both spin-singlet and spin-triplet superconductors, which are degenerate due to the SU(2)$_+ \times$ SU(2)$_-$ symmetry of the interaction.
\beq
f_\k = -\frac{1}{A} \sum_{\k^\prime} V^{c}_{\k, \k^\prime} \left( \frac{\tanh(\beta \xi_+(\k^\prime)/2) }{2 \xi_+(\k^\prime)} \right) f_{\k^\prime}
\eeq
Since $V^{c}_{\k, \k^\prime}$ is maximized at $\k = \k^\prime$ (corresponding to $\q = 0$), we focus on this limit and note that generally such a gap equation does not have a solution, at least for a single Fermi surface.
However, in presence of multiple Fermi surfaces (eg: corresponding to an annular Fermi sea) and/or in presence of other mechanisms for attractive pairing (eg: IVC fluctuations), there may exist a solution which have opposite signs on the inner and outer rings.
In practice, we find that this is indeed the case when both IVC fluctuations and RPA-screened Coulomb repulsion ($V_C \to V_{\rm RPA}$, see Eq.~\ref{eq:HRPASM}) are considered in tandem, as was done for the plots in Fig.~3(a,e) in the main text.

\subsection{Estimate of the coupling strength $g$}
In this section, we provide a rough estimate the phenomenological coupling strength $g$ which appeared in the effective IVC Hamiltonian in Eq.~(6) in the main text.
For this purpose, we start with the long-range Coulomb interaction $H_C$ and consider random phase approximation (RPA) corrections to it by itinerant fermions, as described in Eq.~\eqref{eq:HRPASM}.
We now decouple it into the IVC channel, since strong IVC fluctuations constitute the mechanism of superconductivity in our picture:
\beq
H_{C} = \frac{1}{2A} \sum_\q V_{\rm RPA}(\q) :\rho(\q) \rho(-\q):  \xrightarrow{\text{decoupled to IVC channel}}  = -\frac{1}{2A} \sum_\q V_{\rm RPA}(\q)  \Tr\left[ n^{\rm IV}(\q) [n^{\rm IV}(\q)]^\dagger \right] + \cdots
\eeq
and the $\cdots$ denote terms which involve Coulomb repulsion between electrons in the same valley, and are expected to be unimportant close to a phase transition to an IVC phase. 
Now we assume that the bare interaction is sufficiently gate-screened so that we can approximate $V_C(\q) \approx V_0$, and consequently the RPA screened interaction takes the following form (setting the O(1) constant $c=1$):
\beq
V_{\rm RPA}(\q)= \frac{V_0}{1 + V_0 \chi_{\rho \rho}(\q)} \approx \frac{V_0}{1 - V_0 \, \chi_0(1 - \q^2/k_F^2)} = \frac{k_F^2/\chi_0}{\frac{(1 - V_0 \chi_0)k_F^2}{V_0 \chi_0} + \q^2} \equiv \frac{k_F^2/\chi_0}{\q^2 + \xi_{\rm{IVC}}^{-2}}
\label{eq:V_RPA}
\eeq
where $\chi_0 = - \chi_{\rho \rho}(\bm{0}) \approx \partial_\mu n_e$ is approximately the density of states at the Fermi surface(s), and  $\xi_{\rm{IVC}}^{-2} \equiv [(V_0 \chi_0)^{-1} - 1]k_F^2$ is the squared inverse correlation length set by the distance from the IVC instability that sets in at $V_0 \chi_0 = 1$.
Noting the the Fermi momentum $k_F$ is set by the density, i.e, $k_F \sim \sqrt{n_e}$, we conclude that when IVC fluctuations are large we can approximate:
\beq
H_{C} \approx H^{\rm eff}_{\rm IVC} + \cdots, \text{ where }  H^{\rm eff}_{\rm IVC} = - \frac{1}{A} \sum_\q \frac{n_e/\chi_0}{\q^2 + \xi_{\rm{IVC}}^{-2}}  \Tr\left[ n^{\rm IV}(\q) [n^{\rm IV}(\q)]^\dagger \right]
\eeq
whereby we identify $g = n_e/\chi_0$.

As mentioned in the main text, our analysis does not take into account the frequency dependence of the effective interaction \eqref{eq:V_RPA} and the damping of the electrons by IVC fluctuations. These effects become important sufficiently close to the critical point \cite{abanov2001coherent} (i.e., for sufficiently large $k_F\xi_{\rm{IVC}}$), and our treatment is expected to break down. We leave a full treatment of pairing in the quantum critical regime to future work.

\subsection{Estimate of the IVC correlation length $\xi_{IVC}$} 
\label{sec:xi}

\begin{figure}[htbp]
    \centering
    \includegraphics[width = 0.33\textwidth]{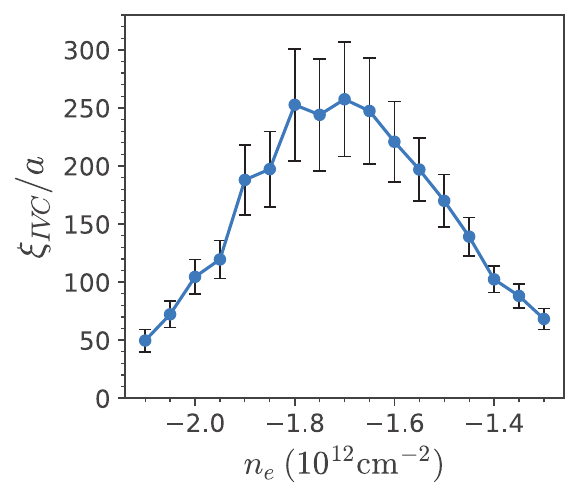}
    \caption{IVC correlation length $\xi_{\rm IVC}$ obtained from self-consistent HF wavefunction using $\chi_0 = \SI{0.16}{eV^{-1}}$ per unit cell, at $u = 30$ meV and $\epsilon = 4.4$. The error bars indicate the least square error obtained from the fitting procedure. $\xi_{\rm IVC}/a$ remains around $100$ for a significant range of carrier density $n_e$ and peaks around $n_e = \SI{-1.7E12}{\cm^{-2}}$.}
    \label{fig:Xi}
\end{figure}

In this section, we numerically calculate the IVC correlation length $\xi_{IVC}$ using the Slater determinant wavefunction obtained in self-consistent Hartree-Fock calculations.
We first prepare a fully symmetric Slater determinant ground state $\ket{\phi_{sym}}$ using self-consistent HF on a $151 \times 151$ momentum grid with $u = 30$ meV, $\epsilon = 4.4$, and $\chi_0 = \SI{0.16}{eV^{-1}}$ per unit cell.
There exists a gapped pseudo-Goldstone mode $\ket{\phi^x_{\mathbf{q}}}$ that corresponds to the breaking of valley U(1)$_v$ symmetry,
\begin{equation}
    \ket{\phi^x_{\mathbf{q}}} = \left (\sum_{\mathbf{k}} c_{+, \mathbf{k}}^{\dagger} c_{-, \mathbf{k+q}} + \mathrm{h.c.} \right ) \ket{\phi_{sym}}
\end{equation}
We compute the dispersion of this mode above the ground state energy using the single-mode approximation:
\begin{equation}
    E_{\mathbf{q}} = \frac{\mel{\phi^x_{\mathbf{q}}}{H}{\phi^x_{\mathbf{q}}}}{\braket{\phi^x_{\mathbf{q}}}} - \mel{\phi_{sym}}{H}{\phi_{sym}}
\end{equation}
where $H$ is the full interacting Hamiltonian. At small momentum, we expect the dispersion to take the form of $E_{\mathbf{q}} = v \sqrt{\mathbf{q}^2+\xi_{\rm IVC}^{-2}}$, where $v$ is the IVC mode velocity, and $\xi^{-1}_{\rm IVC}$ sets the `mass gap' of IVC fluctuations. 
At a second order phase transition, $\xi_{\rm IVC}$ diverges as the fluctuations become truly gapless. However, they can also be nearly gapless for a weakly first order transition, indicated by a very large $\xi_{\rm IVC}$ (relative to lattice spacing $a$).
In practice, we perform all calculations in terms of the projector $P_{\tau,\tau'}(\mathbf{k})$ (see Ref.~\onlinecite{SoftMode} for details) and include the first order symmetry-allowed lattice harmonic in $E_{\mathbf{q}}$ for better fitting, i.e., we fit $E_{\mathbf{q}} =  v \sqrt{\mathbf{q}^2+ \eta E_6(\mathbf{q}) + \xi_{IVC}^{-2}}$, where the lattice harmonic  $E_6(\mathbf{q}) = \Re [(q_x + i q_y)^6]/(q_x^2 + q_y^2)^3 + 1$ is a bounded positive function of $q$ independent of its magnitude. 
Typically, we find $v \approx 10^{4}$ m/s and $\eta \sim 10^{-5}/a^2$, indicating that lattice effects are quite small.  
As a sanity check, we have also confirmed that the actual Goldstone mode $\ket{\phi^z_{\mathbf{q}}} = \sum_{\tau, \mathbf{k}} (-1)^{\tau} c_{\tau, \mathbf{k}}^{\dagger} c_{\tau, \mathbf{k+q}} \ket{\phi_{IVC}}$ is always gapless with linear dispersion once we are within the IVC phase.

Our results are summarized in Supplementary Figure~\ref{fig:Xi}. While the single mode approximation is not expected to be particularly accurate very close to the transition, we nevertheless find that the IVC correlation length $\xi_{IVC}$ remains around $100 a$ in a significant range of hole-density. Based on our numerical solution to the gap equation, the superconducting transition temperature $T_c$ will be around $\SI{0.1}-\SI{1}{\K}$ in such filling range (see Fig.~3(e) in the main text), consistent with experimental observations. 

\subsection{Impact of Hund's coupling on superconductivity}
The Hund's coupling term in Eq.~\eqref{eq:HHundsSM} amplifies certain IVC fluctuations, and thus can aid certain pairing channel.
In this section, we show that a spatially local Hund's coupling does not affect p/f-wave superconductors, but non-local Hund's coupling can indeed aid certain pairing channels in the same manner as IVC fluctuations.
To do so, we start by decoupling the Hund's term in the superconducting channel, to see which channel it aids among the ones induced by IVC fluctuations. 
\beq
\langle  H_{\rm Hund's} \rangle &=& \frac{1}{A} \sum_{\k, \k^\prime} J_H(\q = -\k - \k^\prime) |\lambda^{+-}_{\q = -\k - \k^\prime}(\k)|^2 \Tr[F^*(\k) s^i F(\k^\prime) (s^i)^T]
\label{eq:HHundsOnSC}
\eeq
If the Hund's coupling was local in real space, and the projected form-factor $\lambda^{+-}_\q(\k)$ was featureles (eg: as in the sublattice-polarized case), these would be approximately independent of $\q$.  
In such a scenario, the Hund's coupling cannot affect the degeneracy between spin-singlet and triplet, as the orbital wavefunction (in the preferred channels from IVC fluctuations) vanishes when the two electrons in the Cooper pair approach the same point in space.  
This can be explicitly seen by setting $F(\k) = i s^y f_\k$ (spin-singlet) or $F_\k = i s^y (\d \cdot \s) f_\k$ (spin-triplet) in Eq.~\eqref{eq:HHundsOnSC} implies that $H_{\rm Hund's} \propto J_H |\sum_\k f_\k|^2$, which evaluates to zero as long as $f_{-\k} = - f_{\k}$.

However, in reality the Hund's term need not be so local, and the valence-band wave-functions of the electrons are also extended in real space. 
So we may consider the opposite limit of $\q = 0$ in $J_H(\q)$:
\beq
\Tr[F^\dagger(\k) s^i F(\k) (s^i)^T] = \begin{cases}  6 f_\k^* f_{-\k}, \text{ if } F_\k = i s^y f_\k \\  2 f_\k^* f_{-\k}, \text{ if } F_\k = i s^y (\d \cdot \s) f_\k \end{cases}
\eeq
So for ferromagnetic Hund's ($J_H > 0$) which prefers the SDW IVC, chosen states are  spin-singlets. 
This is not surprising, as $H_{\rm Hund's}$ and $H_{IVC}$ are basically identical near $\q  = \mathbf{0}$ upto constant pre-factors. 
For AF Hund's ($J_H < 0$) which prefers CDW IVC, will choose a non-nodal spin-singlet s-wave state just by itself. 
However, when IVC fluctuations are near-critical, AF Hund's will instead perturbatively prefer a spin-triplet state by amplifying CDW fluctuations, as discussed in the main text.

Next, we consider the spin-polarized case, we have (with $\d = (1,i,0)/\sqrt{2}$):
\beq
\langle  H_{\rm Hund's} \rangle = \frac{1}{A} \sum_{\k, \k^\prime} J_H(\q = - \k - \k^\prime) |\lambda^{+-}_{\q =  - \k - \k^\prime}(\k)|^2 f_\k^* f_{\k^\prime}
\eeq
Analyzing the $\q = \mathbf{0}$ scenario again prefers a gapped chiral or a nodal non-chiral superconductor.

Finally, we comment on the other symmetry allowed Hund's coupling term $\tilde{J}_H$ (as in Eq.~\eqref{eq:HHunds2SM}), although we believe its magnitude is quite small.
This kind of Hund's term has been considered in the context of superconductivity in twisted bilayer graphene \cite{YV2019,Dodaro}.
Such a term can also be decomposed into the pairing channel, but its effects are quite different from $J_H$.
\beq
\langle  \tilde{H}_{\rm Hund's} \rangle &=&- \frac{1}{A} \sum_{\k, \k^\prime} \tilde{J}_H(\q = \k^\prime -\k) |\lambda^{++}_{\q = \k^\prime - \k}(\k)|^2 \Tr[F^\dagger(\k) s^i F(\k^\prime) (s^i)^T]
\eeq
Once again, a local Hund's term cannot affect a pairing wave-function which is spatially antisymmetric. 
However, it can affect a pairing wave-function that is spatially symmetric.
Considering the $\q \to 0$ limit as before, we see:
\beq
\Tr[F^\dagger(\k) s^i F(\k) (s^i)^T] = \begin{cases} - 6 f_\k^* f_\k, \text{ if } F_\k = i s^y f_\k \\  2 f_\k^* f_\k, \text{ if } F_\k = i s^y (\d \cdot \s) f_\k \end{cases}
\eeq
Thus, such a Hund's term will always prefer $f_{-\k} = f_\k$, i.e, s-wave pairing (as it has the least number of nodes).
For ferromagnetic Hund's ($\tilde{J}_H > 0$, weakly prefers the triplet or SDW IVC), $\tilde{H}_{\rm Hund's}$ prefers spin-triplet superconductivity.
For AF Hund's ($\tilde{J}_H < 0$, weakly prefers the singlet or CDW IVC), $\tilde{H}_{\rm Hund's}$ prefers spin-singlet superconductivity.
Finally, FM Hund's $\tilde{J}_H > 0$ can also spin-polarize the system.
In this case, pairing will be mediated between the same spin species (assumed up-spin) with $F(\k) = f_\k \delta_{s,\ua} \delta_{s^\prime,\ua}$ we find that:
\beq
\Tr[F^\dagger(\k) s^i F(\k) (s^i)^T] = 2 f^*_\k f_\k
\eeq
and the resultant superconductivity is a non-unitary s-wave spin-triplet that is aided by this kind of Hund's coupling if only $\tilde{J}_H > 0$ is present.

Thus, in conclusion, ferromagnetic (FM) Hund's ($J_H > 0$) which prefers the spin-triplet SDW IVC, will also choose chiral or nodal superconducting states, just like the IVC fluctuations, but prefer spin-singlets over triplets. 
AF Hund's ($J_H < 0$) which prefers spin-singlet CDW IVC, will perturbatively choose spin-triplet chiral or nodal states when it slightly amplifies CDW flucutations, with a crossover to a spin-singlet non-nodal s-wave states when it is the only term present when it largely magnifies SDW fluctuations.
Finally the spin-polarized IVC is favored by FM Hund's, and thus Hund's would perturbatively favor the spin-polarized chiral or nodal superconductor that arises from spin-polarized IVC fluctuations. 
For the other kind of Hund's term with potentially small $\tilde{J}_H$, the major difference comes for the SDW IVC, where FM $\tilde{J}_H > 0$ picks a conventional s-wave spin-triplet, while FM $J_H > 0$ would pick an unconventional chiral or nodal spin-singlet.

\end{document}